\documentclass[symmetry,article,accept,pdftex,oneauthor]{Definitions/mdpi} 

\firstpage{1} 
\pubvolume{1}
\issuenum{1}
\articlenumber{0}
\pubyear{2024}
\copyrightyear{2024}
\datereceived{ } 
\daterevised{ } 
\dateaccepted{ } 
\datepublished{ } 
\hreflink{https://doi.org/} 

\usepackage{bm}
\usepackage[cal=cm,scr=dutchcal]{mathalfa}
\usepackage{accents}
\usepackage{cancel}
\DeclareFontEncoding{LGR}{}{}
\DeclareSymbolFont{sfgreek}{LGR}{cmss}{m}{n}
\DeclareMathSymbol{\sfbeta}{\mathord}{sfgreek}{`b}
\DeclareMathSymbol{\sfgamma}{\mathord}{sfgreek}{`g}
\DeclareMathSymbol{\sfeta}{\mathord}{sfgreek}{`h}
\DeclareMathSymbol{\sfxi}{\mathord}{sfgreek}{`x}
\DeclareMathSymbol{\sfpi}{\mathord}{sfgreek}{`p}
\DeclareMathSymbol{\sftau}{\mathord}{sfgreek}{`t}
\DeclareMathSymbol{\sfphi}{\mathord}{sfgreek}{`f}
\DeclareMathSymbol{\sfchi}{\mathord}{sfgreek}{`q}
\DeclareMathSymbol{\sfpsi}{\mathord}{sfgreek}{`y}
\DeclareMathSymbol{\sfomega}{\mathord}{sfgreek}{`w}
\DeclareMathSymbol{\sfPhi}{\mathord}{sfgreek}{`F}

\usepackage[eulergreek]{sansmath}
\newcommand{\mymathsf}[1]{\mbox{\sansmath$\mathsf{#1}$}}

\providecommand{\ADD}[1]{{\protect\color{black} #1}}

\Title{A unified perspective on Poincar\'e and Galilei relativity: \\ I. Special relativity}

\TitleCitation{A unified perspective on Poincar\'e and Galilei relativity: I. Special relativity}

\Author{Christian Y. Cardall $^{1,\dagger}$\orcidA{}}

\AuthorNames{Christian Y. Cardall}

\AuthorCitation{Cardall, C. Y.}

\address{%
$^{1}$ \quad Physics Division, Oak Ridge National Laboratory, Oak Ridge, TN 37831-6354, USA; cardallcy@ornl.gov}

\corres{cardallcy@ornl.gov}

\firstnote{This manuscript has been authored by UT-Battelle, LLC, under contract DE-AC05-00OR22725 with the US Department of Energy (DOE). The US government retains and the publisher, by accepting the article for publication, acknowledges that the US government retains a nonexclusive, paid-up, irrevocable, worldwide license to publish or reproduce the published form of this manuscript, or allow others to do so, for US government purposes. DOE will provide public access to these results of federally sponsored research in accordance with the DOE Public Access Plan (\texttt{http://energy.gov/downloads/doe-public-access-plan}).}

\abstract{A semantic adjustment to what physicists mean by the terms `special relativity' and `general relativity' is suggested, which prompts a conceptual shift to a more unified perspective on physics governed by the Poincar\'e group and physics governed by the Galilei group.
After exploring the limits of a unified perspective available in the setting of 4-dimensional spacetime, a particular central extension of the Poincar\'e group---analogous to the Bargmann group that is a central extension of the Galilei group---is presented that deepens a unified perspective on Poincar\'e and Galilei physics in a 5-dimensional spacetime setting.
The immediate focus of this paper is classical physics on affine 4-dimensional and 5-dimensional spacetimes (`special relativity' as redefined here), including the electrodynamics that gave rise to Poincar\'e physics in the first place; but the results here may suggest the existence of a `Galilei general relativity' more extensive than generally known, to be pursued in the sequel.
}


\begin{document}

\section{Introduction}
\label{sec:Introduction}

In present common usage the terms `relativistic physics' and `non-relativistic physics' refer, at least roughly, to what might be called something else---perhaps `Poincar\'e physics' and `Galilei physics' respectively.
These latter terms are intended here as shorthand for `physics governed by the Poincar\'e group' and `physics governed by the Galilei group'. 
The motivation behind such a change in nomenclature, were it socially feasible, is that the essential innovation of what is commonly called `relativistic physics' is the relativity of time, or more precisely, the relativity of simultaneity.
In terms of space, so-called `non-relativistic physics' is in an important sense just as relativistic as `relativistic physics'.

Absolute time---the notion that distinct events either do or do not occur at the same instant---is intuitive to ordinary human experience. 
This is manifest in the everyday expectation that if two observers carrying synchronized clocks each become aware of two distinct events, their clock readings will agree on whether the events occurred at the same time (after correction for finite travel time of light and sound from event to observer), regardless of the events' separation in space or the relative motion of the observers.
Einstein achieves the reconciliation of classical electromagnetism with classical mechanics only by abolishing this intuition; such is the cost of promoting the Maxwell equations, with their specification of the speed of light, to the status of physical law valid for all inertial observers.

True, a consequence of Einstein's theory is that observers in relative motion disagree about intervals in both time and space (`time dilation' and `length contraction' respectively), while in Newton's world all observers agree on both durations measured by ideal clocks and the lengths of straight rulers, regardless of the motion of these clocks and rulers.   
This is presumably a major part of the rationale for nomenclature distinguishing between `relativistic' and `non-relativistic' theories.
But the `relativistic' phenomenon of length contraction is secondary or derivative, in the sense that in both Einstein's and Newton's worlds measurements of length must occur at a single instant, and therefore depend on a notion of simultaneity; while in Newton's world no analogous specification---that measurements occur at the same place---is necessary in order to define measurement of time duration.

That measurements of lengths of objects depend upon a notion of simultaneity is a reminder that, unlike absolute time, absolute space is not as intuitive to ordinary human experience, at least upon reflection. 
(Apparently absolute space \textit{did} seem obvious to Aristotle, and his later followers who dogmatically rejected Copernicus.)
This is manifest in ready acceptance of the fact that if two observers carry triads of mutually orthogonal rulers in matching orientation, the answer to the question of whether two events occur in the `same place' depends very much on both the time interval between the events and the relative motion of the observers.
(Here `place' means the 3-tuple of distances in each of the three dimensions of ordinary space between the events and the origins of the observers' triads.)
This translates into what is appropriately called the `Galilei relativity' of the physics of the everyday human world, made persuasive---to the intelligent person of ordinary experience---by Galilei's thought experiment about the inability to detect uniform motion of a ship based on observation of one's immediate surroundings in a sealed cabin below decks (e.g. \cite{Penrose2004The-Road-to-Rea}). 

That the relativity of time, rather than of space, is the crucial thing distinguishing Einstein circa 1905 from Newton is manifest in the celebrated equations relating the time intervals $\Delta t$, $\Delta t'$ and rectangular space intervals $(\Delta x, \Delta y, \Delta z)$, $(\Delta x', \Delta y', \Delta z')$ between two events measured by two different observers $\mathscr{O}$ and $\mathscr{O}'$. 
Consider events sufficiently close in time and space that the two observers detect no linear acceleration (via spectral shifts) or rotation (via gyroscopes) in their own motion while they observe the two events \cite{Gourgoulhon2013Special-Relativ}.
Orient the observers' $x$-axes along the direction of the observers' uniform relative motion, with $\mathscr{O}'$ moving with velocity $v$ as measured by $\mathscr{O}$, and align their $y$- and $z$-axes.
In Newton's world,
\begin{align*}
\Delta t' & = \Delta t, \\
\Delta x' & = \Delta x - v \, \Delta t, \\
\Delta y' & = \Delta y, \\
\Delta z' & = \Delta z. 
\end{align*}
In Einstein's world,
\begin{align*}
\Delta t' & = \Lambda \left( \Delta t - \frac{v}{c^2} \, \Delta x \right),  \\
\Delta x' & = \Lambda \left( \Delta x - v \, \Delta t \right ),  \\
\Delta y' & = \Delta y,  \\
\Delta z' & = \Delta z, 
\end{align*}
where $c$ is the speed of light and $\Lambda = (1 - v^2 / c^2 )^{-1/2}$.
Newton agrees with Einstein as $v/c \rightarrow 0$.
These transformations come from the homogeneous Galilei group on the one hand and the Lorentz group on the other; these are the linear parts of the Galilei and and Poincar\'e groups respectively, which include translations of the origin that cancel upon taking coordinate differences.
These groups are appropriately named for consequential predecessors of the great synthesists Newton and Einstein.

Newton, while agreeing with Copernicus and Galilei, postulates absolute space in addition to absolute time, in order to formulate his theory given the mathematics of his era, and because of thought experiments about rotation (rotating pails of water, objects joined by a rope rotating about their center of mass) which raise questions not addressed by Galilei. 
Thus overturning both absolutes in a `relativistic' theory comes to be associated with Einstein.
And of course the overthrow of an `absolute space' in the sense of a luminiferous aether as the medium in which light propagates also contributes to a sense that it is Einstein who relativizes space as well as time. 

But it is apparent from the above transformations that the starkest difference is that time intervals are mixed into space intervals in both cases, while space intervals are mixed into time intervals only for Einstein.
Thus the presence of `relativity' in both cases---albeit space only in one case, and both space and time in the other---justifies more careful reference to, for instance, `Galilei relativity' and `Poincar\'e relativity', or `Galilei physics' and `Poincar\'e physics', instead of `non-relativistic physics' and `relativistic physics'.

In the measurement of time and space intervals contemplated above, the stipulation of non-accelerated and non-rotating (i.e. `inertial') observers points toward acknowledgment of another entrenched but unfortunate linguistic fossil: the use of the terms `special relativity' and `general relativity' to distinguish Einstein circa 1905 from Einstein circa 1915. 
Having freed the term `relativity' from specific attachment to the world according to Einstein and recognizing its relevance to the world according to Newton, the terms `special relativity' and `general relativity' must be reconsidered as well. 
The main difference between Einstein circa 1905 and Einstein circa 1915, expressed in terms of the 4-dimensional `spacetime' Minkowski introduces between them, is that the spacetime of Einstein circa 1905 is an affine space, which is flat, while the spacetime of Einstein circa 1915 is a more general pseudo-Riemann manifold whose curvature is determined by the energy and momentum of matter and radiation upon it (and indeed by the gravitational field itself, Einstein's gravitation being a nonlinear theory).
The latter is of course what enabled Einstein to accommodate gravity within a framework governed by the Poincar\'e group, with spectacular observational support from astrophysics and cosmology.

This distinction---between flat and curved spacetime, the latter with curvature determined by the presence of matter and radiation---is what ought to be meant by the terms `special relativity' and `general relativity', without regard for whether the physics is governed by the Poincar\'e group or the Galilei group.
In this perspective the key difference is not between `relativistic physics'---whether `special' or `general'---governed by the Poincar\'e group on the one hand, and `non-relativistic physics' governed by the Galilei group on the other.
Instead, what distinguishes `special relativity' from `general relativity' is whether the group in question---whether Poincar\'e, or Galilei---applies to spacetime \textit{globally}, in which case it is an affine space; or only \textit{locally}, in which case its curvature is determined by its energy and momentum content. 
The proper references, then, would be to `Poincar\'e special relativity' and `Poincar\'e general relativity', and to `Galilei special relativity' and `Galilei general relativity'.

The purpose of this sequence of two papers is to demonstrate that these semantic shifts, unavoidably associated also with conceptual shifts, point toward a unified perspective on Poincar\'e and Galilei physics that may bear fruit in a Galilei general relativity more extensive than generally known, a possibility that has been briefly reported previously \cite{Cardall2023Towards-full-Ga}.
This first paper focuses on `special relativity' as redefined above, that is, physics on spacetimes that are affine spaces (and therefore flat manifolds)---both `Poincar\'e special relativity' and `Galilei special relativity'.
The physics to be addressed in this first paper includes the motion of material particles and electrodynamics, the latter being tied historically both to the emergence of the Poincar\'e group and to the very notion of spacetime.
The second paper will focus on `general relativity' as redefined above, that is, physics on spacetimes that are curved manifolds---both `Poincar\'e general relativity' and `Galilei general  relativity'.
The physics to be addressed in the second paper includes the motion of material continua and gravitation, the latter being tied historically to the reconciliation of the Poincar\'e group with gravitational physics by encoding the latter in the curvature of spacetime. 
Only classical (which here means `non-quantum', not `non-Poincar\'e') physics will be considered.

\begin{figure}
\includegraphics[width=5in]{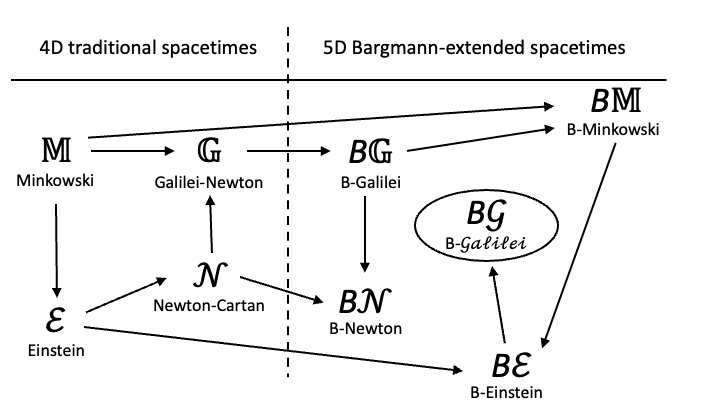}
\caption{Roadmap suggesting historical and/or logical connections between several spacetimes addressed by (or at least mentioned in) this sequence of two papers.
The prefix $B$ indicates a 5-dimensional Bargmann extension of a traditional 4-dimensional spacetime.
The spacetimes $\mathbb{M}$, $\mathbb{G}$, $B\mathbb{G}$ and $B\mathbb{M}$ are affine spaces (flat manifolds) addressed in this first paper focusing on `special relativity'.
The other spacetimes are curved manifolds; a primary goal of the second paper focusing on `general relativity' is the development of a conjectured `Galilei general relativity' more extensive than generally known, labeled $B\mathcal{G}$. 
\label{fig:SpacetimeRoadmap}}
\end{figure}

A conceptual roadmap to the spacetimes addressed by (or at least mentioned in) this sequence of two papers is given in Fig.~\ref{fig:SpacetimeRoadmap}.
The prefix $B$ indicates a 5-dimensional Bargmann extension of a traditional 4-dimensional spacetime.
The affine spacetimes $\mathbb{M}$ and $\mathbb{G}$, and their affine Bargmann extensions $B\mathbb{M}$ and $B\mathbb{G}$, are the primary focus of this paper.
Curved spacetimes will be addressed in the sequel, with a primary goal of developing a conjectured `Galilei general relativity' more extensive than generally known, whose spacetime is labeled labeled $B\mathcal{G}$.

In historical terms, and continuing to refer to Fig.~\ref{fig:SpacetimeRoadmap}, the notion of spacetime begins in earnest with Minkowski's 4-dimensional geometric reformulation of physics according to Poincar\'e and Einstein circa 1905 (traditionally known as `special relativity') on a 4-dimensional affine space $\mathbb{M}$.
Einstein circa 1915 subsequently generalizes to physics on a 4-dimensional curved spacetime $\mathcal{E}$ (traditionally known as `general relativity').
With hindsight, the spacetime perspective of Minkowski and Einstein is backported to the physics of Galilei and Newton in the 1920s with the work of Weyl \cite{Weyl1922Space---Time---} and Cartan \cite{Cartan1923Sur-les-variete,Cartan1924Sur-les-variete,Cartan1986On-manifolds-wi}; examples of subsequent expositions include those of Toupin and Truesdell \cite{Toupin1957World-invariant,Truesdell1960The-Classical-F}, Havas \cite{Havas1964Four-Dimensiona}, Trautmann \cite{Trautman1965Foundations-and,Trautman1966Comparison-of-N}, and K\"untzle \cite{Kunzle1972Galilei-and-Lor}.
Newton gravity is accommodated in the spacetime curvature of the resulting Newton-Cartan spacetime $\mathcal{N}$, though the 3-space leaves of the foliation according to absolute time are flat (traditional 3-dimensional Euclid position space); moreover the geometry is not pseudo-Riemann (there is no non-degenerate spacetime metric), and the field equations with mass density as a source are somewhat ad-hoc in comparison with the tight theoretical structure of physics on $\mathcal{E}$.

The flat (affine space) version of $\mathcal{N}$, Galilei-Newton spacetime $\mathbb{G}$, can be understood as being in effect augmented to a fifth dimension due to quantum mechanical considerations originally articulated by Bargmann \cite{Bargmann1954On-Unitary-ray-} and discussed also for instance by L\'evy-Leblond \cite{Levy-Leblond1971Galilei-Group-a,Levy-Leblond1974The-pedagogical,Levy-Leblond1976Quantum-fact-an} and by Omote et al. \cite{Omote1989Galilean-Covari}.
It is subsequently recognized, for example by Souriau \cite{Souriau1970Structure-des-s,Souriau1997Structure-of-Dy}, by Duval et al. \cite{Duval1985Bargmann-struct,Kunzle1986Relativistic-an,Duval1991Celestial-mecha}, by de~Montigny et al. \cite{Montigny2003Lorentz-like-co,Montigny2003Nonrelativistic}, and by de Saxc\'e and Vall\'ee \cite{de-Saxce2012Bargmann-group-,de-Saxce2016Galilean-Mechan,de-Saxce20175-Dimensional-T}, that the extended affine and curved Bargmann spacetimes $B\mathbb{G}$ and $B\mathcal{N}$ are useful for classical Galilei physics as well.
One bright spot is that these extended spacetimes are now pseudo-Riemann spaces (there is now a non-degenerate spacetime metric, see e.g. \cite{Pinski1968Galilean-Tensor,Duval1985Bargmann-struct,Kunzle1986Relativistic-an,Omote1989Galilean-Covari,Montigny2003Lorentz-like-co,Montigny2003Nonrelativistic,de-Saxce2016Galilean-Mechan,de-Saxce20175-Dimensional-T}), though the comparatively ad-hoc nature of the field equations encoding Newton gravity remains somewhat unsatisfying.

The first few sections of this paper treat Poincar\'e and Galilei relativity in as unified a manner as possible from within a 4-dimensional perspective.
Minkowski spacetime $\mathbb{M}$ and Galilei-Newton spacetime $\mathbb{G}$ are discussed in Sec.~\ref{sec:AffineSpacetimes}, including in particular their foliation into position space 3-slices and corresponding 1+3 (time/space) tensor decompositions.
Spacetime treatments of a material particle and of the electromagnetic field on $\mathbb{M}$ and $\mathbb{G}$ follow in Secs.~\ref{sec:MaterialParticleMG} and \ref{sec:Electrodynamics}.

Subsequent sections of this paper illustrate the more unified perspective on Poincar\'e and Galilei relativity made possible by a 5-dimensional spacetime setting.

A fundamental difference between Poincar\'e and Galilei physics is the unification of inertia and total energy in the Poincar\'e case, in contrast to the invariant nature of inertia and its strict separation from kinetic energy in the Galilei case.
Consideration of the inertia-momentum 4-vector (4-velocity of a material particle multiplied by its rest mass) shows that the Poincar\'e and Galilei groups naturally address the transformations of inertia and 3-momentum.
In the Poincar\'e case the total energy goes along for the ride thanks to its equivalence to inertia, but in the Galilei case kinetic energy is left out in the cold: strictly separated from inertia, kinetic energy is not addressed by Galilei transformations.
The traditional Bargmann group---called in this paper the Bargmann-Galilei or B-Galilei group, and associated with an extended 5-dimensional spacetime $B\mathbb{G}$---is a central extension of the Galilei group that includes the transformation of kinetic energy, enabling a proper understanding of Galilei physics in quantum mechanics \cite{Bargmann1954On-Unitary-ray-,Levy-Leblond1971Galilei-Group-a,Levy-Leblond1974The-pedagogical,Levy-Leblond1976Quantum-fact-an} and a spacetime tensor treatment of material continua that includes kinetic energy and internal energy \cite{de-Saxce2012Bargmann-group-,de-Saxce2016Galilean-Mechan,de-Saxce20175-Dimensional-T}.

What does not seem widely known or appreciated is the existence of what are called here the Bargmann-Poincar\'e (or B-Poincar\'e) group, and its linear part, the Bargmann-Lorentz (or B-Lorentz) group, and their association with a 5-dimensional spacetime called here Bargmann-Minkowski or B-Minkowski spacetime $B\mathbb{M}$.\footnote{These extended Lorentz/Poincar\'e transformations are noticed---perhaps uniquely, and almost as an afterthought---by Omote et al. \cite{Omote1989Galilean-Covari}, a reference unearthed after publication of \cite{Cardall2023Towards-full-Ga}.}
Because it already transforms total energy, a central extension of the Lorentz and Poincar\'e groups is in some sense not \textit{needed}; but it is \textit{allowed}, and provides for a more unified perspective on Poincar\'e and Galilei physics.
Unlike the relationship between the 4-dimensional spacetimes $\mathbb{M}$ and $\mathbb{G}$, these extended Poincar\'e structures associated with $B\mathbb{M}$ limit smoothly to their counterparts associated with $B\mathbb{G}$ as $c \rightarrow \infty$.\footnote{Duval and K{\"u}nzle and collaborators \cite{Duval1985Bargmann-struct,Kunzle1986Relativistic-an} note that Poincar\'e and Galilei physics both can be addressed from a common 5-dimensional setting; but unlike Omote et al. \cite{Omote1989Galilean-Covari} and the present work, do not seem to have noticed and explicitly specified the B-Poincar\'e and B-Lorentz transformations and the even more unified perspective on Poincar\'e and Galilei physics they afford.}
These matters are discussed in Sec.~\ref{sec:ExtendedAffineSpacetimes}, 
including in particular the foliation of $B\mathbb{M}$ and $B\mathbb{G}$ into position space 3-planes and corresponding 1+3+1 (time/space/action) tensor decompositions necessary to make contact with observations.
Treatments of a material particle and of the electromagnetic field on $B\mathbb{M}$ and $B\mathbb{G}$ follow in Secs.~\ref{sec:MaterialParticleBMBG} and \ref{sec:ElectrodynamicsB}.
Reasons supporting a conjectured `Galilei general relativity' more extensive than generally known, whose spacetime is labeled $B\mathcal{G}$, are given in Sec.~\ref{sec:Conclusion}.

Affine spaces and linear tensors are briefly reviewed in the Appendix.
In order to establish notation and the geometric perspective employed here, thorough familiarity with the Appendix is recommended before proceeding to Sec.~\ref{sec:AffineSpacetimes}.
In any case, two grave warnings on notation deserve emphasis.
First, in this work the infix dot operator ($\cdot$) between two tensors will \textit{never} denote a scalar product, if such exists, between two vectors; such will \textit{always} be expressed explicitly in terms of the metric tensor defining the scalar product. 
Here the dot operator will instead denote only tensor evaluation, or contraction, via an obvious `pairing between lower and upper indices'.
Second, index notation will be used sparingly, so that when a tensor or tensor field is introduced, careful attention should be paid to its type.

\section{The affine spacetimes $\mathbb{M}$ and $\mathbb{G}$}
\label{sec:AffineSpacetimes}

Minkowski spacetime $\mathbb{M}$ and Galilei-Newton spacetime $\mathbb{G}$ are affine spaces of dimension 4 whose points are called `events'.
These spacetimes are endowed with additional structure (related to causality) on their respective underlying vector spaces $\mathrm{V}_\mathbb{M}$ and $\mathrm{V}_\mathbb{G}$ whose preservation requires restriction to particular subgroups of $\mathrm{GL}(4)$.
For $\mathrm{V}_\mathbb{M}$ the required subgroup of $\mathrm{GL}(4)$ is the Lorentz group $\mathrm{O}(1,3)$;
for $\mathrm{V}_\mathbb{G}$ it is the homogeneous Galilei group $\mathrm{Gal}_0$.
These are the linear (that is, homogeneous) parts of the Poincar\'e group (or inhomogeneous Lorentz group) $\mathrm{IO}(1,3)$ and Galilei group $\mathrm{Gal}$ respectively; the latter are the subgroups of $\mathrm{Aff}(4)$ comprising the symmetries of $\mathbb{M}$ and $\mathbb{G}$ respectively, which add translations to the Lorentz group $\mathrm{O}(1,3)$ and to the homogeneous Galilei group $\mathrm{Gal}_0$.

\subsection{Minkowski spacetime $\mathbb{M}$}
\label{sec:SpacetimeM}

The causal structure on $\mathrm{V}_\mathbb{M}$ to be preserved is the null cone, embodied in a metric tensor $\bm{g}$ of $\mathbb{M}$, a symmetric and non-degenerate bilinear form that defines a scalar product on $\mathrm{V}_\mathbb{M}$.
The metric $\bm{g}$ is such that for any vectors $\bm{a}, \bm{b} \in \mathrm{V}_\mathbb{M}$, there exists a basis $( \bm{e}_0, \bm{e}_1, \bm{e}_2, \bm{e}_3 )$ of $\mathrm{V}_\mathbb{M}$ such that
\begin{equation}
\bm{g}(\bm{a},\bm{b}) = - c^2 \, a^0 b^0 + a^1 b^1 + a^2 b^2 + a^3 b^3,   
\label{eq:ScalarProduct}
\end{equation}
where the scalar $c$ is the speed of light, and $\bm{a} = \bm{e}_\alpha \, a^\alpha$ and $\bm{b} = \bm{e}_\alpha \, b^\alpha$ define the components $(a^\mu) = (a^0, a^1, a^2, a^3)$ and $(b^\mu) = (b^0, b^1, b^2, b^3)$ of $\bm{a}$ and $\bm{b}$ respectively.
Note the summation convention, with Greek indices taking values in $\{0, 1, 2, 3\}$, with letters $\alpha, \beta, \dots$ near the beginning of the Greek alphabet preferred for dummy indices, and letters $\mu, \nu, \dots$ from later in the alphabet preferred for free indices.
Let $\mathrm{V}_{\mathbb{M}*}$ be the dual space of $\mathrm{V}_\mathbb{M}$ and $( \bm{e}_*^0, \bm{e}_*^1, \bm{e}_*^2, \bm{e}_*^3 )$ the basis dual to $( \bm{e}_0, \bm{e}_1, \bm{e}_2, \bm{e}_3 )$.
With respect to this dual basis, it is apparent from Eq.~(\ref{eq:ScalarProduct}) that
\begin{equation}
\bm{g} = - \, c^2 \, \bm{e}_*^0 \otimes \bm{e}_*^0 
		\, + \, \bm{e}_*^1 \otimes \bm{e}_*^1 
		\, + \, \bm{e}_*^2 \otimes \bm{e}_*^2
		\, + \, \bm{e}_*^3 \otimes \bm{e}_*^3.
\label{eq:MetricMinkowski}
\end{equation}
The null cone of $\mathrm{V}_\mathbb{M}$ is the set of all vectors $\bm{a} \in \mathrm{V}_\mathbb{M}$ such that $\bm{g}(\bm{a},\bm{a}) = 0$.

The particular basis $( \bm{e}_0, \bm{e}_1, \bm{e}_2, \bm{e}_3 )$ considered above is not the only one for which the scalar product $\bm{g}(\bm{a},\bm{a})$ takes not only the value, but the algebraic form exhibited in Eq.~(\ref{eq:ScalarProduct}).
Let $\sfeta \in \mathbb{R}^{4\times4}$ be the Minkowski matrix
\begin{equation}
\sfeta 
	= \begin{bmatrix} 
		-c^2 & \mymathsf{0} \\
		\mymathsf{0} & \mymathsf{1}
	\end{bmatrix}
	= \begin{bmatrix} 
		-c^2 & 0_j \\
		0_i & 1_{i j}
	\end{bmatrix},
\label{eq:MinkowskiMatrix}
\end{equation}
and $\mathsf{g} \in \mathbb{R}^{4\times4}$ the matrix
\begin{equation}
\mathsf{g} 
	= \begin{bmatrix} g_{\mu\nu} \end{bmatrix}
	= \begin{bmatrix} 
			g_{00} & g_{0 j} \\
			g_{i 0} & g_{i j}
	\end{bmatrix}
\label{eq:MetricRepresentation}
\end{equation}
collecting the components $g_{\mu \nu} = \bm{g}(\bm{e}_\mu, \bm{e}_\nu)$ of $\bm{g}$.
Note that latin indices take values in $\{1,2,3\}$, and that letters $i, j, \dots $ near the middle of the alphabet are preferred for free indices; letters $a, b, \dots $ near the beginning of the alphabet will be preferred for dummy indices.
With respect to the particular basis considered above, it follows from Eq.~(\ref{eq:MetricMinkowski}) that the matrix representing $\bm{g}$ is the Minkowski matrix:
\begin{equation}
\mathsf{g} = \sfeta.
\label{eq:MinkowskiRepresentation}
\end{equation}
Let $\mathsf{a}$ and $\mathsf{b}$ be the $n$-column representations of $\bm{a}, \bm{b} \in \mathrm{V}_\mathbb{M}$ with respect to the considered basis (see the Appendix).
Then Eq.~(\ref{eq:ScalarProduct}) is expressed by the matrix equation
\begin{equation}
\bm{g}(\bm{a},\bm{b}) = \mathsf{a}^\mathrm{T} \, \sfeta \, \mathsf{b}.
\label{eq:ScalarProductMatrixEquation}
\end{equation}
Lorentz transformations are the invertible linear transformations $\bm{P}_\mathbb{M}$ of $\mathrm{V}_\mathbb{M}$ that preserve the scalar product defined by $\bm{g}$ (without also transforming $\bm{g}$):
\begin{equation}
\bm{g}\left( \bm{P}_\mathbb{M} \left( \bm{a} \right), \bm{P}_\mathbb{M}\left( \bm{b} \right) \right) = \bm{g}(\bm{a},\bm{b}).
\label{eq:LorentzScalarProduct}
\end{equation}
The Lorentz transformations constitute a subgroup of $\mathrm{GL}\left( V_\mathbb{M} \right)$. 
With respect to the considered basis, this preservation of the scalar product reads
\[
 \mathsf{a}^\mathrm{T}\, ( \mathsf{P}_\mathbb{M}^\mathrm{T} \, \sfeta\, {\mathsf{P}_\mathbb{M}}) \, \mathsf{b} 
	= \mathsf{a}^\mathrm{T} \, \sfeta \, \mathsf{b}. 
\]
(Recall from the Appendix that $\bm{P}_\mathbb{M}$ is taken to act on the basis elements rather than on the $n$-columns collecting the vector components.)
With slight ambiguity, refer to both the set of Lorentz transformations, and the set of matrices whose elements $\mathsf{P}_\mathbb{M} \in \mathrm{GL}(4)$ faithfully represent them and are such that the Minkowski matrix is preserved according to the relation
\begin{equation}
\mathsf{P}_\mathbb{M}^\mathrm{T}\, \sfeta\, {\mathsf{P}_\mathbb{M}} = \sfeta,
\label{eq:MinkowskiInvariance}
\end{equation}
as the Lorentz group $\mathrm{O}(1,3)$.
Under (suitable representations of) Lorentz transformations, the matrix expression of Eq.~(\ref{eq:ScalarProductMatrixEquation}) for the scalar product is indifferent to a change of basis $\begin{bmatrix} \bm{e}'_0 & \bm{e}'_1 & \bm{e}'_2 & \bm{e}'_3 \end{bmatrix} = \begin{bmatrix} \bm{e}_0 & \bm{e}_1 & \bm{e}_2 & \bm{e}_3 \end{bmatrix} \mathsf{P}_\mathbb{M}$:
\[
\bm{g}(\bm{a},\bm{b}) = \mathsf{a}^\mathrm{T} \, \sfeta \, \mathsf{b} 
	= ( {\mathsf{P}_\mathbb{M}}\, \mathsf{a}' )^\mathrm{T}\, \sfeta\, ( {\mathsf{P}_\mathbb{M}}\, \mathsf{b}' )
	= {\mathsf{a}'}^\mathrm{T} \, \sfeta \, \mathsf{b}'. 
\]
Call a `Minkowski basis' any basis of $\mathrm{V}_\mathbb{M}$ for which Eq.~(\ref{eq:MinkowskiRepresentation}) holds, so that the inner product $\bm{g}(\bm{a},\bm{b})$ is given by Eq.~(\ref{eq:ScalarProductMatrixEquation}) with $\sfeta$ the Minkowski matrix of Eq.~(\ref{eq:MinkowskiMatrix}), which yields the arithmetic form of Eq.~(\ref{eq:ScalarProduct}).

The definition of the null cone of $\mathrm{V}_\mathbb{M}$ as the set of all vectors $\bm{a} \in \mathrm{V}_\mathbb{M}$ such that $\bm{g}(\bm{a},\bm{a}) = 0$, together with the invariance of the scalar product as the defining property of Lorentz transformations (Eq.~(\ref{eq:LorentzScalarProduct})), implies that the null cone is preserved under Lorentz transformations. 

Elements $\bm{P}_\mathbb{M}^+$ of the identity component $\mathrm{SO}^+(1,3)$ of the Lorentz group (the connected component containing the identity), also called the restricted Lorentz group or proper orthochronous Lorentz group, can be uniquely factored into a `boost' and a `rotation'. 
With respect to a Minkowski basis $( \bm{e}_0, \bm{e}_1, \bm{e}_2, \bm{e}_3 )$ of $\mathrm{V}_\mathbb{M}$,
\[
\mathsf{P}_\mathbb{M}^+ = \mathsf{L}_\mathbb{M} \, \mathsf{R}.
\]
Here 
\begin{equation}
\mathsf{R}
	= \begin{bmatrix}
		1 & \mymathsf{0} \\
		\mymathsf{0} & \mathsf{R}_\mathbb{S}
	\end{bmatrix},
\label{eq:Rotation}
\end{equation}
with $\mathsf{R}_\mathbb{S} \in \mathrm{SO}(3)$ a rotation of the subspace $\mathrm{V}_\mathbb{S}$ of $\mathrm{V}_\mathbb{M}$ spanned by $( \bm{e}_1, \bm{e}_2, \bm{e}_3 )$;  $\mathrm{V}_\mathbb{S}$ is the orthogonal complement (relative to $\bm{g}$) of the 1-dimensional subspace spanned by $\bm{e}_0$.
Moreover, $\mathsf{L}_\mathbb{M}$ is a boost that can be parametrized as
\begin{equation}
\mathsf{L}_\mathbb{M} 
	= \begin{bmatrix}
		\Lambda_\mathsf{u} & \frac{1}{c^2} \, \Lambda_\mathsf{u} \, \mathsf{u}^\mathrm{T} \\[5pt]
		\Lambda_\mathsf{u} \, \mathsf{u} & \mymathsf{1} + \frac{1}{\lVert \mathsf{u} \rVert^2}( \Lambda_\mathsf{u}  - 1 )\, \mathsf{u} \, \mathsf{u}^\mathrm{T}
	\end{bmatrix}
\label{eq:MinkowskiBoost}
\end{equation}
where the 3-column $\mathsf{u} \in \mathbb{R}^{3 \times 1}$ is the boost velocity parameter, 
\[
\Lambda_\mathsf{u} = \left( 1 - \frac{\lVert \mathsf{u} \rVert^2}{c^2} \right)^{-1/2}
\]
is the Lorentz factor associated with $\mathsf{u}$, and $\lVert \mathsf{u} \rVert^2 = \mathsf{u}^\mathrm{T} \mathsf{u}$ is the squared Euclid norm with respect to an orthonormal basis of $\mathrm{V}_\mathbb{S}$ (naturally appropriate to a Minkowski basis of $\mathrm{V}_\mathbb{M}$).
Thus
\begin{equation}
\mathsf{P}_\mathbb{M}^+
	= \begin{bmatrix}
		\Lambda_\mathsf{u} 
		& \frac{1}{c^2} \, \Lambda_\mathsf{u} \, \mathsf{u}^\mathrm{T} \, \mathsf{R}_\mathbb{S} \\[5pt]
		\Lambda_\mathsf{u} \, \mathsf{u} & \mathsf{R}_\mathbb{S} + \frac{1}{\lVert \mathsf{u} \rVert^2}( \Lambda_\mathsf{u}  - 1 )\, \mathsf{u} \, \mathsf{u}^\mathrm{T} \mathsf{R}_\mathbb{S}
	\end{bmatrix}.
\label{eq:LorentzTransformation}
\end{equation}
The inverse is
\[
{\mathsf{P}_\mathbb{M}^+ }^{-1} = \mathsf{R}^\mathrm{T}\, \mathsf{L}_\mathbb{M}^{-1}
	= \begin{bmatrix}
		\Lambda_\mathsf{u} 
		& - \frac{1}{c^2} \, \Lambda_\mathsf{u} \, \mathsf{u}^\mathrm{T}  \\[5pt]
		- \Lambda_\mathsf{u} \, \mathsf{R}_\mathbb{S}^\mathrm{T}  \mathsf{u} 
		& \mathsf{R}_\mathbb{S}^\mathrm{T} + \frac{1}{\lVert \mathsf{u} \rVert^2}( \Lambda_\mathsf{u}  - 1 )\, \mathsf{R}_\mathbb{S}^\mathrm{T} \mathsf{u} \, \mathsf{u}^\mathrm{T} 
	\end{bmatrix},
\]
where $\mathsf{L}_\mathbb{M}^{-1}$ is obtained from $\mathsf{L}_\mathbb{M}$ via $\mathsf{u} \mapsto -\mathsf{u}$.

Because $\bm{g}$ is non-degenerate, its matrix representation $\mathsf{g}$ of Eq.~(\ref{eq:MetricRepresentation}) has an inverse
\[
\overleftrightarrow{\mathsf{g}} 
	= \begin{bmatrix} g^{\mu\nu} \end{bmatrix}
	= \begin{bmatrix} 
			g^{00} & g^{0 j} \\
			g^{i 0} & g^{i j}
	\end{bmatrix},
\]
written here in a way suggestive of the fact that $\overleftrightarrow{\mathsf{g}}$ collects the components $g^{\mu \nu} = \overleftrightarrow{\bm{g}}(\bm{e}_*^\mu, \bm{e}_*^\nu)$ resulting from the evaluation of the inverse metric $\overleftrightarrow{\bm{g}}$ on elements of a basis of ${\mathrm{V}_\mathbb{M}}_*$.
The inverse Minkowski matrix is
\begin{equation}
\overleftrightarrow{\sfeta}
	= \begin{bmatrix} 
		-\frac{1}{c^2} & \mymathsf{0} \\[5pt]
		\mymathsf{0} & \mymathsf{1}
	\end{bmatrix}
	= \begin{bmatrix} 
		-\frac{1}{c^2} & \mathsf{0}^j \\[5pt]
		\mathsf{0}^i & \mathsf{1}^{i j}
	\end{bmatrix},
\label{eq:InverseMetricMinkowskiBasis}
\end{equation}
so that
\begin{equation}
\overleftrightarrow{\bm{g}} = - \, \frac{1}{c^2} \, \bm{e}_0 \otimes \bm{e}_0 
		\, + \, \bm{e}_1 \otimes \bm{e}_1 
		\, + \, \bm{e}_2 \otimes \bm{e}_2
		\, + \, \bm{e}_3 \otimes \bm{e}_3
\label{eq:InverseMetricMinkowski}
\end{equation}
with respect to a Minkowski basis---say, the same basis used to obtain Eq.~(\ref{eq:MinkowskiInvariance}) from Eq.~(\ref{eq:LorentzScalarProduct}).
Given a Lorentz transformation $\bm{P}_\mathbb{M}$ acting on $\mathrm{V}_\mathbb{M}$, the dual space ${\mathrm{V}_\mathbb{M}}_*$ is transformed by the algebraic adjoint of the inverse transformation, $\left( \bm{P}_\mathbb{M}^{-1}\right)_*$; see the Appendix.
The inverse metric $\overleftrightarrow{\bm{g}}$ defines a scalar product on ${\mathrm{V}_\mathbb{M}}_*$, also Lorentz-invariant, in the sense that for any $\bm{\psi}, \bm{\omega} \in {\mathrm{V}_\mathbb{M}}_*$,
\[
\overleftrightarrow{\bm{g}}\left( \left( \bm{P}_\mathbb{M}^{-1} \right)_* (\bm{\psi}), \left( \bm{P}_\mathbb{M}^{-1} \right)_* (\bm{\omega}) \right) 
	= \overleftrightarrow{\bm{g}}\left( \bm{\psi}, \bm{\omega} \right).  
\]
This implies the preservation of the inverse Minkowski matrix,
\begin{equation}
\mathsf{P}_\mathbb{M}^{-1} \, \overleftrightarrow{\sfeta}\, \mathsf{P}_\mathbb{M}^{-\mathrm{T}} = \overleftrightarrow{\sfeta},
\label{eq:InverseMinkowskiInvariance}
\end{equation}
consistent with Eq.~(\ref{eq:MinkowskiInvariance}).

Equipped with a metric $\bm{g}$ and its inverse $\overleftrightarrow{\bm{g}}$, the affine spacetime (and flat differentiable manifold) $\mathbb{M}$ with its underlying vector space $\mathrm{V}_\mathbb{M}$ enjoys a fulness of the apparatus of tensor algebra (and tensor calculus).

The tensor algebra includes metric duality between vectors and linear forms manifest in conventions for the raising and lowering of indices.
Associated with a vector $\bm{a} \in \mathrm{V}_\mathbb{M}$, which is a $(1,0)$ tensor, is a linear form $\underline{\bm{a}} = \bm{g}(\bm{a},\cdot) \in {\mathrm{V}_\mathbb{M}}_*$, which is a $(0,1)$ tensor.
This is expressed in matrix notation as ${\underline{\mathsf{a}}}^\mathrm{T} = \mathsf{g} \, \mathsf{a}$ (so that $\underline{\bm{a}}$ is represented as a 4-row $\underline{\mathsf{a}}$), and in terms of indexed components as $a_\mu = g_{\mu \alpha}\, a^\alpha$. 
Associated with a covector $\bm{\omega} \in {\mathrm{V}_\mathbb{M}}_*$, which is a $(0,1)$ tensor, is a vector $\overleftarrow{\bm{\omega}} = \overleftrightarrow{\bm{g}}(\bm{\omega},\cdot) \in \mathrm{V}_\mathbb{M}$, which is a $(1,0)$ tensor.
This is expressed in matrix notation as ${\overleftarrow{\sfomega}}^\mathrm{T} = \sfomega \, \overleftrightarrow{\mathsf{g}}$ (so that $\overleftarrow{\bm{\omega}}$ is represented as a 4-column $\overleftarrow{\sfomega}$), and in terms of indexed components as $\omega^\mu = \omega_\alpha \, g^{\alpha \mu}$. 
Consider also a bilinear form $\bm{F} \in {\mathrm{V}_\mathbb{M}}_* \otimes {\mathrm{V}_\mathbb{M}}_*$, which is a $(0,2)$ tensor, taking values $\bm{F}(\bm{a},\bm{b})$ for $\bm{a},\bm{b} \in \mathrm{V}_\mathbb{M}$.
It is associated by metric duality with a $(1,1)$ tensor $\overleftarrow{\bm{F}} \in \mathrm{V}_\mathbb{M} \otimes {\mathrm{V}_\mathbb{M}}_*$ defined by $\overleftarrow{\bm{F}}(\bm{\omega},\bm{b}) = \bm{F}\left( \overleftarrow{\bm{\omega}},\bm{b} \right)$ and a $(2,0)$ tensor $\overleftrightarrow{\bm{F}} \in \mathrm{V}_\mathbb{M} \otimes \mathrm{V}_\mathbb{M}$ defined by $\overleftrightarrow{\bm{F}}(\bm{\omega},\bm{\psi})  = \bm{F}\left( \overleftarrow{\bm{\omega}}, \overleftarrow{\bm{\psi}} \right)$, where in both cases $\bm{\omega}, \bm{\psi} \in {\mathrm{V}_\mathbb{M}}_*$.
In matrix notation, $\overleftarrow{\mathsf{F}} = \overleftrightarrow{\mathsf{g}} \, \mathsf{F}$ and $\overleftrightarrow{\mathsf{F}} = \overleftarrow{\mathsf{F}} \, \overleftrightarrow{\mathsf{g}} = \overleftrightarrow{\mathsf{g}} \, \mathsf{F} \, \overleftrightarrow{\mathsf{g}}$.
In terms of indexed components, ${F^\mu}_\nu = g^{\mu \alpha} \, F_{\alpha \nu}$ and $F^{\mu \nu} = g^{\mu \alpha} \, F_{\alpha \beta} \, g^{\beta \nu}$.
In terms of the infix dot operator ($\cdot$),
\[
\begin{array}{rlllll}
	\underline{\bm{a}} & = & \bm{g} \cdot \bm{a} & = &  \bm{a} \cdot \bm{g} & \mathrm{for \ } \bm{a} \in \mathrm{V}_\mathbb{M}, \\[5pt]
	\overleftarrow{\bm{\omega}} & = & \bm{\omega} \cdot \overleftrightarrow{\bm{g}} & = & \overleftrightarrow{\bm{g}} \cdot \bm{\omega} & \mathrm{for \ } \bm{\omega} \in {\mathrm{V}_\mathbb{M}}_*, \\[5pt]
	\overleftarrow{\bm{F}} & = & \overleftrightarrow{\bm{g}} \cdot \bm{F} & & & \mathrm{for \ } \bm{F} \in {\mathrm{V}_\mathbb{M}}_* \otimes  {\mathrm{V}_\mathbb{M}}_*, \\
	\overleftrightarrow{\bm{F}} & = & \overleftrightarrow{\bm{g}} \cdot \bm{F} \cdot \overleftrightarrow{\bm{g}} & = & \overleftarrow{\bm{F}} \cdot \overleftrightarrow{\bm{g}} & \mathrm{for \ } \bm{F} \in {\mathrm{V}_\mathbb{M}}_* \otimes  {\mathrm{V}_\mathbb{M}}_*
\end{array}
\]
express the above relations describing metric duality.

As for tensor calculus, $\mathbb{M}$ is a pseudo-Riemann manifold; the natural affine connection $\bm{\nabla}$ possessed by an affine space mentioned at the end of the Appendix becomes a Levi-Civita connection associated with $\bm{g}$.
An orientation on $\mathbb{M}$ is specified with a volume form on $\mathrm{V}_\mathbb{M}$, the Levi-Civita tensor $\bm{\varepsilon}$ defined such that
\[
\bm{\varepsilon}( \bm{e}_0, \bm{e}_1, \bm{e}_2, \bm{e}_3 ) = 1
\]
with components
\[
\varepsilon_{ \mu \nu \rho \sigma } = \left[ \mu \nu \rho \sigma \right]
\]
for a right-handed Minkowski basis.
With respect to another right-handed but otherwise arbitrary basis $( \bm{e}'_0, \bm{e}'_1, \bm{e}'_2, \bm{e}'_3 )$, Eq.~(\ref{eq:VolumeFormTransformed}), together with the matrix relation $\mathsf{g}' = \mathsf{P}^\mathrm{T} \, \sfeta \, \mathsf{P}$, show that in the more general basis the components are given by
\[
\varepsilon'_{ \mu \nu \rho \sigma } = \frac{\sqrt{ -g' }}{c} \, \left[ \mu \nu \rho \sigma \right]
	\quad (\mathbb{M} \text{ only}),
\]
where $g' = \det \mathsf{g'}$.
Raising all four indices yields
\[
\varepsilon'^{ \mu \nu \rho \sigma } = - \frac{1}{c \, \sqrt{ -g' }} \, \left[ \mu \nu \rho \sigma \right]
	\quad (\mathbb{M} \text{ only}),
\]
with respect to a general basis, or
\[
\varepsilon^{ \mu \nu \rho \sigma } = - \frac{1}{c^2} \, \left[ \mu \nu \rho \sigma \right]
	\quad (\mathbb{M} \text{ only})
\]
with respect to a Minkowski basis.

A metric $\bm{g}$ that makes the volume form $\bm{\varepsilon}$ also a Levi-Civita tensor makes available the Hodge star operator that provides a bijection between $p$-forms and $(4-p)$-forms.
In particular, a 2-form $\bm{F}$ is related to a complementary 2-form $\bm{\star} \bm{F}$ by
\begin{equation}
\bm{\star} \bm{F} = \frac{1}{2} \, \overleftrightarrow{\bm{F}} : \bm{\varepsilon}
	\quad (\mathbb{M} \text{ only}),
\label{eq:HodgeStar}
\end{equation}
where $\overleftrightarrow{\bm{F}} = \overleftrightarrow{\bm{g}} \cdot \bm{F} \cdot \overleftrightarrow{\bm{g}}$ is the version of $\bm{F}$ with both indices raised, and the expression with a double contraction $(:)$ reads
$\star F_{\mu\nu} = F^{\alpha\beta} \, \varepsilon_{\alpha\beta\mu\nu} / 2$.
The bijective nature of the Hodge duality relation is manifest in its `invertibility', in the sense that (again for a 2-form)
\begin{equation}
\bm{\star} {\bm{\star} \bm{F}} = -\frac{1}{c^2} \, \bm{F}
	\quad (\mathbb{M} \text{ only}).
\label{eq:DoubleHodgeStar}
\end{equation}
One application concerns 4-vector fields $\bm{a}$ and $\bm{b}$, for which the identities
\begin{equation}
\bm{\star} \left( \underline{\bm{a}} \wedge \underline{\bm{b}} \right)
	= \bm{\varepsilon} ( \bm{a}, \bm{b}, \, . \, , \, . \, )
		\quad (\mathbb{M} \text{ only})
\label{eq:HodgeWedge1}
\end{equation}
and
\begin{equation}
\bm{\star} \bm{\varepsilon} ( \bm{a}, \bm{b}, \, . \, , \, . \, )
	=  -\frac{1}{c^2}   \left( \underline{\bm{a}} \wedge \underline{\bm{b}} \right)
		\quad (\mathbb{M} \text{ only})
\label{eq:HodgeWedge2}
\end{equation}
hold, where $\underline{\bm{a}} = \bm{g} \cdot \bm{a}$ and $\underline{\bm{b}} = \bm{g} \cdot \bm{b}$.

\subsection{Galilei-Newton spacetime $\mathbb{G}$}
\label{sec:SpacetimeG}

Galilei-Newton spacetime $\mathbb{G}$ might in a more or less literal sense be regarded as a  `degeneration' of Minkowski spacetime $\mathbb{M}$ as $c \rightarrow \infty$.
In many respects one obtains a smooth limit, but crucially the limit of the metric $\bm{g}$ does not exist, so that a qualitatively different geometric structure results. 
In particular, Eq.~(\ref{eq:MetricMinkowski}) asymptotes as 
\[
\bm{g} \sim -c^2 \, \bm{e}_*^0 \otimes \bm{e}_*^0.
\]
This indicates that the covector (linear form) 
\begin{equation}
\bm{\tau} = \bm{e}_*^0
\label{eq:TimeForm}
\end{equation} 
becomes the fundamental causal structure on $\mathrm{V}_\mathbb{G}$.
With respect to what once was a Minkowski basis---now to be called a Galilei basis---it is represented by the 4-row
\[
\sftau = \begin{bmatrix} 1 & \mymathsf{0} \end{bmatrix} = \begin{bmatrix} 1 & 0_i \end{bmatrix}.
\]
The other remnant that survives from Minkowski spacetime is a limit that does exist as $c \rightarrow \infty$, namely
\[
\overleftrightarrow{\bm{g}} \rightarrow \overleftrightarrow{\bm{\gamma}} = \bm{e}_1 \otimes \bm{e}_1 
		+ \bm{e}_2 \otimes \bm{e}_2
		+ \bm{e}_3 \otimes \bm{e}_3
\]
from Eq.~(\ref{eq:InverseMetricMinkowski}).
With respect to a Galilei basis it is represented by
\begin{equation}
\overleftrightarrow{\sfgamma}
	= \begin{bmatrix} 
		0 & \mymathsf{0} \\
		\mymathsf{0} & \mymathsf{1}
	\end{bmatrix}
	= \begin{bmatrix} 
		0 & 0^j \\
		0^i & 1^{i j}
	\end{bmatrix},
\label{eq:InverseMetricGalileiBasis}
\end{equation}
and does not qualify as an inverse metric on $\mathrm{V}_\mathbb{G}$ because for $\bm{\tau}$ (and similarly for any scalar multiple thereof)
\begin{equation}
\overleftrightarrow{\bm{\gamma}} ( \bm{\tau}, \bm{\omega} ) = 0
\label{eq:DegeneracyCondition}
\end{equation}
for any $\bm{\omega} \in {\mathrm{V}_\mathbb{G}}_*$, that is, it is degenerate in the technical sense.

The homogeneous Galilei group $\mathrm{Gal}_0$ consists of the linear transformations $\bm{P}_\mathbb{G}$ of $\mathrm{V}_\mathbb{G}$ that preserve these structures, and it turns out that they are the $c \rightarrow \infty$ limit of the Lorentz transformations.
Require first that
\[
\bm{\tau}\left(\bm{P}_\mathbb{G}(\bm{a}) \right) = \bm{\tau}\left( \bm{a} \right)
\]
for any $\bm{a} \in \mathrm{V}_\mathbb{G}$.
When expressed with respect to a Galilei basis, this requirement implies
\[
\begin{bmatrix} 1 & 0_i \end{bmatrix} \mathsf{P}_\mathbb{G} =  \begin{bmatrix} 1 & 0_i \end{bmatrix}.
\]
Require also that for any $\bm{\psi}, \bm{\omega} \in {\mathrm{V}_\mathbb{G}}_*$,
\[
\overleftrightarrow{\bm{\gamma}}\left( \left( \bm{P}_\mathbb{G}^{-1} \right)_* (\bm{\psi}), \left( \bm{P}_\mathbb{G}^{-1} \right)_* (\bm{\omega}) \right) 
	= \overleftrightarrow{\bm{\gamma}}\left( \bm{\psi}, \bm{\omega} \right),  
\]
which implies
\[
\mathsf{P}_\mathbb{G}^{-1} \, \begin{bmatrix} 
		0 & \mathsf{0}^j \\
		\mathsf{0}^i & \mathsf{1}^{i j}
	\end{bmatrix} \, \mathsf{P}_\mathbb{G}^{-\mathrm{T}}
	= \begin{bmatrix} 
		0 & \mathsf{0}^j \\
		\mathsf{0}^i & \mathsf{1}^{i j}
	\end{bmatrix}
\]
when expressed with respect to a Galilei basis (compare Eq.~(\ref{eq:InverseMinkowskiInvariance})).

As with the restricted Lorentz group, the elements $\bm{P}_\mathbb{G}^+$ of the identity component $\mathrm{Gal}_0^+$ of $\mathrm{Gal}_0$ can be uniquely factored into a boost and a rotation.
With respect to a Galilei basis, these read
\[
\mathsf{P}_\mathbb{G}^+ = \mathsf{L}_\mathbb{G} \, \mathsf{R},
\]
where $\mathsf{R}$ is the same as in Eq.~(\ref{eq:Rotation}), and the Galilei boost
\begin{equation}
\mymathsf{L}_\mathbb{G} 
	= \begin{bmatrix}
		1 & \mymathsf{0} \\
		\mathsf{u} & \mymathsf{1}
	\end{bmatrix}
\label{eq:GalileiBoost}
\end{equation}
is the $c \rightarrow \infty$ limit of Eq.~(\ref{eq:MinkowskiBoost}), so that
\begin{equation}
\mathsf{P}_\mathbb{G}^+
	= \begin{bmatrix}
		1 & \mymathsf{0} \\
		\mathsf{u} & \mathsf{R}_\mathbb{S}
	\end{bmatrix}
\label{eq:GalileiTransformation}
\end{equation}
with $\mathsf{u} \in \mathbb{R}^{3 \times 1}$ and $\mathsf{R}_\mathbb{S} \in \mathrm{SO}(3)$.
The inverse is
\[
{\mathsf{P}_\mathbb{G}^+ }^{-1} = \mathsf{R}^\mathrm{T}\, \mathsf{L}_\mathbb{G}^{-1}
	= \begin{bmatrix}
		1 & 0 \\[5pt]
		- \mathsf{R}_\mathbb{S}^\mathrm{T}  \mathsf{u} & \mathsf{R}_\mathbb{S}^\mathrm{T}
	\end{bmatrix},
\]
where $\mathsf{L}_\mathbb{G}^{-1}$ is obtained from $\mathsf{L}_\mathbb{G}$ via $\mathsf{u} \mapsto -\mathsf{u}$.
It is easy to see that the matrices $\mathsf{P}_\mathbb{G}^+$ satisfy the above conditions for the invariance of $\bm{\tau}$ and $\overleftrightarrow{\bm{\gamma}}$.

Without a spacetime metric, the tensor algebra and tensor calculus on the affine spacetime $\mathbb{G}$ are more limited.
In particular there is no metric duality, no `raising and lowering of indices', for tensors on $\mathrm{V}_\mathbb{G}$ or tensor fields on $\mathbb{G}$: the type of a particular tensor is fixed.
As will be discussed shortly, there is metric duality on a subspace $\mathrm{V}_\mathbb{S}$ of $\mathrm{V}_\mathbb{G}$, and for later notational consistency a double arrow adorns the degenerate inverse `metric' $\overleftrightarrow{\bm{\gamma}}$.
For $\overleftrightarrow{\bm{\gamma}}$ regarded as a tensor on $\mathrm{V}_\mathbb{G}$, however, this must not be associated with metric duality, but simply as an integral part of the symbol denoting this particular tensor of fixed type $(2,0)$.

A volume form exists\ADD{, given by Eq.~(\ref{eq:VolumeForm}) with the `preferred basis' being a Galilei basis,} but is not the traditional Levi-Civita tensor associated with a metric. 
There is no Hodge star operator \ADD{because there is no inverse metric}, though one can define something partly comparable---a `slash-star operator'---using the Galilei-invariant $\overleftrightarrow{\bm{\gamma}}$ instead of $\overleftrightarrow{\bm{g}}$ available only on $\mathbb{M}$.
For a 2-form $\bm{F}$, the analogue of Eq.~(\ref{eq:HodgeStar}) is
\begin{equation}
\cancel{\bm{\star}} \bm{F} =  \frac{1}{2} \, \overleftrightarrow{\bm{F}} : \bm{\varepsilon}
	\quad (\mathbb{G} \text{ only}),
\label{eq:SlashStar}
\end{equation}
where now the raised-index object is interpreted as $\overleftrightarrow{\bm{F}} = \overleftrightarrow{\bm{\gamma}} \cdot \bm{F} \cdot \overleftrightarrow{\bm{\gamma}}$ and has had its time components projected out, and
\begin{equation}
\cancel{\bm{\star}} \cancel{\bm{\star}} \bm{F} = 0
	\quad (\mathbb{G} \text{ only})
\label{eq:DoubleSlashStar}
\end{equation}
in contrast to Eq.~(\ref{eq:DoubleHodgeStar}) as a consequence of the degeneracy of $\overleftrightarrow{\bm{\gamma}}$.

\subsection{Spacetime foliation and tensor decomposition}
\label{sec:SpacetimeFoliation}

Humans and their measuring instruments do not apprehend spacetime directly, but only perceive happenings in nearby `space' at successive instants of `time'. 
Thus if a physical theory is formulated in terms of tensor fields on spacetime, comparison with human observations requires a means of decomposing spacetime and tensor fields thereon into structures compatible with perceptions experienced and recorded in this way.
The key tensor structures on the vector spaces $\mathrm{V}_\mathbb{M}$ and $\mathrm{V}_\mathbb{G}$ underlying the affine spacetimes $\mathbb{M}$ and $\mathbb{G}$, along with the symmetry groups compatible with them, enable such decompositions.
The manner in which these structures describe time implies a notion of space: given an event $\mathbf{a}$ in $\mathbb{M}$ or $\mathbb{G}$, the subset of events `simultaneous' to $\mathbf{a}$ constitutes `space' according to an observer at $\mathbf{a}$.
A notion of time also embodies `causality': if the value of a physical variable at event $\mathbf{a}$ in $\mathbb{M}$ or $\mathbb{G}$ is to influence the value of a physical variable at event $\mathbf{b}$, event $\mathbf{a}$ must `precede' event $\mathbf{b}$. 

Affine spacetimes permit `inertial observers' with straight worldlines and no rotation, and the splitting of spacetime into space and time as perceived by a single inertial observer is formally similar on $\mathbb{M}$ and $\mathbb{G}$.
Select an event $\mathbf{O}$ of $\mathbb{M}$ or $\mathbb{G}$ as origin, and a Minkowski basis of $\mathrm{V}_\mathbb{M}$ or a Galilei basis of $\mathrm{V}_\mathbb{G}$ accordingly, designated $\left( \bm{e}_0, \bm{e}_1, \bm{e}_2, \bm{e}_3 \right)$ in either case.
Such bases are determined by the metric $\bm{g}$ and the Lorentz transformations which preserve it in the case of $\mathbb{M}$, or by the covector $\bm{\tau}$ and the $(2,0)$ tensor $\overleftrightarrow{\bm{\gamma}}$ and the homogeneous Galilei transformations which preserve them in the case of $\mathbb{G}$, as discussed in Secs.~\ref{sec:SpacetimeM} and \ref{sec:SpacetimeG}.
For these affine spacetimes, such choices of origin and vector basis determine a global coordinate system, as in Eq.~(\ref{eq:PointExpansion}); call these coordinates $\left( X^0, X^1, X^2, X^3 \right) = \left( t, x^i \right)$, with $t$ the time coordinate and $(x^i)$ the space coordinates.
The $t$ coordinate curve passing through $\mathbf{O}$ is the straight line 
\[
\mathbb{T} = \{\mathbf{O} +  \bm{e}_0 \, t \mid t \in \mathbb{R} \}.
\]
Interpret $\mathbb{T}$ as as the worldline of a fiducial (and inertial) observer, whose location in $\mathbb{M}$ or $\mathbb{G}$ when her ideal clock (which marks time at a constant rate) reads time $t$ is the event $\mathbf{O} + \bm{e}_0 \, t$. 
Let $\mathrm{V}_\mathbb{S}$ be the subspace of $\mathrm{V}_\mathbb{M}$ or $\mathrm{V}_\mathbb{G}$ spanned by $\left( \bm{e}_1, \bm{e}_2, \bm{e}_3 \right)$.
For a given time $t \in \mathbb{R}$, consider a one-to-one mapping 
\begin{align*}
V_\mathbb{S} & \rightarrow  \mathbb{M} \mathrm{\ or \ } \mathbb{G}  \\
\bm{x} & \mapsto  \mathbf{O} +  \bm{e}_0 \, t + \bm{x}.
\end{align*}
The image of this mapping, 
\[
\mathbb{S}_t = \{ \mathbf{O} + \bm{e}_0 \, t + \bm{e}_i \, x^i \mid ( x^i ) \in \mathbb{R}^3 \},
\]
is a hyperplane (a 3-dimensional affine subspace) of $\mathbb{M}$ or $\mathbb{G}$ through the event $\mathbf{O} +  \bm{e}_0 \, t $.
Interpret $\mathbb{S}_t$ as `space', that is, position space, according to the fiducial observer with straight worldline $\mathbb{T}$ at her time $t$: each of its points also has time coordinate $t$, and together they constitute a surface of simultaneity with the fiducial observer.  
Each hypersurface $\mathbb{S}_t$ is a level surface of (abusing notation) the coordinate function $t$; these hypersurfaces partition spacetime, and the complete collection $\left( \mathbb{S}_t \right)_{t \in \mathbb{R}}$ is said to be a foliation of $\mathbb{M}$ or $\mathbb{G}$.

For a given inertial observer---one with a straight worldline $\mathbb{T}$---the structure of position space, that is, of the leaf $\mathbb{S}_t$ of the foliation of spacetime she encounters at time $t$, is the same for $\mathbb{M}$ or $\mathbb{G}$: 
it is a three-dimensional affine space whose underlying vector space $\mathrm{V}_\mathbb{S}$ is rotationally invariant.
This is apparent from the expressions for Lorentz transformations $\mathsf{P}_\mathbb{M}^+$ on $\mathrm{V}_\mathbb{M}$ and homogeneous Galilei transformations $\mathsf{P}_\mathbb{G}^+$ on $\mathrm{V}_\mathbb{G}$ exhibited in Secs.~\ref{sec:SpacetimeM} and \ref{sec:SpacetimeG} respectively: 
the symmetry transformations $\mathsf{P}_\mathbb{M}^+$ and $\mathsf{P}_\mathbb{G}^+$ both reduce to a rotation of $\mathrm{V}_\mathbb{S}$ for vanishing boost parameter $\mathsf{u}$.
As a rotationally invariant vector space, $\mathrm{V}_\mathbb{S}$ is naturally endowed with a flat Euclid metric defining the usual scalar product; call it $\bm{\gamma}$.

While much is the same for the split of $\mathbb{M}$ and $\mathbb{G}$ into space and time for a single inertial observer, an important difference becomes apparent in comparing these splits for different inertial observers.
In a conventional spacetime diagram for $\mathrm{V}_\mathbb{M}$, the fiducial time and space axes---here aligned with $\bm{e}_0$ and $\bm{e}_1$---are vertical and horizontal respectively, and for $c=1$ the trace of the null cone makes a $45^\circ$ angle midway between them.
Under a pure boost of magnitude $u$ aligned with $\bm{e}_1$, and temporarily setting $c = 1$, the basis relation (see the Appendix) $\bm{\mathsf{B}}' = \bm{\mathsf{B}} \, \mathsf{P}_\mathbb{M}^+$ yields
\begin{align*}
\bm{e}'_0 &= \Lambda_u \left( \bm{e}_0 + \bm{e}_1  u \right), \\
\bm{e}'_1 &= \Lambda_u \left( \bm{e}_0 u + \bm{e}_1 \right), \\
\bm{e}'_2 &= \bm{e}_2, \\
\bm{e}'_3 &= \bm{e}_3
\end{align*}
for the transformation of the basis vectors.
According to these equations, the time axis and the first space axis of another inertial observer moving with speed $u$ relative to the fiducial observer undergo a pseudo-rotation governed by $\bm{g}$, each tilting towards the null cone by an equal amount so as to maintain (pseudo-)orthogonality (see for instance Figs.~1 and 2 of Ref.~\cite{Cardall2019Minkowski-and-G}).
That the worldline $\mathbb{T}'$ of the second observer is tilted relative to $\mathbb{T}$ according to its velocity is expected; the new result of Poincar\'e physics is that the hyperplane $\mathbb{S}'_{0}$, which reflects simultaneity according to the second observer at $t = t' = 0$, is tilted relative relative to $\mathbb{S}_0$.
This is the geometric origin of the relativity of simultaneity.
For $\mathrm{V}_\mathbb{G}$, the degeneration of the spacetime metric $\bm{g}$ and its inverse $\overleftrightarrow{\bm{g}}$ into the covector $\bm{\tau}$ and the $(2,0)$ tensor $\overleftrightarrow{\bm{\gamma}}$, related by the degeneracy condition of Eq.~(\ref{eq:DegeneracyCondition}), can be understood by returning $c$ to its value in, say, SI units; this is a large number representing the rapid speed of light propagation as perceived in ordinary human experience.
Then the trace of the null cone opens wide (large distance for small time) until it nearly coincides with the horizontal axis; indeed this coincidence is complete in the limit $c \rightarrow \infty$.
What was the invariant null cone for $\mathrm{V}_\mathbb{M}$ is now the invariant hypersurface $\mathbb{S}'_{0} = \mathbb{S}_0$, corresponding to the invariant covector $\bm{\tau}$ on $\mathrm{V}_\mathbb{G}$.
The basis vectors
\begin{align*}
\bm{e}'_0 &= \bm{e}_0 + \bm{e}_1  u,  \\
\bm{e}'_1 &= \bm{e}_1, \\
\bm{e}'_2 &= \bm{e}_2, \\
\bm{e}'_3 &= \bm{e}_3 
\end{align*}
transformed by a pure Galilei boost $\bm{\mathsf{B}}' = \bm{\mathsf{B}} \, \mathsf{P}_\mathbb{G}^+$ confirm that $\mathbb{T}'$ is tilted but $\mathbb{S}'_{0}$ is not (see for instance Figs.~3 and 4 of Ref.~\cite{Cardall2019Minkowski-and-G}).
This is the geometric origin of absolute simultaneity in Galilei physics, and the corresponding `floppiness' of straight inertial observer worldlines relative to a fixed surface of simultaneity results in the degeneracy of $\overleftrightarrow{\bm{\gamma}}$.

Having split spacetime into space and time for a given observer, a means of decomposing tensor fields on $\mathbb{M}$ or $\mathbb{G}$ into pieces `pointing along $\mathbb{T}$' and `tangent to $\mathbb{S}_t$' is needed.
On $\mathbb{M}$ these are `orthogonal decompositions' thanks to the spacetime metric $\bm{g}$; this allows flexibility in the raising and lowering of indices of decomposed pieces, but is not the fundamental source of the uniqueness of the decomposition.
On $\mathbb{G}$ unique decompositions are still possible even though they are not `orthogonal',
because the uniqueness that matters is the uniqueness inherent to expansion with respect to a particular basis.
As will be seen explicitly below, the point is that even without a metric, one always has an identity operator $\bm{\delta}$ that preserves an entire vector and a dual basis that can be used to pick off particular pieces. 

What is needed is a projection operator $\overleftarrow{\bm{\gamma}}$ that subtracts off the portion of a vector field lying along $\mathbb{T}$, which is parallel to $\bm{e}_0$; the result is necessarily a vector tangent to $\mathbb{S}_t$, because $\mathrm{V}_\mathbb{S}$ is spanned by the remaining basis vectors $\left( \bm{e}_1, \bm{e}_2, \bm{e}_3 \right)$. 
To emphasize the status of $\bm{e}_0$ as the value of the 4-velocity field of the fiducial observer associated with the selected Minkowski or Galilei basis, label it $\bm{n} = \bm{e}_0$ and call it the `fiducial observer vector'.
(The notion of 4-velocity will be introduced in Sec.~\ref{sec:Kinematics}.)
The other key element is the dual basis covector $\bm{e}_*^0$, for which $\bm{e}_*^0 \left( \bm{e}_0 \right) = 1$ and $\bm{e}_*^0 \left( \bm{e}_i \right) = 0$.
Thus $\bm{e}_*^0$ corresponds to a covector field `pointing completely away from' $\mathbb{S}_t$, in the sense that it vanishes when evaluated on any vector tangent to $\mathbb{S}_t$.
Because $\mathbb{S}_t$ is a level surface of the coordinate function $t$, the covector $\bm{e}_*^0$ also corresponds to the exterior derivative or (covariant) gradient of this function.
Thus at each point of $\mathbb{S}_t$ the relation $\bm{e}_*^0 = \mathbf{d} t = \bm{\nabla} t$ holds, with components $\begin{bmatrix} 1 & 0_i \end{bmatrix}$ in the selected basis.
With $\bm{e}_*^0 = \bm{\nabla} t $ and $\bm{e}_0 = \bm{n}$, the dual basis relationship reads
\begin{align*}
\bm{\nabla} t \cdot \bm{n} &= 1, \\
\bm{\nabla} t \cdot \bm{e}_i &= 0. 
\end{align*}
Thus any vector field $\bm{a}$ on $\mathbb{M}$ or $\mathbb{G}$ can be uniquely decomposed as 
\[
\bm{a} = a_{\bm{\nabla} t} \, \bm{n} + \bm{a}_{\overleftarrow{\bm{\gamma}}},
\]
with
\begin{align*}
a_{\bm{\nabla} t} &= \bm{\nabla} t \cdot \bm{a}, \\
 \bm{a}_{\overleftarrow{\bm{\gamma}}} &= \overleftarrow{\bm{\gamma}} \cdot \bm{a}. 
\end{align*}
Here
\begin{equation}
\overleftarrow{\bm{\gamma}} = \bm{\delta} - \bm{n} \otimes \bm{\nabla} t,
\label{eq:ProjectionOperator}
\end{equation}
where $\bm{\delta}$ is the identity tensor.
This is the desired projection operator: the second term in Eq.~(\ref{eq:ProjectionOperator}) removes the part along $\mathbb{T}$, leaving a vector field tangent to $\mathbb{S}_t$.

The same projection operator can be used to decompose covector fields on $\mathbb{M}$ and $\mathbb{G}$.
Writing 
\[
\bm{\omega} = \omega_{\bm{n}} \, \bm{\nabla} t  + \bm{\omega}_{\overleftarrow{\bm{\gamma}}}
\]
with
\begin{align*}
\omega_{\bm{n}} &= \bm{\omega} \cdot \bm{n},  \\
\bm{\omega}_{\overleftarrow{\bm{\gamma}}} &= \bm{\omega} \cdot \overleftarrow{\bm{\gamma}} 
\end{align*}
decomposes $\bm{\omega}$ into pieces that do and do not vanish when evaluated on a vector parallel to $\bm{n}$, namely $\bm{\omega}_{\overleftarrow{\bm{\gamma}}}$ and $\omega_{\bm{n}} \, \bm{\nabla} t $ respectively.

For $\mathbb{M}$, the `covector pointing away' $\bm{\nabla} t$ can be characterized in terms of a `dual observer vector' 
\[
\bm{\chi} = \overleftarrow{\bm{\nabla} t} \quad (\mathbb{M} \text{ only}),
\]
the metric dual of 
\begin{equation}
\underline{\bm{\chi}} = \bm{g} \cdot \bm{\chi} = \bm{\nabla} t   \quad (\mathbb{M} \text{ only}),
\label{eq:DualObserverCovector}
\end{equation} 
characterized by 
\[
\underline{\bm{\chi}} \left( \bm{n} \right) = \underline{\bm{\chi}} \cdot \bm{n} = 1 
	\quad (\mathbb{M} \text{ only}).
\]
This relation is what motivates the names `dual observer vector' for $\bm{\chi}$ and `dual observer covector' for $\underline{\bm{\chi}}$.
Moreover, because $\bm{g} \left( \bm{n}, \bm{n} \right) = -c^2$ while $\bm{g} \left( \bm{\chi}, \bm{n} \right) = 1$, it is clear from the non-degeneracy of $\bm{g}$ that
\begin{equation}
\bm{\chi} = -\frac{1}{c^2} \, \bm{n} \quad (\mathbb{M} \text{ only}),
\label{eq:DualFiducialObserverVector}
\end{equation}
that is, that the dual observer vector is a rescaled and oppositely-directed version of the observer vector, and unlike a 4-velocity is characterized by
\[
\bm{g} \left( \bm{\chi}, \bm{\chi} \right) = -\frac{1}{c^2} 
	\quad (\mathbb{M} \text{ only}).
\]
In terms of the dual observer covector $\underline{\bm{\chi}}$,  Eq.~(\ref{eq:ProjectionOperator}) becomes
\[
\overleftarrow{\bm{\gamma}} = \bm{\delta} - \bm{n} \otimes \underline{\bm{\chi}} 
	\quad (\mathbb{M} \text{ only})
\]
as follows from Eqs.~(\ref{eq:ProjectionOperator}) and (\ref{eq:DualObserverCovector}).

For $\mathbb{G}$, the `covector pointing away' $\bm{\nabla} t$ is already a fundamental structure, the previously-encountered invariant covector $\bm{\tau}$:
any Galilei basis must conform to this fundamental structure by having $\bm{e}_*^0 = \bm{\tau}$.
With $\bm{e}_0 = \bm{n}$ and $\bm{e}_*^0 = \bm{\tau}$, the dual basis relationship requires
\[
\bm{\tau} ( \bm{n} ) = \bm{\tau} \cdot \bm{n} = 1 \quad (\mathbb{G} \text{ only}),
\]
and the projection operator reads 
\[
\overleftarrow{\bm{\gamma}} = \bm{\delta} - \bm{n} \otimes \bm{\tau} 
	\quad (\mathbb{G} \text{ only})
\]
as follows from Eq.~(\ref{eq:ProjectionOperator}).

Naturally one has a different projection operator $\overleftarrow{\bm{\gamma}}'$ relative to a different Minkowski or Galilei observer vector $\bm{n}' = \bm{e}_0'$.
In the case of $\mathbb{M}$,
\[
\overleftarrow{\bm{\gamma}}' = \bm{\delta} - \bm{n}' \otimes \underline{\bm{\chi}}' 
	\quad (\mathbb{M} \text{ only}),
\]
with the dual observer vector $\bm{\chi}$ 
pseudo-rotating along with $\bm{n}$ to maintain the pseudo-orthogonality of $\mathbb{T}'$ and $\mathbb{S}'_{t'}$. 
In contrast, for $\mathbb{G}$ the projection operator relative to a different Galilei observer includes the same invariant covector $\bm{\tau}$:
\[
\overleftarrow{\bm{\gamma}}' = \bm{\delta} - \bm{n}' \otimes \bm{\tau} 
	\quad (\mathbb{G} \text{ only}).
\]
That is, a \textit{different} projection is made to the \textit{same} invariant hypersurface $\mathbb{S}_t$ embodied by the covector $\bm{\tau}$ pointing away from it.
Despite this `degeneracy', the decompositions relative to $\bm{n}$ and $\bm{n}'$ are unique.

In summary, appropriate contractions project out desired parts of decomposed tensors.
Contraction of vectors and the `vector-like parts' (contravariant indices) of more general tensors with $\underline{\bm{\chi}}$ or $\bm{\tau}$ projects out the `time' parts parallel to $\bm{n}$, while contraction with $\overleftarrow{\bm{\gamma}}$ projects out the `space' parts belonging to $\mathrm{V}_\mathbb{S}$.
Contraction of covectors and the `covector-like parts' (covariant indices) of more general tensors with $\bm{n}$ projects out the `time' parts that do not vanish when evaluated on vectors parallel to $\bm{n}$, while contraction with $\overleftarrow{\bm{\gamma}}$ projects out the `space' parts that vanish when evaluated on vectors parallel to $\bm{n}$.

Another issue is the question of how to extend a multilinear form originally defined only on $\mathrm{V}_\mathbb{S}$ to $\mathrm{V}_\mathbb{M}$ or $\mathrm{V}_\mathbb{G}$.
A case in point is the Euclid metric $\bm{\gamma}$ defined on $\mathrm{V}_\mathbb{S}$ by virtue of its rotational invariance, as described above.
An extension to $\mathrm{V}_\mathbb{M}$ or $\mathrm{V}_\mathbb{G}$, denoted by the same symbol $\bm{\gamma}$, is defined with the help of the projection operator defined above, which is used to enforce tangency to $\mathbb{S}_t$.
For $\bm{a}, \bm{b}$ in $\mathrm{V}_\mathbb{M}$ or $\mathrm{V}_\mathbb{G}$, the $(0,2)$ tensor $\bm{\gamma}$ on $\mathrm{V}_\mathbb{M}$ or $\mathrm{V}_\mathbb{G}$ is defined by
\begin{equation}
\bm{\gamma} \left( \bm{a}, \bm{b} \right) 
	= \bm{\gamma} \left( \overleftarrow{\bm{\gamma}} ( \bm{a} ),\overleftarrow{\bm{\gamma}}( \bm{b} ) \right).
\label{eq:ThreeMetricExtension}
\end{equation}
The $\bm{\gamma}$ on the left is the tensor extended to $\mathrm{V}_\mathbb{M}$ or $\mathrm{V}_\mathbb{G}$, and the $\bm{\gamma}$ on the right is the original tensor on $\mathrm{V}_\mathbb{S}$.
The notation may seem a bit odd, but it evades a proliferation of symbols, and the meaning is generally clear from the context.

The notational subtleties of the various tensors $\bm{\gamma}$, $\overleftarrow{\bm{\gamma}}$,  and $\overleftrightarrow{\bm{\gamma}}$ can now be explained.
When denoting tensors on $\mathrm{V}_\mathbb{S}$, these are simply the 3-metric; the 3-metric with first index raised, that is, the identity tensor on $\mathrm{V}_\mathbb{S}$; and the inverse 3-metric.
When denoting tensors on $\mathrm{V}_\mathbb{M}$, it turns out that
\[
\bm{\gamma} = \bm{g} - \underline{\bm{n}} \otimes \underline{\bm{\chi}} 
	\quad (\mathbb{M} \text{ only}),
\]
and the versions adorned with arrows reflect index raising with $\bm{g}$.
(On $\mathrm{V}_\mathbb{M}$ the identity tensor is related to the metric and its inverse by raising an index of the metric itself: $\bm{\delta} = \overleftarrow{\bm{g}} = \overleftrightarrow{\bm{g}} \cdot \bm{g}$ or $\bm{\delta} = \overrightarrow{\bm{g}} = \bm{g} \cdot \overleftrightarrow{\bm{g}}$.)
When denoting tensors on $\mathrm{V}_\mathbb{G}$, each of the tensors $\bm{\gamma}$, $\overleftarrow{\bm{\gamma}}$, and $\overleftrightarrow{\bm{\gamma}}$ can be defined as distinct projection tensors, but they are not related by metric duality; the arrows must simply be considered integral to the symbols defining those particular tensors.

A word on a unified presentation of the volume form $\bm{\varepsilon}$ on $\mathbb{M}$ and $\mathbb{G}$ is in order.
This is defined in terms of a right-handed Minkowski or Galilei basis respectively, with respect to which the components are
\[
\varepsilon_{\mu \nu \rho \sigma} = \left[ \mu \nu \rho \sigma \right]
\]
in either case, where the right hand side is the alternating (permutation) symbol.
Given some fiducial Minkowski or Galilei basis, it may be useful to employ coordinates that include curvilinear space coordinates on $\mathbb{S}_t$ and/or observers in (generally non-inertial) motion relative to the fiducial observer, with (generally curved) worldlines exhibiting a 3-velocity $\bm{\beta} \in \mathrm{V}_\mathbb{S}$ with rectangular components given by $\bm{\beta} = \bm{e}_i \, b^i$ according to the fiducial Minkowski or Galilei observer.
The matrix representing a basis change governing this case on either $\mathrm{V}_\mathbb{M}$ or $\mathrm{V}_\mathbb{G}$ is of the form
\[
\mathsf{P} = \begin{bmatrix} 1 & \mymathsf{0} \\[5pt] \mathsf{b} & \mathsf{C} \end{bmatrix} = \begin{bmatrix} 1 & 0_j \\[5pt] b^i & {C^i}_j \end{bmatrix},
\] 
with $\mathsf{b} \in \mathbb{R}^{3 \times 1}$ and $\mathsf{C} \in \mathbb{R}^{3 \times 3}$.
Since the 3-metric $\bm{\gamma}$ on $\mathrm{V}_\mathbb{S}$ is represented by the identity matrix with respect to a Minkowski or Galilei basis, the matrix $\sfgamma$ representing it in the curvilinear/moving basis has components $\gamma_{ij} = {C^a}_i \, 1_{a b} \, {C^b}_j$, so that $\sfgamma = \mathsf{C}^\mathrm{T} \mathsf{C}$.
Thus $\det \mathsf{P} = \det \mathsf{C} = \sqrt{\gamma}$, where $\gamma = \det \sfgamma$, and the components of the volume form on either $\mathbb{M}$ or $\mathbb{G}$ become 
\[
\varepsilon_{\mu \nu \rho \sigma} = \sqrt{\gamma} \, \left[ \mu \nu \rho \sigma \right]
\]
according to Eq.~(\ref{eq:VolumeFormTransformed}).\footnote{As an aside, in the case of $\mathbb{M}$ it is interesting to note what the matrix $\mathsf{g} = \mathsf{P}^\mathrm{T} \, \sfeta \, \mathsf{P}$ representing $\bm{g}$ becomes under such a transformation.
It may help to put $\mathsf{1} = \mathsf{C}^{-\mathrm{T}}\, \sfgamma \, \mathsf{C}^{-1}$ in the space part of $\sfeta$, and note that the 3-column $\sfbeta = \mathsf{C}^{-1} \, \mathsf{b}$ collects the components of $\bm{\beta}$ relative to the curvilinear basis, with the 3-row $\underline{\sfbeta} = (\sfgamma \, \sfbeta)^\mathrm{T}$ representing the covector $\underline{\bm{\beta}} = \bm{\gamma} \cdot \bm{\beta}$ on $\mathrm{V}_\mathbb{S}$.
The result is
\[
\mymathsf{g} 
	= \begin{bmatrix} -c^2 + \underline{\sfbeta} \, \sfbeta & \underline{\sfbeta} \\[5pt]
		\underline{\sfbeta}^\mathrm{T} & \sfgamma \end{bmatrix}
	= \begin{bmatrix} -c^2 + \beta_a \, \beta^a & \beta_i \\[5pt]
		\beta_j & \gamma_{i j} \end{bmatrix},
\]  
which up to setting the lapse function $\alpha = 1$ is of the same form as in the $1+3$ (traditionally, $3+1$) formalism of Poincar\'e general relativity.}

It is useful to consider further consequences of spacetime foliation for the spacetime volume form $\bm{\varepsilon}$ and the spacetime exterior derivative operator $\mathbf{d}$.
Because $\bm{n} = \bm{e}_0$, the contraction
\begin{equation}
\underaccent{\check}{\bm{\varepsilon}} = \bm{\varepsilon}(\bm{n}, \, . \, , \, . \, , \, . \, ) 
	= \bm{n} \cdot \bm{\varepsilon}
\label{eq:VolumeContraction} 
\end{equation}
yields the space volume form $ \underaccent{\check}{\bm{\varepsilon}}$ on $\mathbb{S}_t$,
with components
\[
\underaccent{\check}{\varepsilon}_{i j k} = \varepsilon_{0 i j k} = \sqrt{\gamma} \left[ i j k \right].
\]
Conversely, because $\underline{\bm{\chi}} = \bm{e}_*^0$ and $\bm{\tau} = \bm{e}_*^0$ on $\mathbb{M}$ and $\mathbb{G}$ respectively,
\begin{equation}
\bm{\varepsilon} = 
	\begin{cases}
		\underline{\bm{\chi}} \wedge \underaccent{\check}{\bm{\varepsilon}}
			& (\text{on } \mathbb{M}) \\[5pt]
		\bm{\tau} \wedge \underaccent{\check}{\bm{\varepsilon}}
			& (\text{on } \mathbb{G})
	\end{cases}
\label{eq:VolumeDecomposition}
\end{equation}
is a useful factorization of the spacetime volume form $\bm{\varepsilon}$.
For vectors $\bm{a}, \bm{b} \in \mathrm{V}_\mathbb{S}$, the cross product $\bm{a} \times \bm{b}$ familiar from $\mathbb{R}^3$ with Euclid metric $\bm{\gamma}$---or more precisely, the covector $\underline{\bm{a} \times \bm{b}} = \bm{\gamma} \cdot (\bm{a} \times \bm{b})$ that is the metric dual of $\bm{a} \times \bm{b}$---is defined by
\[
 \underaccent{\check}{\bm{\varepsilon}} ( \bm{a}, \bm{b}, \, . \, ) = \underline{\bm{a} \times \bm{b}}.
\] 
The spacetime exterior derivative, represented symbolically as
\[
\mathbf{d} = \bm{e}_*^\alpha \wedge \frac{\partial}{\partial x^\alpha}
\]
as in Eq.~(\ref{eq:ExteriorDerivativeOperator}), breaks naturally into
\begin{equation}
\mathbf{d} = 
	\left\{
	\begin{aligned}
		\underline{\bm{\chi}} \wedge \frac{\partial}{\partial t} 
		+ \mathrm{\underaccent{\check}{\mathbf{d}}}
			& & (\text{on } \mathbb{M}) \\[5pt]
		\bm{\tau} \wedge \frac{\partial}{\partial t} +  \mathrm{\underaccent{\check}{\mathbf{d}}}
			& & (\text{on } \mathbb{G}),
	\end{aligned}
	\right.
\label{eq:ExteriorDerivativeDecomposition}
\end{equation}
where
\[
 \mathrm{\underaccent{\check}{\mathbf{d}}} = \bm{e}_*^i \wedge \frac{\partial}{\partial x^i}
\]
is the exterior derivative operator on $\mathbb{S}_t$.
Combining the volume form and the exterior derivative on $\mathbb{S}_t$ enables contact with the vector calculus familiar on $\mathbb{R}^3$ with Euclid metric $\bm{\gamma}$.
For a vector field $\bm{a}$ tangent to $\mathbb{S}_t$, the expression
\[
 \mathrm{\underaccent{\check}{\mathbf{d}}} ( \bm{a} \cdot  \underaccent{\check}{\bm{\varepsilon}} ) 
	= ( \mathrm{\underaccent{\check}{\bm{\nabla}}} \cdot \bm{a} ) \,  \underaccent{\check}{\bm{\varepsilon}}
\]
defines $\mathrm{\underaccent{\check}{\bm{\nabla}}} \cdot \bm{a}$, the 3-divergence of $\bm{a}$; and the expression
\[
 \mathrm{\underaccent{\check}{\mathbf{d}}} \underline{\bm{a}} 
	=  \underaccent{\check}{\bm{\varepsilon}} ( \mathrm{\underaccent{\check}{\bm{\nabla}}} \times \bm{a}, \, . \, , \, . \, )
	= ( \mathrm{\underaccent{\check}{\bm{\nabla}}} \times \bm{a} ) \cdot  \underaccent{\check}{\bm{\varepsilon}},
\]
where $\underline{\bm{a}} = \bm{\gamma} \cdot \bm{a}$, defines $\mathrm{\underaccent{\check}{\bm{\nabla}}} \times \bm{a}$, the curl of $\bm{a}$.

Finally, a word about causality.
Recall that for $\mathrm{V}_\mathbb{M}$, vectors $\bm{a} \ne \bm{0}$ are classified as timelike for $\bm{g}(\bm{a},\bm{a}) < 0$, spacelike for $\bm{g}(\bm{a},\bm{a}) > 0$, and null for $\bm{g}(\bm{a},\bm{a}) = 0$.
These correspond to vectors `inside' the null cone, `outside' the null cone, and `on' the null cone respectively.
It is well known that for two events separated by a spacelike vector it is possible for the sign of the time interval between them to be reversed by a Lorentz transformation.
In contrast, while simultaneity is relative for two events separated by a timelike vector, the time ordering of the events is invariant.
Particles, and signals transmitted by field disturbances, must have timelike worldlines (curves in $\mathbb{M}$ with tangent vectors everywhere timelike) or straight null worldlines (curves in $\mathbb{M}$ with unchanging null tangent vector) directed toward the future.
However, in the case of $\mathrm{V}_\mathbb{G}$ the distinction between spacelike and null vectors vanishes as the past and future light cones merge with the invariant surface of simultaneity.
Time intervals between events are invariant under Galilei transformations.
There is no upper limit to the speed of particles or of signals transmitted by field disturbances, and indeed forces effecting instantaneous action at a distance are not excluded.
For $\mathrm{V}_\mathbb{G}$, vectors $\bm{a} \ne \bm{0}$ are classified as timelike for $\bm{\tau} \left( \bm{a} \right) \ne 0$ and null/spacelike for $\bm{\tau} \left( \bm{a} \right) = 0$.

\section{A material particle on $\mathbb{M}$ or $\mathbb{G}$}
\label{sec:MaterialParticleMG}

Having described the spacetimes $\mathbb{M}$ and $\mathbb{G}$, a discussion of classical (that is, non-quantum) physics thereon begins with a description of a material particle.
First, its kinematics: where is the particle, and how fast is it moving?
This is given by a worldline---a parametrized curve in spacetime---and the tangent vector to that worldline, the particle's 4-velocity. 
Second, its dynamics: what determines the particle's worldline?
This is described by momentum---a covector related to velocity via the particle mass and a metric---and the force acting on the particle, a covector that relates the values of particle momentum on neighboring points of the worldline. 

\subsection{Kinematics}
\label{sec:Kinematics}

Call a `material particle' any effectively pointlike entity whose history in spacetime $\mathbb{M}$ or $\mathbb{G}$ is represented by a timelike worldline, that is, a parametrized curve whose tangent vector is timelike at each of its points.
Let the particle be at spacetime event $\mathbf{X}(\tau) \in \mathbb{M}, \mathbb{G}$ at proper time $\tau \in \mathbb{R}$. 
Increments $\mathrm{d}\tau$ of proper time are measured by an ideal clock carried by an observer riding along with the particle.
Consider the 4-vector 
\[
\mathrm{d}\mathbf{X} = \overrightarrow{\mathbf{X}(\tau) \, \mathbf{X}(\tau + \mathrm{d}\tau)}
\] 
connecting two points on the worldline separated by the proper time increment $\mathrm{d}\tau$.
In the limit $\mathrm{d}\tau \rightarrow 0$ one has the tangent vector
\[
\bm{U} = \frac{\mathrm{d}\mathbf{X}}{\mathrm{d}\tau},
\]
the 4-velocity of the particle.
While the spacetime position $\mathbf{X}(\tau)$ and the 4-velocity $\bm{U}$ can both be represented by 4-columns as discussed in Sec.~\ref{sec:AffineSpaces}, they are of a fundamentally different nature: $\mathbf{X}(\tau)$ is a point of $\mathbb{M}$ or $\mathbb{G}$, and the tangent vector $\bm{U}$ is an element of $\mathrm{V}_\mathbb{M}$ or $\mathrm{V}_\mathbb{G}$.

In order to relate the 4-velocity $\bm{U}$ to something operationally measurable, select a fiducial observer with global coordinates $( t, x^i )$ associated with a choice of origin $\mathbf{O}$ of $\mathbb{M}$ or $\mathbb{G}$ and a Minkowski or Galilei basis $( \bm{n}, \bm{e}_i )$ for $\mathrm{V}_\mathbb{M}$ or $\mathrm{V}_\mathbb{G}$ respectively, as discussed in Sec.~\ref{sec:SpacetimeFoliation}.
Associated with the fiducial observer is a straight time axis $\mathbb{T} = \left\{ \mathbf{O} + \bm{n} \, t \mid t \in \mathbb{R}\right\}$ and a foliation of spacetime into position space slices, the affine hyperplanes $(\mathbb{S}_t)$.
(In fact, $\mathbb{T}$ is the worldline of the fiducial observer, parametrized by the fiducial observer's proper time $t$ and with constant 4-velocity $\bm{n}$ as the tangent vector at each point of $\mathbb{T}$.)
Decompose the particle 4-velocity $\bm{U}$ into pieces parallel to $\mathbb{T}$ and tangent to $\mathbb{S}_t$ by writing
\begin{equation}
\begin{aligned}
\bm{U} &= \frac{\mathrm{d}t}{\mathrm{d}\tau} \, \frac{\mathrm{d}\mathbf{X}}{\mathrm{d}t} \\[5pt]
	&= \frac{\mathrm{d}t}{\mathrm{d}\tau} \left( \bm{n} + \bm{v} \right).
\end{aligned}
\label{eq:FourVelocityDecomposed}
\end{equation}
This follows from representing the particle position $\mathbf{X}$ by the 4-column
\begin{equation}
\mathsf{X} = \begin{bmatrix} t \\ \mathsf{x} ( t )  \end{bmatrix} 
	= \begin{bmatrix} t \\ x^i ( t ) \end{bmatrix},
\label{eq:PointFourColumn} 
\end{equation}
the time axis direction $\bm{n}$ by the 4-column
\[
\mymathsf{n} = \begin{bmatrix} 1 \\ \mymathsf{0} \end{bmatrix} = \begin{bmatrix} 1 \\ 0^i \end{bmatrix}, \ \ \ 
\]
 and the particle 3-velocity $\bm{v}$ (tangent to $\mathbb{S}_t$) by the 4-column 
\[
\mymathsf{v} = \begin{bmatrix} 0 \\ \mathsf{v} \end{bmatrix} 
		= \begin{bmatrix} 0 \\ \mathrm{d} \mathsf{x} / \mathrm{d}t \end{bmatrix}
		= \begin{bmatrix} 0 \\ \mathrm{d} x^i / \mathrm{d}t \end{bmatrix}
\]
relative to the fiducial observer.
This last expression for 3-velocity calls for comment, as the symbol $\mathsf{v} \in \mathbb{R}^{4 \times 1}$ and $\mathsf{v} \in \mathbb{R}^{3 \times 1}$ is being used in two different ways.
For a vector field tangent to $\mathbb{S}_t$, such as the 3-velocity satisfying
\[
\bm{v} = \overleftarrow{\bm{\gamma}} \cdot \bm{v},
\]
use the same symbol $\bm{v}$ to denote the vector field on $\mathbb{S}_t$ and the vector field on $\mathbb{M}$ or $\mathbb{G}$ that happens to be tangent to $\mathbb{S}_t$, and similiarly use $\mathsf{v}$ for the 3-column and and 4-column representing them.

The leading factor $\mathrm{d}t / \mathrm{d}\tau$ in Eq.~(\ref{eq:FourVelocityDecomposed}) is given by the fundamental structures governing causality, namely $\bm{g}$ on $\mathbb{M}$ and $\bm{\tau}$ on $\mathbb{G}$.
A basic postulate of physics on $\mathbb{M}$ is that the proper time increment $\mathrm{d}\tau$ is given in terms of the spacetime distance between $\mathbf{X}(\tau)$ and $\mathbf{X}(\tau + \mathrm{d}\tau)$:
\begin{equation}
c \, \mathrm{d}\tau 
	= \sqrt{ - \bm{g} \left( \mathrm{d}\mathbf{X}, \mathrm{d}\mathbf{X} \right)}
	= c \, \Lambda_{\bm{v}}^{-1} \, \mathrm{d}t \quad (\mathbb{M} \text{ only}),
\label{eq:ProperTimeM}
\end{equation}
where the Lorentz factor $\Lambda_{\bm{v}}$ is given by 
\[
 \Lambda_{\bm{v}}^{-1} = \sqrt{ 1 - \frac{\bm{\gamma} \left( \bm{v}, \bm{v} \right) }{c^2} }
 	 \quad (\mathbb{M} \text{ only}),
\]
found with the help of the Minkowski matrix $\mathsf{g} = \sfeta$ relative to a Minkowski basis, and expressed in terms of the Euclid 3-metric $\bm{\gamma}$ on $\mathbb{S}_t$.
The analogous postulate on $\mathbb{G}$ is that
\begin{equation}
\mathrm{d}\tau = \bm{\tau} \left( \mathrm{d}\mathbf{X} \right) = \mathrm{d} t
	\quad (\mathbb{G} \text{ only}),
\label{eq:ProperTimeG}
\end{equation}
consistent with $\Lambda_{\bm{v}} \rightarrow 1$ as $c \rightarrow \infty$.
Therefore
\begin{equation}
\bm{U} =
	\begin{cases}
		 \Lambda_{\bm{v}} \left( \bm{n} + \bm{v} \right) & (\text{on } \mathbb{M})
		 \\[5pt]
	  \phantom{ \Lambda_{\bm{v}} } \ \left. \bm{n} + \bm{v} \right. \ & (\text{on } \mathbb{G}),
	\end{cases}
\label{eq:FourVelocity}
\end{equation}
represented by the 4-column 
\begin{equation}
\mathsf{U} =
	\begin{cases}
		 \Lambda_\mathsf{v} \begin{bmatrix} 1 \\ \mathsf{v}\end{bmatrix} 
			& (\text{on } \mathbb{M}) \\[15pt]
		  \phantom{\Lambda_\mathsf{v}} \begin{bmatrix} 1 \\ \mathsf{v} \end{bmatrix}
			& (\text{on } \mathbb{G})
	\end{cases}
\label{eq:FourVelocityFiducial}
\end{equation}
relative to the fiducial observer.

This account of the 4-velocity $\bm{U}$ is an example of a principle mentioned in Sec.~\ref{sec:SpacetimeFoliation}: tensors on spacetime are not measured directly, and must be time-space decomposed in order to acquire operational physical significance.
In this case measurement of the 3-velocity components $\left( v^i \right)$ at each $t$ allows reconstruction of all components $\left( U^\mu \right)$ and therefore of $\bm{U}$ itself along the worldline.

That four spacetime components can be determined from three measured space components at each instant indicates that some constraint, characteristic of 4-velocities and involving the fundamental structures on $\mathbb{M}$ and $\mathbb{G}$, is at work.
For $\mathbb{M}$ the constraint characterizing a 4-velocity is that its squared length as given by $\bm{g}$ is equal to $-c^2$, as exemplified by $\bm{g} \left( \bm{n}, \bm{n} \right) = -c^2$ and $\bm{g} \left( \bm{U}, \bm{U} \right) = -c^2$.
For $\mathbb{G}$ the constraint characterizing a 4-velocity is that it yields the value $1$ when evaluated by $\bm{\tau}$, exemplified by $\bm{\tau} \left( \bm{n} \right) = 1$ and $\bm{\tau} \left( \bm{U} \right) = 1$.

The condition for a vector to qualify as a 4-velocity on $\mathbb{M}$ looks more similar to 
\[
\bm{\tau} \left( \bm{U} \right) = 1 \quad (\mathbb{G} \text{ only}) 
\]
when one defines a dual observer vector $\bm{\chi}$ whose metric dual $\underline{\bm{\chi}}$ (`dual observer covector') plays a role partly like that of $\bm{\tau}$ on $\mathbb{G}$, as discussed in Sec.~\ref{sec:SpacetimeFoliation}:
defining $\bm{\chi} = \overleftarrow{\bm{\nabla} t} = - \bm{n} / c^2$, the condition $\bm{g} \left( \bm{n}, \bm{n} \right) = -c^2$ is equivalent to $\underline{\bm{\chi}} \left( \bm{n} \right) = 1$.
In partial similarity one can define 
\begin{equation}
\bm{\chi}_{\bm{U}} = - \frac{1}{c^2} \, \bm{U} \quad (\mathbb{M} \text{ only}), 
\label{eq:DualComovingObserverVector}
\end{equation}
so that the condition $\bm{g} \left( \bm{U}, \bm{U} \right) = -c^2$ is equivalent to 
\[
\underline{\bm{\chi}}_{\bm{U}} \left( \bm{U} \right) = 1 \quad (\mathbb{M} \text{ only}).
\]
Note however that because $\bm{U}$ is not the constant direction of a straight line in $\mathbb{M}$ when the particle is accelerated (worldline with curvature, $\mathrm{d}\bm{U} / \mathrm{d}\tau \ne 0$), the affine hyperplanes $\mathbb{S}_{\bm{U}(\tau)}$ orthogonal to $\bm{U}(\tau)$ are not parallel for different values of $\tau$, and therefore do not constitute a foliation of $\mathbb{M}$.
This is why the similarity is only partial: unlike $\underline{\bm{\chi}} = \bm{\nabla} t$ for the fiducial inertial observer, for an accelerated particle on $\mathbb{M}$ the covector field $\underline{\bm{\chi}}_{\bm{U}}$ is not in general equal to the (covariant) gradient of a global time coordinate function.

These constraints on the 4-velocity of a material particle on $\mathbb{M}$ or $\mathbb{G}$ encode two assumptions built into this discussion. 
First, the timelike character of the worldline is invariant: no boost can make a tangent vector null or spacelike relative to $\bm{g}$ on $\mathbb{M}$ or $\bm{\tau}$ on $\mathbb{G}$.
Second, a `comoving observer' always exists, so that proper time $\tau$ can be used to parametrize the worldline.
There is always a local boost $\mathsf{L}$, Minkowski or Galilei as appropriate, for which
\begin{equation}
\mymathsf{U} = \mymathsf{L} \, \mymathsf{U}' = \mymathsf{L} \, \begin{bmatrix} 1 \\ \mymathsf{0} \end{bmatrix},
\label{eq:FourVelocityBoost}
\end{equation}
that is, there is always a local boost (here $\mathsf{L}^{-1}$) which results in a vanishing 3-velocity.

Moreover, instead of decomposing tensors relative to the fiducial observer with 4-velocity $\bm{n}$, one can locally decompose tensor fields as measured by a comoving observer with 4-velocity $\bm{U}$.
(For example, a vector may be decomposed into a piece parallel to $\bm{U}$, and a piece tangent to $\mathbb{S}_{\bm{U}(\tau)}$ on $\mathbb{M}$ or the invariant $\mathbb{S}_t$ on $\mathbb{G}$.)
On $\mathbb{M}$ this is accomplished with $\underline{\bm{\chi}}_{\bm{U}}$ and the projection operator
\[
\overleftarrow{\bm{\gamma}_{\bm{U}}} 
	= \bm{\delta} - \bm{U} \otimes \underline{\bm{\chi}}_{\bm{U}}
		\quad (\mathbb{M}\text{ only}), 
\]
while $\bm{\tau}$ and
\[
\overleftarrow{\bm{\gamma}_{\bm{U}}} 
	= \bm{\delta} - \bm{U} \otimes \bm{\tau}
		\quad (\mathbb{G}\mathrm{\ only}),
\]
are used on $\mathbb{G}$.

As far as a spacetime description goes, so far so good on both $\mathbb{M}$ and $\mathbb{G}$: a spacetime description of particle kinematics---specifying where a particle is (a point $\mathbf{X}(\tau)$ on its worldline), and how fast it is moving (the 4-velocity $\bm{U}$ tangent to the worldline)---is unproblematic in either case.

\subsection{Dynamics}
\label{sec:Dynamics}

If the kinematics of a material particle is the \textit{description} of its motion (specification of its worldline), the dynamics of the particle is the \textit{prescription} of that motion (that which determines the shape of the worldline).
The spacetime formulation of Newton's first law on affine spacetimes $\mathbb{M}$ and $\mathbb{G}$ is that, in the absence of an external force, the worldline of a material particle is a straight timelike line with constant tangent vector $\bm{U}$.
As for a spacetime formulation of Newton's second law, which produces worldline curvature, there are both vector and covector (or 1-form) versions.
In different ways both versions ultimately require invocation of a metric.
Indeed metric duality might be regarded as the geometric embodiment of the conjugate relationship, inherent in any dynamical scheme, between position and velocity (represented by vectors) on the one hand, and momentum and force (represented by covectors) on the other.    
Thus the absence of a spacetime metric limits the nature of a spacetime version of Newton's second law on $\mathbb{G}$: a spacetime account of dynamics is more problematic than a spacetime account of mere kinematics.

Define on both $\mathbb{M}$ and $\mathbb{G}$ an inertia-momentum 4-vector
\[
\bm{I} = M \, \bm{U},
\]
where the mass $M$ quantifies the particle's resistance to a bending of its worldline by an external force.
Its components relative to the fiducial observer, 
\begin{equation}
\mymathsf{I} = M \, \mymathsf{U}
	= \begin{cases} 
		\begin{bmatrix} M \Lambda_\mathsf{v}  \, \phantom{\mathsf{v}} 
				\\ M \Lambda_\mathsf{v} \, \mathsf{v}
			\end{bmatrix} 
			& (\mathrm{on \ } \mathbb{M}) \\[15pt]
		\begin{bmatrix} M \, \phantom{\mathsf{v}} 
				\\ M  \, \mathsf{v}
			\end{bmatrix} 
			& (\mathrm{on \ } \mathbb{G})
	\end{cases}
\label{eq:InertiaMomentumFiducial}
\end{equation}
as obtained from Eq.~(\ref{eq:FourVelocityFiducial}), are the inertia and vector 3-momentum.

Postulate also a 4-force covector $\bm{\Upsilon}_{\!\bm{I}}$.
A couple of reasons from experience with a 3-force $\bm{f}$ in physics according to Newton motivate the fundamental covector nature of a force.
First, in many cases it is given as the gradient of a scalar potential, for instance with components $f_i = \partial_i \phi = \partial \phi / \partial x^i$, coming naturally with a covariant lower index.
Second, given a particle displacement in position space with components $\mathrm{d} x^i$ effected by a 3-force, the work done by the force on the particle is given directly by the contraction $f_a \, \mathrm{d} x^a$, without any need for a metric; for force regarded as a vector, one would have to write instead $f^a \, \gamma_{a b} \, \mathrm{d}x^b$, presupposing and interposing a metric as an additional structure.

The vector version of a spacetime formulation of Newton's second law requires an index raising of the force:
\begin{equation}
\frac{\mathrm{d} \bm{I}}{\mathrm{d} \tau} = \overleftarrow{\bm{\Upsilon}_{\!\bm{I}}}.
\label{eq:NewtonSecondVector}
\end{equation}
In order that this equation apply on both $\mathbb{M}$ and $\mathbb{G}$, interpret the right-hand side as
\begin{equation}
\overleftarrow{\bm{\Upsilon}_{\!\bm{I}}} = 
	\begin{cases}
		\bm{\Upsilon}_{\!\bm{I}} \cdot \overleftrightarrow{\bm{g}} 
			& (\text{on  } \mathbb{M}) \\[5pt]
		\bm{\Upsilon}_{\!\bm{I}} \cdot \overleftrightarrow{\bm{\gamma}} 
			& (\text{on } \mathbb{G}),
	\end{cases}
\label{eq:FourForceVector}
\end{equation}
recalling that the fundamental structure $\overleftrightarrow{\bm{\gamma}}$ on $\mathbb{G}$ is the $c \rightarrow \infty$ limit of the inverse metric on $\mathbb{M}$.
Comparing Eq.~(\ref{eq:InverseMetricGalileiBasis}) with Eq.~(\ref{eq:InverseMetricMinkowskiBasis}), consistent with $\bm{\tau} \cdot \overleftrightarrow{\bm{\gamma}} = 0$ as opposed to $\bm{g} \cdot \overleftrightarrow{\bm{g}} = 1$, it is apparent that $\overleftrightarrow{\bm{\gamma}}$ has a projective character that will prove consequential.

Consider first a decomposition relative to the comoving observer of the vector version of Newton's second law.
On the left-hand side,
\[
\frac{\mathrm{d} \bm{I}}{\mathrm{d} \tau} 
	= \frac{\mathrm{d} M}{\mathrm{d} \tau} \, \bm{U} 
		+ M  \bm{a},
\]
where the 4-acceleration
\[
\bm{a} = \frac{\mathrm{d} \bm{U}}{\mathrm{d} \tau} 
\]
has been defined.
This vector is tangent to $\mathbb{S}_{\bm{U}(\tau)}$ on $\mathbb{M}$ or the invariant $\mathbb{S}_t$ on $\mathbb{G}$, and therefore may be regarded as the 3-acceleration measurable by the comoving observer (as each person knows from experiencing the start and stop of her own motion).
On $\mathbb{M}$ tangency to $\mathbb{S}_t$ is apparent from
\[
\underline{\bm{\chi}}_{\bm{U}} \cdot \bm{a} 
	= - \frac{1}{c^2} \, \bm{g} \left( \bm{U},  \frac{\mathrm{d} \bm{U}}{\mathrm{d} \tau} \right)
	= - \frac{1}{2 c^2} \, \frac{\mathrm{d}}{\mathrm{d} \tau} \, \bm{g} \left( \bm{U}, \bm{U} \right)
	= 0.
\]
On $\mathbb{G}$,
\[
\bm{\tau} \cdot \bm{a} = \sftau \, \mathsf{a} = 0
\]
is apparent from evaluation in terms of the 4-row and 4-column representations $\sftau$ and $\mathsf{a} = \mathrm{d} \mathsf{U} / \mathrm{d} \tau$ with respect to the Galilei basis of the fiducial observer.
As for the right-hand side of the vector version of Newton's second law, write the 4-force covector or 1-form as
\[
\bm{\Upsilon}_{\!\bm{I}} = 
	\begin{cases}
		- \theta \, \underline{\bm{\chi}}_{\bm{U}} + \bm{\mathscr{f}} 
			&  (\text{on } \mathbb{M})  \\[5pt]
 		 - \theta \, \bm{\tau}\phantom{_{\bm{U}}} + \bm{\mathscr{f}} 
			&  (\text{on } \mathbb{G}). 
	\end{cases}
\]
Here 
\[
\bm{\Upsilon}_{\!\bm{I}} \cdot \overleftarrow{\bm{\gamma}_{\bm{U}}} = \bm{\mathscr{f}},
\]
with $\bm{\mathscr{f}} \cdot \bm{U} = 0$, is the 3-force covector according to the comoving observer.
Soon it will become apparent that the scalar
\begin{equation}
- \bm{\Upsilon}_{\!\bm{I}} \cdot \bm{U} = \theta
\label{eq:ComovingHeating}
\end{equation}
is a heating rate affecting only the particle's internal energy.
Raising the index and noting Eq.~(\ref{eq:DualComovingObserverVector}), the 4-force vector is
\[
\overleftarrow{\bm{\Upsilon}_{\!\bm{I}}} =
	\left\{
	\begin{aligned}
		 \frac{ \theta }{ c^2 } \, \bm{U} + \overleftarrow{\bm{\mathscr{f}}} 
			& & (\text{on } \mathbb{M}) \\
 		\overleftarrow{\bm{\mathscr{f}}} 
			& & (\text{on } \mathbb{G}).
	\end{aligned}
	\right. 
\]
On $\mathbb{G}$ the projective character of the degenerate inverse `metric' ($\bm{\tau} \cdot \overleftrightarrow{\bm{\gamma}} = 0$), or a direct $c \rightarrow \infty$ limit, cause the heating rate to disappear from the vector version of the 4-force.
The comoving observer time projection of Eq.~(\ref{eq:NewtonSecondVector}) obtained by contraction with $\underline{\bm{\chi}}_{\bm{U}}$ on $\mathbb{M}$ or $\bm{\tau}$ on $\mathbb{G}$ is
\begin{equation}
 \frac{\mathrm{d} M}{\mathrm{d} \tau} =
 	\left\{ 
 	\begin{aligned}
		\frac{ \theta }{ c^2 } & & (\text{on } \mathbb{M}) \\[5pt]
		0 & & (\text{on } \mathbb{G}).
	\end{aligned}
	\right.
\label{eq:MassRate}
\end{equation}
On $\mathbb{M}$, Poincar\'e physics allows in principle the particle mass to be changed by virtue of an external heating rate $\theta$ that alters the particle internal energy $M c^2$.
(\ADD{Consider for example an infinitesimal element of a simple material continuum, a `composite particle' composed of $N$ fundamental particles of rest mass $m$; this composite particle would be characterized by a `composite mass' $M = N m + E/c^2$, where $E$ is the mutual internal energy of the fundamental particles in the element, excluding rest mass energy.} 
Conversely, for a single fundamental particle with no internal structure and constant mass, of necessity $\theta = 0$.\footnote{\ADD{A treatment of the relationship between the equation for a material particle and the equations governing a material continuum is beyond the scope of this paper, but see for example Ref.~\cite{Cardall2020Combining-3-Mom} for a discussion along these lines.}})
On $\mathbb{G}$, Galilei physics maintains a strict distinction between inertia and energy and enforces conservation of mass.\footnote{\ADD{The vanishing time derivative of mass, interpreted here as an expression or even derivation of the fundamental principle of mass conservation in Galilei physics (see also Ref.~\cite{Cardall2020Combining-3-Mom} in the case of a material continuum), arises as a mathematical consequence of the assumption that force---like momentum---is fundamentally a 4-covector, together with the degeneracy of the inverse `metric' on $\mathbb{G}$ used to obtain a 4-vector version of the force.
For the purpose of modeling in a simple way a rocket subject to thrust by mass ejection (for example), this consequence can be relaxed by taking force to be a 4-vector from the outset, with time component equal to time rate of change of mass \cite{de-Saxce2016Galilean-Mechan}.
Such a treatment of thrust is highly phenomenological, as manifest for example in the fact that constitutive relations for the mass loss rate and velocity of mass ejection are now required, and the reality that a single worldline is continuously being split into many worldlines is ignored.
The point of view taken in this paper is that beginning with force as a 4-covector, yielding mass conservation on a single unsplit worldline, is a more fundamental expression of Galilei physics.  
}}
As for the spatial part, on both $\mathbb{M}$ and $\mathbb{G}$, the comoving observer space projection obtained by contraction with $\overleftarrow{\bm{\gamma}_{\bm{U}}}$ is 
\begin{equation}
M \bm{a} = \overleftarrow{\bm{\mathscr{f}}},
\label{eq:ThreeAccelerationComoving}
\end{equation}
the familiar 3-vector form of Newton's second law.

Consider next a decomposition relative to the fiducial observer of the vector version of Newton's second law.
On the left-hand side,
\[
\frac{\mathrm{d} \bm{I}}{\mathrm{d} \tau} 
	= \frac{ \mathrm{d} t }{ \mathrm{d} \tau } \frac{ \mathrm{d} \bm{I} }{ \mathrm{d} t } =
	\left\{
	\begin{aligned}
		\Lambda_{\bm{v}} \, \frac{ \mathrm{d} \bm{I} }{ \mathrm{d} t } 
			& & (\text{on } \mathbb{M}) \\[5pt]
		 \frac{ \mathrm{d} \bm{I} }{ \mathrm{d} t }  
			& & (\text{on } \mathbb{G})
	\end{aligned}
	\right.
\]
exhibits the rate of change $\mathrm{d} \bm{I} / \mathrm{d} t $ according to the fiducial observer. Thanks to Eq.~(\ref{eq:FourVelocity}) the latter reads
\[
\frac{ \mathrm{d} \bm{I} }{ \mathrm{d} t } =
	\left\{
	\begin{aligned}
		\frac{ \mathrm{d} \left( M  \Lambda_{\bm{v}} \right) }{ \mathrm{d} t } 
		 			\left( \bm{n} + \bm{v} \right) 
			&+ M  \Lambda_{\bm{v}} \, \frac{ \mathrm{d} \bm{v} }{ \mathrm{d} t }
		& & (\text{on } \mathbb{M}) \\[5pt]
		\frac{ \mathrm{d} M }{ \mathrm{d} t }  \left( \bm{n} + \bm{v} \right) 
			&+ \phantom{ \Lambda_{\bm{v}} } M  \, \frac{ \mathrm{d} \bm{v} }{ \mathrm{d} t }
		& & (\text{on } \mathbb{G}).
	\end{aligned}
	\right.
\]
As for the right-hand side of the vector version of Newton's second law, write the 4-force covector as
\[
\bm{\Upsilon}_{\!\bm{I}} =
	\begin{cases} 
		\Lambda_{\bm{v}} \left( - \Theta \, \underline{\bm{\chi}} + \bm{\mathscr{F}} \right)
			& (\text{on } \mathbb{M}) \\[5pt]
 		 \phantom{ \Lambda_{\bm{v}} ( } - \Theta \, \bm{\tau} + \bm{\mathscr{F}} 
			& (\text{on } \mathbb{G}). 
	\end{cases}
\]
Here 
\[
\bm{\Upsilon}_{\!\bm{I}} \cdot \overleftarrow{\bm{\gamma}} =
	\begin{cases}
		\Lambda_{\bm{v}} \, \bm{\mathscr{F}} & (\text{on } \mathbb{M}) \\[5pt]
		\phantom{\Lambda_{\bm{v}} \, } \bm{\mathscr{F}} & (\text{on } \mathbb{G}),
	\end{cases}
\]
with $\bm{\mathscr{F}} \cdot \bm{n} = 0$, is the 3-force covector according to the fiducial observer.
It is useful to write $\Theta$, the rate of energy input according to the fiducial observer, given by
\[
- \bm{\Upsilon}_{\!\bm{I}} \cdot \bm{n} =
	\begin{cases}
		\Lambda_{\bm{v}} \, \Theta & (\text{on } \mathbb{M})  \\[5pt]
		\phantom{\Lambda_{\bm{v}}} \, \Theta & (\text{on } \mathbb{G}),
	\end{cases}
\]
in terms of the heating rate $\theta$ tied to the internal energy of the particle.
This is accomplished by using Eq.~(\ref{eq:FourVelocity}) in Eq.~(\ref{eq:ComovingHeating}), with the result
\begin{equation}
\Theta = 
	\left\{
	\begin{aligned}
		\frac{ \theta }{ \Lambda_{\bm{v}}^2 } + \bm{\mathscr{F}} \cdot \bm{v}
			& & (\text{on } \mathbb{M})  \\[5pt]
		\theta \phantom{ ^2 } + \bm{\mathscr{F}} \cdot \bm{v}
		 	& & (\text{on } \mathbb{G}).
	\end{aligned}
	\right.
\label{eq:FiducialEnergyInput}
\end{equation}
Raising the index and noting Eq.~(\ref{eq:DualFiducialObserverVector}), the 4-force vector is
\[
\overleftarrow{\bm{\Upsilon}_{\!\bm{I}}} =
	\left\{
	\begin{aligned}
		\Lambda_{\bm{v}} 
			\left ( \frac{ \Theta }{ c^2 } \, \bm{n} + \overleftarrow{\bm{\mathscr{F}}} \right)
		& & (\text{on } \mathbb{M}) \\
 	\phantom{ \Lambda_{\bm{v}} ( \ } 
		 	\phantom{ \frac{ \Theta }{ c^2 } \, \bm{n} + } \ \overleftarrow{\bm{\mathscr{F}}} 
	\phantom{ ) \ \ } & &  (\text{on } \mathbb{G}).
	\end{aligned}
	\right.
\]
The fiducial observer time projection of Eq.~(\ref{eq:NewtonSecondVector}), obtained by contraction with $\underline{\bm{\chi}}$ on $\mathbb{M}$ or $\bm{\tau}$ on $\mathbb{G}$ and using Eq.~(\ref{eq:FiducialEnergyInput}), is
\begin{align*}
\frac{ \mathrm{d} \left( M  \Lambda_{\bm{v}} \right) }{ \mathrm{d} t } 
 	&= \frac{1}{c^2} \left( \frac{ \theta }{ \Lambda_{\bm{v}}^2 } 
						+ \bm{\mathscr{F}} \cdot \bm{v} \right) 
 	\quad  (\text{on } \mathbb{M})  \\[5pt]
\frac{ \mathrm{d} M }{ \mathrm{d} t } 	&= 0 \phantom{ \left( \frac{ \theta }{ {\Lambda_{\bm{v}}}^2 } 
						+ \bm{\mathscr{F}} \cdot \bm{v} \right)  }
 	\quad \quad (\text{on } \mathbb{G}).
\end{align*}
On $\mathbb{M}$ this is an equation for the evolution of what may be regarded as the inertia $M  \Lambda_{\bm{v}}$ measured by the fiducial observer (see below). 
On $\mathbb{G}$ there is no new information, but only a confirmation of mass conservation.
The fiducial observer space projection obtained by contraction with $\overleftarrow{\bm{\gamma}}$ is 
\begin{align*}
M \Lambda_{\bm{v}} \, \frac{ \mathrm{d} \bm{v} }{ \mathrm{d} t } 
 	& =  \overleftarrow{\bm{\mathscr{F}}}  
		- \frac{\bm{v}}{c^2} \left( \frac{ \theta }{ \Lambda_{\bm{v}}^2 } 
						+ \bm{\mathscr{F}} \cdot \bm{v} \right) 
		\quad (\text{on } \mathbb{M}) \\[5pt]
M \, \frac{ \mathrm{d} \bm{v} }{ \mathrm{d} t } 
 	& =  \overleftarrow{\bm{\mathscr{F}}} 
			\phantom{ - \frac{\bm{v}}{c^2} \left( \frac{ \theta }{ {\Lambda_{\bm{v}}}^2 } 
						+ \bm{\mathscr{F}} \cdot \bm{v} \right) }  
		\quad \ (\text{on } \mathbb{G}). 
\end{align*}
Note that $\mathrm{d} \bm{v} / \mathrm{d} t$ is the 3-acceleration measured by the fiducial observer; this justifies the interpretation of $M \Lambda_{\bm{v}}$ as the inertia measured by the fiducial observer on $\mathbb{M}$.
In comparing these relations with Eq.~(\ref{eq:ThreeAccelerationComoving}) for the comoving observer on $\mathbb{M}$ or $\mathbb{G}$, one finds that on $\mathbb{G}$ they are precisely the same: $\bm{a} = \mathrm{d} \bm{v} / \mathrm{d} t$ and $\overleftarrow{\bm{\mathscr{f}}} = \overleftarrow{\bm{\mathscr{F}}}$.
That is, the 3-vector version of Newton's law is Galilei invariant; this well-known fact is perhaps not surprising since both the comoving and fiducial observers project to the same position space $\mathbb{S}_t$.
(Beware however that for the covector 3-forces, $\bm{\mathscr{f}} \ne \bm{\mathscr{F}}$ even on $\mathbb{G}$!)
On $\mathbb{M}$ the relations between $\bm{a}$ and $\mathrm{d} \bm{v} / \mathrm{d} t$, and $\overleftarrow{\bm{\mathscr{f}}}$ and $\overleftarrow{\bm{\mathscr{F}}}$, are more complicated, not least because the first elements of these pairs belong to a different affine hyperplane than the second elements of these pairs ($\mathbb{S}_{\bm{U}(\tau)}$ and $\mathbb{S}_t$ respectively).

The covector or 1-form version of a spacetime formulation of Newton's second law naturally accommodates the covector 4-force $\bm{\Upsilon}_{\!\bm{I}}$; but because the inertia-momentum $\bm{I}$ is a 4-vector, the `total-energy--momentum 4-covector'  
\begin{equation}
\underline{\bm{I}} = \bm{g} \cdot \bm{I} = M \, \underline{\bm{U}} 
	= - M c^2 \, \Lambda_{\bm{v}} \, \underline{\bm{\chi}} 
		+  M \, \Lambda_{\bm{v}} \, \underline{\bm{v}}
		\quad (\mathbb{M} \text{ only}),
\label{eq:TotalEnergyMomentum}
\end{equation}
represented relative to the fiducial observer basis by the 4-row
\begin{equation}
\underline{\mathsf{I}} 
	= \begin{bmatrix}
		- M c^2 \, \Lambda_{\mathsf{v}} & M \, \Lambda_{\mathsf{v}} \, \underline{\mathsf{v}}
	\end{bmatrix}
	\quad (\mathbb{M} \text{ only})
\label{eq:EnergyMomentumFiducial}
\end{equation}
where $\underline{\mathsf{v}} = \mathsf{v}^{\mathrm{T}}$ is a 3-row, is available on $\mathbb{M}$ but not on $\mathbb{G}$.
Because $\underline{\bm{\chi}}$ on $\mathbb{M}$ corresponds to $\bm{\tau}$ on $\mathbb{G}$, the first term would read $- M c^2  \, \Lambda_{\bm{v}} \, \bm{\tau}$ on $\mathbb{G}$, which would make no sense as $c \rightarrow \infty$: Galilei physics must exclude a notion of `total energy' that includes `rest mass energy'.
This leads to a conceptual and notational difference from Eq.~(\ref{eq:FourForceVector}), in which the index-raising of a 4-covector is allowed on $\mathbb{G}$ via $\overleftrightarrow{\bm{\gamma}}$.
This is because $\overleftrightarrow{\bm{\gamma}}$ is a fundamental invariant structure on $\mathbb{G}$, the natural $c \rightarrow \infty$ limit of $\overleftrightarrow{\bm{g}}$ on $\mathbb{M}$, with the same matrix representation $\overleftrightarrow{\sfgamma}$ of Eq.~(\ref{eq:InverseMetricGalileiBasis}) with respect to any Galilei basis.
There is a temptation to allow similarly the notation $\underline{\bm{I}} = \bm{\gamma} \cdot \bm{I}$ for a 4-vector $\bm{I}$ on $\mathbb{G}$, where $\bm{\gamma}$ on $\mathbb{G}$ is defined in terms of the Euclid 3-metric $\bm{\gamma}$ on $\mathbb{S}_t$ by Eq.~(\ref{eq:ThreeMetricExtension}).
This temptation is to be resisted for generic 4-vectors on $\mathbb{G}$ because $\bm{\gamma}$ is not a fundamental invariant object on $\mathbb{G}$: its matrix representation $\sfgamma$ differs with respect to different Galilei bases.
Actually, however, the temptation may be indulged for vectors tangent to $\mathbb{S}_t$, that is, 3-vectors, with no time component when regarded as a vector on $\mathbb{G}$, because the purely space part of $\sfgamma$ is the same with respect to any Galilei basis.
Thus it is acceptable on both $\mathbb{M}$ and $\mathbb{G}$ to write for example
\[
\underline{\bm{v}}  =  \bm{\gamma} \cdot \bm{v}
\]
for the 3-velocity field $\bm{v}$ tangent to $\mathbb{S}_t$.

The covector or 1-form version of a spacetime formulation of Newton's second law on $\mathbb{M}$,
\begin{equation}
\frac{\mathrm{d}}{\mathrm{d}\tau} \, \underline{\bm{I}} = \bm{\Upsilon}_{\!\bm{I}} 
	\quad (\mathbb{M} \text{ only}),
\label{eq:CovectorNewtonSecond}
\end{equation}
contains the same information as the vector version, but suggests a different perspective that focuses on energy and 3-momentum rather than inertia and 3-velocity.
Write
\[
\underline{\bm I} = - \mathscr{E}_{\bm{p}} \, \underline{\bm{\chi}} + \bm{p}
	\quad (\mathbb{M} \text{ only}),
\]
where
\[
\bm{p} =  M \, \Lambda_{\bm{v}} \, \underline{\bm{v}} \quad (\mathbb{M} \text{ only})
\]
is the 3-momentum covector, and
\[
\begin{aligned}
 \mathscr{E}_{\bm{p}}  
 	&= M c^2  \Lambda_{\bm{v}}  \\[5pt]
 	&= \sqrt{ c^2 \, \overleftrightarrow{\bm{\gamma}} \left( \bm{p}, 
												\bm{p} \right) 
			 + M^2 c^4}
\end{aligned}
 	\quad (\mathbb{M} \mathrm{\ only})
\]
is the total particle energy.
Then the time projection according to the fiducial observer of the covector version of Newton's second law reads
\[
\frac{ \mathrm{d} \mathscr{E}_{\bm{p}} }{ \mathrm{d} t} 
	= \Theta 
	=   \frac{ \theta }{ \Lambda_{\bm{v}}^2 } + \bm{\mathscr{F}} \cdot \bm{v}
		\quad  (\mathbb{M} \text{ only}),
\]
while
\begin{equation}
\frac{ \mathrm{d} \bm{p} }{ \mathrm{d} t} 
	= \bm{\mathscr{F}}
\label{eq:ThreeMomentumEvolution}
\end{equation}
gives the space projection.

While not fully satisfying, there is a covector or 1-form version of a spacetime formulation of Newton's second law that can be used on $\mathbb{G}$ \cite{Cardall2020Combining-3-Mom}.
On $\mathbb{M}$, form a `relative-energy--momentum 4-covector' or `kinetic-energy--momentum 4-covector' $\bm{\Pi}$ from the total energy-momentum 4-covector $\underline{\bm{I}}$ of Eq.~(\ref{eq:TotalEnergyMomentum}) by the following combination:
\begin{equation}
\bm{\Pi} =  \underline{\bm{I}} - M \, \underline{\bm{n}} 
	=\underline{\bm{I}} + M c^2 \, \underline{\bm{\chi}} \quad (\text{on } \mathbb{M}).
\label{eq:RelativeEnergyMomentum}
\end{equation}
This does have a reasonable limit as $c \rightarrow \infty$.
It can be expressed
\begin{equation}
\bm{\Pi} =
	\begin{cases}
		- \epsilon_{\bm{p}} \, \underline{\bm{\chi}} + \bm{p}
			& (\text{on } \mathbb{M}) \\[5pt]
		- \epsilon_{\bm{p}} \, \bm{\tau} + \bm{p}
			& (\text{on } \mathbb{G}).
	\end{cases}
\label{eq:RelativeEnergyMomentumExplicit}
\end{equation}
Here the 3-momentum covector $\bm{p}$ is related to the 3-velocity vector $\bm{v}$ by
\begin{equation}
\bm{p} =
	\begin{cases}
		M \, \Lambda_{\bm{v}} \, \underline{\bm{v}} 
			& (\text{on } \mathbb{M}) \\[5pt]
	 	\phantom{\, \Lambda_{\bm{v}}} M  \, \underline{\bm{v}} 
			& (\text{on } \mathbb{G}), 
	\end{cases}
\label{eq:ThreeMomentum}
\end{equation}
which satisfies Eq.~(\ref{eq:ThreeMomentumEvolution}).
More notable is the kinetic energy $\epsilon_{\bm{p}}$, which can be expressed in terms of the 3-velocity,
\begin{equation}
\epsilon_{\bm{p}} = 
	\left\{ 
	\begin{aligned}
		M c^2 \left( \Lambda_{\bm{v}} - 1 \right) 
			&= \frac{ M \, {\Lambda_{\bm{v}}^2} \, \bm{\gamma} ( \bm{v}, \bm{v} ) }
				{ \Lambda_{\bm{v}} + 1 }
				&& (\text{on } \mathbb{M}) \\[5pt]
			&= \frac{ M \, \bm{\gamma} ( \bm{v}, \bm{v} ) } { 2 }
		 && (\text{on } \mathbb{G}),
	\end{aligned}
	\right.
\label{eq:KineticEnergy}
\end{equation}
or in terms of the 3-momentum,
\begin{equation}
\epsilon_{\bm{p}} =
	\left\{
	\begin{aligned}
	\mathscr{E}_{\bm{p}} - M c^2 
		&= \frac{ \overleftrightarrow{\bm{\gamma}} 
			( \bm{p}, \bm{p} ) }
			{ M \left( \Lambda_{\bm{v}} ( \bm{p} ) + 1 \right) }
			&&  (\text{on } \mathbb{M}) \\[5pt]
		&= \frac{ \overleftrightarrow{\bm{\gamma}} ( \bm{p}, \bm{p} ) }
			{ 2 \, M  }
		 	&&  (\text{on } \mathbb{G}),  
	\end{aligned}
	\right.
\label{eq:KineticEnergy2}
\end{equation}
where
\[
 \Lambda_{\bm{v}} ( \bm{p} ) 
 	= \sqrt{ 1 + \frac{ 1 }{ M^2 c^2 } \overleftrightarrow{\bm{\gamma}} 
			( \bm{p}, \bm{p} )}
		\quad (\text{on } \mathbb{M}).
\]
In terms of $\bm{\Pi}$, Eq.~(\ref{eq:CovectorNewtonSecond}) becomes 
\begin{equation}
\frac{\mathrm{d} \bm{\Pi} }{\mathrm{d}\tau} = \bm{\Upsilon},
\label{eq:CovectorNewtonSecondRelative}
\end{equation}
where the `relative 4-force' 
\begin{equation}
\bm{\Upsilon}
	= \bm{\Upsilon}_{\!\bm{I}} + c^2 \, \frac{\mathrm{d} M }{\mathrm{d}\tau} \, \underline{\bm{\chi}} 
	\quad (\text{on } \mathbb{M})
\label{eq:RelativeFourForce}
\end{equation}
has a reasonable limit as $c \rightarrow \infty$:
\[
\bm{\Upsilon} = 
	\left\{
	\begin{aligned}
		&- \Lambda_{\bm{v}} \left( \bm{\mathscr{F}} \cdot \bm{v} 
			- \frac{ \left( \Lambda_{\bm{v}} - 1 \right) }{ \Lambda_{\bm{v}}^2 } \, \theta \right) 
				\underline{\bm{\chi}} 
		+ \Lambda_{\bm{v}} \, \bm{\mathscr{F}} 	
		& & (\mathrm{on \ } \mathbb{M}) \\[5pt]	
		&-  \left( \bm{\mathscr{F}} \cdot \bm{v} \right) \bm{\tau} + \bm{\mathscr{F}}
		& & (\mathrm{on \ } \mathbb{G}).
	\end{aligned}
	\right.
\]
Thus the time projection of Eq.~(\ref{eq:CovectorNewtonSecondRelative}) according to the fiducial observer yields
\begin{equation}
\frac{\mathrm{d} \epsilon_{\bm{p}}  }{\mathrm{d} t} =
	\left\{
	\begin{aligned}  
		 & \bm{\mathscr{F}} \cdot \bm{v} 
			- \frac{ \left( \Lambda_{\bm{v}} - 1 \right) }{ \Lambda_{\bm{v}}^2 } \, \theta
		& & (\mathrm{on \ } \mathbb{M}) \\[5pt]
		 & \bm{\mathscr{F}} \cdot \bm{v} 
		& & (\mathrm{on \ } \mathbb{G}).
	\end{aligned}
	\right.
\label{eq:WorkEnergyTheorem}
\end{equation}
This is the work-energy theorem.
In the spacetime formulation, there is no need to take a scalar product of $\bm{v}$ with the 3-vector version of Newton's second law to obtain it; it is already contained in the time component of the tensor formulation.

This 4-covector version of Newton's second law, based on kinetic energy rather than total energy  so as to accommodate $\mathbb{G}$ as well as $\mathbb{M}$, is not fully satisfying because the notion of kinetic energy (energy of motion) inherently depends on a choice of observer (motion relative to whom?). 
Thus in Eq.~(\ref{eq:RelativeEnergyMomentum}), the fiducial observer covector $\underline{\bm{n}}$ is built into the definition of the 4-covector $\bm{\Pi}$ whose time component is the kinetic energy relative to the fiducial observer.
The unsatisfying result is that Lorentz or homogeneous Galilei transformations cannot transform the components of $\bm{\Pi}$ in such a way as to demonstrate the transformation rule of kinetic energy.
The reality to be faced is that the transformation of kinetic energy is not directly addressed by the Lorentz and homogeneous Galilei groups.
Note that these groups are oriented towards the transformation of time and space, compatible also with the transformation of 4-velocity and of inertia-momentum; and that the Lorentz group only manages the transformation of energy, sneaking it in by the back door as it were, thanks to the equivalence (up to a factor of $c^2$) of inertia and \textit{total} energy. 
This is manifest for instance in the fact that a local Lorentz or Galilei boost can give the 4-velocity components or the inertia-momentum 4-vector components relative to the fiducial observer in terms of the components relative to a comoving observer, for whom the 3-velocity vanishes; see for instance Eq.~(\ref{eq:FourVelocityBoost}). 
However, a `comoving relative energy-momentum covector' $\bm{\Pi}_{\bm{U}} = \underline{\bm{I}} - M \, \underline{\bm{U}}$ would vanish identically: kinetic energy vanishes when 3-velocity vanishes.
The zero vector having vanishing components in every basis, there is no Lorentz or Galilei boost that could give it non-zero components.
Extensions of the Lorentz and homogeneous Galilei groups that address the transformation of kinetic energy, and the extended spacetimes on which these groups act, are the subject of Sec.~\ref{sec:ExtendedAffineSpacetimes}.

\section{Electrodynamics on $\mathbb{M}$ and $\mathbb{G}$}
\label{sec:Electrodynamics}

Before turning to `extended' flat spacetimes, it is appropriate to recall from the perspective of `normal' flat spacetimes the physics that motivated the introduction of Poincar\'e physics and Minkowski spacetime $\mathbb{M}$ in the first place: electrodynamics.
This also provides closure to the discussion in Sec.~\ref{sec:MaterialParticleMG} by introducing a concrete example of a 4-force $\bm{\Upsilon}$ acting on a (here, electrically charged) material particle.
In rough parallel with the division in Sec.~\ref{sec:MaterialParticleMG} between kinematics (a \textit{description} of a material particle via introduction of a worldline) and dynamics (a \textit{prescription} that determines the worldline), so also electrodynamics divides into two parts: first, a \textit{description} of the electromagnetic field, in the sense of an operational definition in terms of the 4-force it exerts on a charged material particle; and second, a \textit{prescription} for how the electromagnetic field arises from sources.
The full marriage of these two halves of electrodynamics is most at home---in particular, can only be performed in an invariant manner---on $\mathbb{M}$.
As recognized by Le Bellac and L\'evy-Leblond \cite{Le-Bellac1973Galilean-electr}, it is in what are best understood as the constitutive relations that close the system of electrodynamics equations that the Galilei invariance of full electrodynamics founders.
As also shown by Le Bellac and L\'evy-Leblond, insistence upon Galilei invariance requires a partial truncation of electrodynamics in one of two different ways. 
The geometric spacetime perspective employed here---which differs substantially from the approach taken by Le Bellac and L\'evy-Leblond---affords a fresh and insightful perspective on these matters.   
(For 4-dimensional and 5-dimensional spacetime descriptions of Galilei electrodynamics that differ in certain respects from the presentation here and in Sec.~\ref{sec:ElectrodynamicsB}, see for example \cite{Kunzle1976Covariant-Newto,Duval1991Celestial-mecha,Montigny2003Nonrelativistic}.)

\subsection{The electromagnetic force equations}
\label{sec:ElectromagneticForce}

Describe the electromagnetic field in terms of the 4-force it exerts on a material particle with an electric charge.
Note three conditions characterizing this force.
First, it is a `pure force', meaning that it does not induce any heating of the particle, that is, any change in its internal energy or its mass.
In terms of Eqs.~(\ref{eq:ComovingHeating}) and (\ref{eq:MassRate}),
\[
\theta = c^2 \, \frac{\mathrm{d}M}{\mathrm{d}\tau} = - \bm{\Upsilon}_{\!\bm{I}} \cdot \bm{U} = 0,
\]
which by Eq.~(\ref{eq:RelativeFourForce}) implies that $\bm{\Upsilon}_{\!\bm{I}} = \bm{\Upsilon}$.
Second, the electromagnetic 4-force is assumed to be linear in the particle 4-velocity $\bm{U}$, and is therefore expressible in terms of a bilinear form $\bm{F}$, the electromagnetic force tensor:
\begin{equation}
\bm{\Upsilon} = q \, \bm{F}( \, . \, , \bm{U}) = q \, \bm{F} \cdot \bm{U},
\label{eq:ElectromagneticForce}
\end{equation}
where the scalar $q$ is the electric charge of the particle.
Combining these two conditions yields $\bm{F}(\bm{U},\bm{U}) = 0$; this implies that $\bm{F}$ is antisymmetric, that is, a 2-form.
Third, the exterior derivative of the 2-form $\bm{F}$ is taken to satisfy
\[
\mathbf{d}\bm{F} = 0.
\]
Since a 3-form on 4-dimensional spacetime has four independent components (3-forms are in one-to-one correspondence with 4-vectors via the spacetime volume form), this yields four independent equations, which turn out to be the scalar and 3-vector homogeneous Maxwell equations.

As usual, contact with measurements requires 1+3 (time/space) decompositions according to the foliation of spacetime corresponding to a fiducial (here, inertial) observer.
The discussion toward the end of Sec.~\ref{sec:SpacetimeFoliation} on volume forms, exterior derivatives, and vector calculus in the context of foliated spacetime will be particularly relevant in what follows.

Select a fiducial inertial observer and decompose $\bm{F}$ in terms of a 1-form $\underline{\bm{E}}$ and a 2-form $\bm{\mathcal{B}}$, both of which are tangent to $\mathbb{S}_t$:
\begin{equation}
\bm{F} = 
	\begin{cases}
		- \underline{\bm{\chi}}\wedge \underline{\bm{E}} + \bm{\mathcal{B}} 
			& (\text{on } \mathbb{M}) \\[5pt]
		- \bm{\tau}\wedge \underline{\bm{E}} + \bm{\mathcal{B}} 
			& (\text{on } \mathbb{G}).
	\end{cases}
\label{eq:ElectromagneticForceTensor31}
\end{equation}
For $\underline{\bm{E}}$, tangency to the space slice $\mathbb{S}_t$ means that $\underline{\bm{E}} \cdot \bm{n}= 0$. 
For the 2-form $\bm{\mathcal{B}}$, tangency to $\mathbb{S}_t$ means, first, that $\bm{n} \cdot \bm{\mathcal{B}} = 0$ and $\bm{\mathcal{B}} \cdot \bm{n} = 0$.
By way of Eq.~(\ref{eq:VolumeContraction}), it also means that $\bm{\mathcal{B}}$ can be related to a vector $\bm{B}$ tangent to $\mathbb{S}_t$ by means of the volume form $ \underaccent{\check}{\bm{\varepsilon}}$ of $\mathbb{S}_t$:
\[
\bm{\mathcal{B}} = \bm{\varepsilon}( \bm{n}, \bm{B}, \, . \, , \, . \, ) 
	=  \underaccent{\check}{\bm{\varepsilon}}( \bm{B}, \, . \, , \, . \, )
	= \bm{B} \cdot  \underaccent{\check}{\bm{\varepsilon}}
\]
or $\mathcal{B}_{ij} = \varepsilon_{0 a i j} B^a = B^a \underaccent{\check}{\varepsilon}_{a i j}$ in components.
(This can be understood as a Hodge dual relationship between $\bm{\mathcal{B}}$ and $\underline{\bm{B}}$ on $\mathbb{S}_t$ via the 3-volume form $ \underaccent{\check}{\bm{\varepsilon}}$ and the inverse 3-metric $\overleftrightarrow{\bm{\gamma}}$.) 
The electromagnetic force tensor $\bm{F}$ is represented by the matrix
\begin{equation}
\mymathsf{F}
	=  	\begin{bmatrix}
			0 & -E_j \\[5pt] 
			E_i & \mathcal{B}_{ij}
		\end{bmatrix}
	=	\begin{bmatrix}
			0 & -E_1 & -E_2 & -E_3 \\[5pt]
			E_1 & 0 & B^3 & -B^2 \\[5pt]
			E_2 & -B^3 & 0 & B^1 \\[5pt]
			E_3 & B^2 & -B^1 & 0
		\end{bmatrix}
\label{eq:ElectromagneticForceTensorComponents}
\end{equation}
with respect to a Minkowski or Galilei basis.

With the decompositions of $\bm{F}$ in Eq.~(\ref{eq:ElectromagneticForceTensor31}) and $\bm{U}$ in Eq.~(\ref{eq:FourVelocity}), the electromagnetic 4-force of Eq.~(\ref{eq:ElectromagneticForce}) reads
\begin{equation}
\bm{\Upsilon} = 
	\begin{cases}
		-q \, \Lambda_{\bm{v}} \left( \underline{\bm{E}} \cdot \bm{v} \right) \underline{\bm{\chi}}
			\, + \, q \, \Lambda_{\bm{v}} \left( \underline{\bm{E}}  
				+ \underline{ \bm{v} \times \bm{B} } \right)  
					& (\text{on } \mathbb{M}) \\[5pt]
		\quad \ \, -q \left( \underline{\bm{E}} \cdot \bm{v} \right) \bm{\tau}
			\, + \, \quad \ q \left( \underline{\bm{E}}  
				+ \underline{ \bm{v} \times \bm{B} } \right)  
					& (\text{on } \mathbb{G}).
	\end{cases}
\label{eq:ElectromagneticForce31}
\end{equation}
Considering Eq.~(\ref{eq:RelativeEnergyMomentumExplicit}) for the decomposition of the relative 4-momentum $\bm{\Pi}$, the time and space projections of Newton's second law in the form of Eq.~(\ref{eq:CovectorNewtonSecondRelative}) are
\begin{align*}
\frac{\mathrm{d} \epsilon_{\bm{p}}  }{\mathrm{d} t}
	&=  q \left( \underline{\bm{E}} \cdot \bm{v} \right), \\[5pt]
\frac{ \mathrm{d} \bm{p} }{ \mathrm{d} t}
	&= q \left( \underline{\bm{E}}  + \underline{ \bm{v} \times \bm{B} } \right)
\end{align*}
on both $\mathbb{M}$ and $\mathbb{G}$, where the expressions for 3-momentum $\bm{p}$ and kinetic energy $\epsilon_{\bm{p}}$ are given by Eqs.~(\ref{eq:ThreeMomentum}) and (\ref{eq:KineticEnergy}) or (\ref{eq:KineticEnergy2}).
The familiar Lorentz force law is evident, expressed (up to metric duality relative to $\bm{\gamma}$) in terms of the electric field strength $\bm{E}$ and magnetic flux density $\bm{B}$, vector fields tangent to $\mathbb{S}_t$ measured by the fiducial observer.

Finally, turn to the condition $\mathbf{d}\bm{F} = 0$.
The decompositions of $\mathbf{d}$ in Eq.~(\ref{eq:ExteriorDerivativeDecomposition}) and of $\bm{F}$ in Eq.~(\ref{eq:ElectromagneticForceTensor31}) result in
\[
\mathbf{d}\bm{F} =
	\left\{
	\begin{aligned}
		\underline{\bm{\chi}} \wedge 
			\left(  \mathrm{\underaccent{\check}{\mathbf{d}}} \underline{\bm{E}} 
			+ \frac{\partial \bm{\mathcal{B}}}{\partial t} \right) 
		+  \mathrm{\underaccent{\check}{\mathbf{d}}} \bm{\mathcal{B}} & & (\text{on } \mathbb{M}) \\[5pt]
		\bm{\tau} \wedge 
			\left(  \mathrm{\underaccent{\check}{\mathbf{d}}} \underline{\bm{E}} 
			+ \frac{\partial \bm{\mathcal{B}}}{\partial t} \right) 
		+  \mathrm{\underaccent{\check}{\mathbf{d}}} \bm{\mathcal{B}} & & (\text{on } \mathbb{G}).
	\end{aligned}
	\right.
\]
The term tangent to $\mathbb{S}_t$ and the term that is not must separately vanish, so that
\begin{align*}
 \mathrm{\underaccent{\check}{\mathbf{d}}} \bm{\mathcal{B}} &= 0, \\[5pt]
 \mathrm{\underaccent{\check}{\mathbf{d}}} \underline{\bm{E}} 
			+ \frac{\partial \bm{\mathcal{B}}}{\partial t} &= 0,
\end{align*}
which correspond to the familiar scalar and vector homogeneous Maxwell equations
\begin{equation}
\begin{aligned}
\mathrm{\underaccent{\check}{\bm{\nabla}}} \cdot \bm{B} &= 0, \\[5pt]
\mathrm{\underaccent{\check}{\bm{\nabla}}} \times \bm{E} + \frac{\partial \bm{B}}{\partial t} &= 0 
\end{aligned}
\label{eq:HomogeneousMaxwell}
\end{equation}
on both $\mathbb{M}$ and $\mathbb{G}$.

\subsection{The electromagnetic source equations}
\label{sec:ElectromagneticSource}

Prescribe the electromagnetic field by giving an equation that determines how it arises from a source field, the electric current.
Postulate another 2-form $\bm{\mathcal{F}}$, the electromagnetic source tensor, and an electric current 3-form $\bm{\mathcal{J}}$, related by
\[
\mathbf{d} \bm{\mathcal{F}} = \bm{\mathcal{J}}.
\] 
As was the case with $\mathbf{d}\bm{F} = 0$ satisfied by the electromagnetic force tensor $\bm{F}$, this yields four independent equations, which in this case turn out to be the scalar and 3-vector inhomogeneous Maxwell equations.
The four independent components of $\bm{\mathcal{J}}$ can be related to those of an electric current 4-vector $\bm{J}$ by
\begin{equation}
\bm{\mathcal{J}} = \bm{\varepsilon} ( \bm{J}, \, . \, , \, . \, , \, . \, ) = \bm{J} \cdot \bm{\varepsilon}.
\label{eq:ElectricCurrent3Form}
\end{equation}
(On $\mathbb{M}$ this corresponds to a Hodge dual relationship $\bm{\mathcal{J}} = \bm{\star} \underline{\bm{J}}$.)
Conservation of electric charge in the form $\mathbf{d} \bm{\mathcal{J}} = 0$ is immediate from $\mathbf{d}^2 = 0$, and corresponds to the vanishing 4-divergence $\bm{\nabla} \cdot \bm{J} = 0$ thanks to Eq.~(\ref{eq:DivergenceDefinition}).

Decompose $\bm{\mathcal{F}}$ in terms of a 1-form $\underline{\bm{H}}$ and a 2-form $\bm{\mathcal{D}}$, both of which are tangent to $\mathbb{S}_t$:
\begin{equation}
\bm{\mathcal{F}} = 
	\begin{cases}
		   \underline{\bm{\chi}} \wedge \underline{\bm{H}} + \bm{\mathcal{D}} 
			& (\text{on } \mathbb{M}) \\[5pt]
		  \bm{\tau} \wedge \underline{\bm{H}} + \bm{\mathcal{D}} 
			& (\text{on } \mathbb{G}).
	\end{cases}
\label{eq:ElectromagneticSourceTensor31}
\end{equation}
For $\underline{\bm{H}}$, tangency to the space slice $\mathbb{S}_t$ means that $\underline{\bm{H}} \cdot \bm{n}= 0$. 
For the 2-form $\bm{\mathcal{D}}$, tangency to $\mathbb{S}_t$ means, first, that $\bm{n} \cdot \bm{\mathcal{D}} = 0$ and $\bm{\mathcal{D}} \cdot \bm{n} = 0$.
By way of Eq.~(\ref{eq:VolumeContraction}), it also means that $\bm{\mathcal{D}}$ can be related to a vector $\bm{D}$ tangent to $\mathbb{S}_t$ by means of the volume form $ \underaccent{\check}{\bm{\varepsilon}}$ of $\mathbb{S}_t$:
\[
\bm{\mathcal{D}} = \bm{\varepsilon}( \bm{n}, \bm{D}, \, . \, , \, . \, ) 
	=  \underaccent{\check}{\bm{\varepsilon}}( \bm{D}, \, . \, , \, . \, )
	= \bm{D} \cdot  \underaccent{\check}{\bm{\varepsilon}}
\]
or $\mathcal{D}_{ij} = \varepsilon_{0 a i j} D^a = D^a  \underaccent{\check}{\varepsilon}_{a i j}$ in components.
The electromagnetic source tensor $\bm{\mathcal{F}}$ is represented by the matrix
\begin{equation}
\mathcal{F}
	=  	\begin{bmatrix}
			0 & H_j \\[5pt] 
			-H_i & \mathcal{D}_{ij}
		\end{bmatrix}
	=	\begin{bmatrix}
			0 & H_1 & H_2 & H_3 \\[5pt]
			-H_1 & 0 & D^3 & -D^2 \\[5pt]
			-H_2 & -D^3 & 0 & D^1 \\[5pt]
			-H_3 & D^2 & -D^1 & 0
		\end{bmatrix}
\label{eq:ElectromagneticSourceTensorComponents}
\end{equation}
with respect to a Minkowski or Galilei basis.

Turn next to the decomposition of the electric current 3-form $\bm{\mathcal{J}}$.
Decompose the electric current 4-vector as
\begin{equation}
\bm{J} = \rho \, \bm{n} + \bm{j},
\label{eq:ElectricCurrent4Vector31}
\end{equation}
where $\rho$ is the charge density and the 3-current $\bm{j}$ is tangent to $\mathbb{S}_t$.
Then Eq.~(\ref{eq:ElectricCurrent3Form}) yields
\begin{equation}
\bm{\mathcal{J}} = 
	\begin{cases}
		- \underline{\bm{\chi}} \wedge \bm{\mathscr{J}} + \bm{\mathscr{C}} 
			& (\text{on } \mathbb{M}) \\[5pt]
		- \bm{\tau} \wedge \bm{\mathscr{J}} + \bm{\mathscr{C}} 
			& (\text{on } \mathbb{G}),
	\end{cases}
\label{eq:ElectricCurrent3FormForm31}
\end{equation}
where the electric charge 3-form $\bm{\mathscr{C}}$ is related to the charge density by 
\begin{equation}
\bm{\mathscr{C}} = \rho \,  \underaccent{\check}{\bm{\varepsilon}},
\label{eq:ElectricCharge3Form}
\end{equation}
and the electric current 2-form $\bm{\mathscr{J}}$ is related to the current density 3-vector by
\[
\bm{\mathscr{J}} =  \underaccent{\check}{\bm{\varepsilon}} ( \bm{j}, \, . \, , \, . \, ) 
	= \bm{j} \cdot  \underaccent{\check}{\bm{\varepsilon}}.
\]
Both $\bm{\mathscr{C}}$ and $\bm{\mathscr{J}}$ are tangent to $\mathbb{S}_t$.

Finally, turn to the condition $\mathbf{d}\bm{\mathcal{F}} = \bm{\mathcal{J}}$.
The decompositions of $\mathbf{d}$ in Eq.~(\ref{eq:ExteriorDerivativeDecomposition}) and of $\bm{\mathcal{F}}$ in Eq.~(\ref{eq:ElectromagneticSourceTensor31}) result in
\[
\mathbf{d}\bm{\mathcal{F}} =
	\left\{
	\begin{aligned}
		- \underline{\bm{\chi}} \wedge 
		\left( \mathrm{\underaccent{\check}{\mathbf{d}}} \underline{\bm{H}} 
			- \frac{\partial \bm{\mathcal{D}}}{\partial t} \right) 
		+  \mathrm{\underaccent{\check}{\mathbf{d}}} \bm{\mathcal{D}} & & (\text{on } \mathbb{M}) \\[5pt]
		-  \bm{\tau} \wedge \left( \mathrm{\underaccent{\check}{\mathbf{d}}} \underline{\bm{H}} 
			- \frac{\partial \bm{\mathcal{D}}}{\partial t} \right)
		+  \mathrm{\underaccent{\check}{\mathbf{d}}} \bm{\mathcal{D}} & & (\text{on } \mathbb{G}).
	\end{aligned}
	\right.
\]
Putting this together with the decomposition of $\bm{\mathcal{J}}$ in Eq.~(\ref{eq:ElectricCurrent3FormForm31}), the term tangent to $\mathbb{S}_t$ and the term that is not must separately vanish, so that
\begin{align}
 \mathrm{\underaccent{\check}{\mathbf{d}}} \bm{\mathcal{D}} &= \bm{\mathscr{C}}, \\[5pt]
 \mathrm{\underaccent{\check}{\mathbf{d}}} \underline{\bm{H}} 
			- \frac{\partial \bm{\mathcal{D}}}{\partial t} &= \bm{\mathscr{J}},
\label{eq:InhomogeneousMaxwellForms}
\end{align}
which correspond to the familiar scalar and vector inhomogeneous Maxwell equations
\begin{equation}
\begin{aligned}
\mathrm{\underaccent{\check}{\bm{\nabla}}} \cdot \bm{D} &= \rho, \\[5pt]
\mathrm{\underaccent{\check}{\bm{\nabla}}} \times \bm{H} - \frac{\partial \bm{D}}{\partial t} &= \bm{j} 
\end{aligned}
\label{eq:InhomogeneousMaxwell}
\end{equation}
on both $\mathbb{M}$ and $\mathbb{G}$, in terms of the electric displacement field $\bm{D}$ and the magnetic field strength $\bm{H}$.

\subsection{Full electrodynamics: Poincar\'e invariant but not Galilei-invariant}
\label{sec:FullElectrodynamics}

The electrodynamics equations presented thus far do not constitute a closed system.
As 2-forms on 4-dimensional spacetime $\mathbb{M}$ or $\mathbb{G}$, the electromagnetic force tensor $\bm{F}$ and the electromagnetic source tensor $\bm{\mathcal{F}}$ have six independent components each, which correspond to pairs of vector fields tangent to $\mathbb{S}_t$: the electric field strength $\bm{E}$ and the magnetic flux density $\bm{B}$ in the case of $\bm{F}$, and the electric displacement field $\bm{D}$ and the magnetic field strength $\bm{H}$ in the case of $\bm{\mathcal{F}}$.
Thus there are a total of twelve fields.
The field equations $\mathbf{d} \bm{F} = 0$ and $\mathbf{d} \bm{\mathcal{F}} = \bm{\mathcal{J}}$ each provide one scalar equation without time derivatives and one 3-vector equation with a time derivative.
The scalar equations are best regarded as constraints on the initial conditions of $\bm{B}$ and $\bm{D}$, constraints which the structure of the 3-vector equations enforces for all time (divergence of a curl vanishes).
The 3-vector equations then give the time evolution of $\bm{B}$ and $\bm{D}$, but these provide only six evolution equations.

The familiar way to close this system of twelve electromagnetic fields governed by six evolution equations is to posit the following relations between the two pairs of 3-vector fields:
\begin{align*}
\bm{D} &= \epsilon \, \bm{E} \\
\bm{B} &= \mu \, \bm{H}.
\end{align*}
A priori, these are to be regarded as constitutive relations which hold only in the frame of some (here, isotropic) medium, just as an equation of state that closes the equations governing a perfect fluid holds only in a `material frame' comoving with the fluid. 
For present purposes set aside all `normal' matter capable of polarization and magnetization in the region occupied by the electromagnetic field, so that the only `medium' in question is a supposed `luminiferous aether', in which the permittivity $\epsilon$ and permeability $\mu$ have the constant values $\epsilon = \epsilon_0$ and $\mu = \mu_0$.

In this matter-free case, the celebrated result on $\mathbb{M}$ is that the a priori assumption of a `luminiferous aether' as a necessary medium can be discarded.
To see how this comes about, consider the matrix representations of $\bm{F}$ and $\bm{\mathcal{F}}$ with respect to a Minkowski basis and examine the transformations
\begin{align*}
\mathsf{F}' &= \mathsf{L}_\mathbb{M}^\mathrm{T} \, \mathsf{F} \, \mathsf{L}_\mathbb{M}, \\[5pt]
\mathcal{F}' &= \mathsf{L}_\mathbb{M}^\mathrm{T} \, \mathcal{F} \, \mathsf{L}_\mathbb{M}
\end{align*}
under a Lorentz boost (see the Appendix).
For this purpose it is convenient to write Eq.~(\ref{eq:MinkowskiBoost}) as
\[
\mathsf{L}_\mathbb{M} =
	\begin{bmatrix}
		\Lambda_\mathsf{u} & \frac{1}{c^2} \, \Lambda_\mathsf{u} \, u_j \\[5pt]
		\Lambda_\mathsf{u} \, u^i & {\perp^i}_j + \Lambda_\mathsf{u} \, \hat{u}^i \hat{u}_j
	\end{bmatrix},
\]
where $\hat{u}^i = u^i / \lVert \mathsf{u} \rVert$ is a unit 3-vector component and ${\perp^i}_j = {\delta^i}_j - \hat{u}^i \hat{u}_j$ projects to the plane perpendicular to the boost velocity $\bm{u}$.
In the case of $\mathsf{F}$ given by Eq.~(\ref{eq:ElectromagneticForceTensorComponents}), $\underline{\bm{E}}$ and $\bm{B}$ can be decomposed as
\begin{align*}
E_i &= ( E_\parallel )_i + ( E_\perp )_i, &  B^i &= ( B_\parallel )^i + ( B_\perp )^i, \\[5pt]
( E_\parallel )_i &= ( E_a \hat{u}^a ) \hat{u}_i, &  ( B_\parallel )^i &= ( \hat{u}_a B^a ) \hat{u}^i, \\[5pt]
( E_\perp )_i &= E_a {\perp^a}_i, & ( B_\perp )^i &= {\perp^i}_a B^a, 
\end{align*}
and their transformations are
\[
\begin{aligned}
E'_i &= ( E_\parallel )_i + \Lambda_\mathsf{u} \left( ( E_\perp )_i 
		\, + \phantom{\frac{1}{1}}\!\!\!\!\! \left( \bm{u} \times \bm{B} \right)_i \right), \\[5pt]
B'^i &= ( B_\parallel )^i + \Lambda_\mathsf{u} \left( ( B_\perp )^i 
		- \frac{1}{c^2} \left( \bm{u} \times \bm{E} \right)^i \right)
\end{aligned}
\quad (\mathbb{M} \text{ only}).
\]
With a relative sign change and swapped roles of the electric and magnetic fields in $\mathcal{F}$ of Eq.~(\ref{eq:ElectromagneticSourceTensorComponents}) relative to $\mathsf{F}$, the transformations of $\bm{D}$ and $\underline{\bm{H}}$ are
\[
\begin{aligned}
D'^i &= ( D_\parallel )^i + \Lambda_\mathsf{u} \left( ( D_\perp )^i 
		+ \frac{1}{c^2} \left( \bm{u} \times \bm{H} \right)^i \right), \\[5pt]
H'_i &= ( H_\parallel )_i + \Lambda_\mathsf{u} \left( ( H_\perp )_i 
		\, - \phantom{\frac{1}{1}}\!\!\!\!\! \left( \bm{u} \times \bm{D} \right)_i \right)
\end{aligned}
\quad (\mathbb{M} \text{ only}).
\]
Noting that in a Minkowski basis covariant space components are equal to the contravariant space components, the miracle of Poincar\'e physics is that, thanks to the empirical relation 
\[
\epsilon_0 \mu_0 = \frac{1}{c^2}, 
\]
the components $D^i$ transform in the same manner as the $E_i$, and the components $H_i$ transform in the same manner as the $B^i$, so that 
\[
\begin{aligned}
\bm{D}' &= \epsilon_0 \, \bm{E}', \\
\bm{B}' &= \mu_0 \, \bm{H}'
\end{aligned}
\quad (\mathbb{M} \text{ only})
\]
hold in all frames related by Lorentz boosts.
One is therefore led to set aside the a priori interpretation of a constitutive relation valid only in a particular frame and dispense with the notion of a luminiferous aether.

The same conclusion does not hold for electrodynamics on $\mathbb{G}$.
Under a Galilei boost of Eq.~(\ref{eq:GalileiBoost}), the transformations 
\begin{align*}
\mathsf{F}' &= \mathsf{L}_\mathbb{G}^\mathrm{T} \, \mathsf{F} \, \mathsf{L}_\mathbb{G}, \\[5pt]
\mathcal{F}' &= \mathsf{L}_\mathbb{G}^\mathrm{T} \, \mathcal{F} \, \mathsf{L}_\mathbb{G}
\end{align*}
yield
\[
\begin{aligned}
E'_i &= E_i + \left( \bm{u} \times \bm{B} \right)_i, \\[5pt]
B'^i &= B^i
\end{aligned}
\quad (\mathbb{G} \text{ only})
\]
and
\[
\begin{aligned}
D'^i &= D^i, \\[5pt]
H'_i &= H_i - \left( \bm{u} \times \bm{D} \right)_i
\end{aligned}
\quad (\mathbb{G} \text{ only}).
\]
The constitutive relations can only hold for $\bm{u} = 0$.
The hypothesis of a medium, i.e. the luminiferous aether, defining the frame (modulo rotations and translations) in which the closed system of electrodynamics equations are valid in this form, cannot be discarded.
In this respect full electrodynamics is not Galilei invariant.

\subsection{Galilei-invariant partial electrodynamics}
\label{sec:PartialElectrodynamics}

It is interesting to consider how much, or what forms of, electrodynamics remain on $\mathbb{G}$ if Galilei invariance is insisted upon.
To prepare for this it is useful to recall three additional aspects of full electrodynamics on $\mathbb{M}$.

First, consider another perspective on electrodynamics in a vacuum that makes use of the Hodge star operator on $\mathbb{M}$, discussed at the end of Sec.~\ref{sec:SpacetimeM}.
Noting the decomposition in Eq.~(\ref{eq:ElectromagneticForceTensor31}),
consider the Hodge dual 
\[
\bm{\star} \bm{F} = \bm{\star} \left( - \underline{\bm{\chi}} \wedge \underline{\bm{E}} \right)
	+ \bm{\star}\bm{\mathcal{B}} 
		\quad (\mathbb{M} \text{ only}). 
\]
Using Eq.~(\ref{eq:DualFiducialObserverVector}) along with the identities of Eqs.~(\ref{eq:HodgeWedge1}) and (\ref{eq:HodgeWedge2}), 
\[
\bm{\star} \bm{F} = \underline{\bm{\chi}} \wedge \underline{\bm{B}} 
	+ \frac{1}{c^2} \, \bm{\mathcal{E}} 
		\quad (\mathbb{M} \text{ only}), 
\]
where
\[
\bm{\mathcal{E}} = \bm{\varepsilon}( \bm{n}, \bm{E}, \, . \, , \, . \, ) 
	=  \underaccent{\check}{\bm{\varepsilon}}( \bm{E}, \, . \, , \, . \, )
	= \bm{E} \cdot  \underaccent{\check}{\bm{\varepsilon}}
\]
defines the 2-form $\bm{\mathcal{E}}$ tangent to $\mathbb{S}_t$ in terms of $\bm{E}$.
Comparison with the decomposition in Eq.~(\ref{eq:ElectromagneticSourceTensor31}) and the vacuum closure relations $\bm{D} = \epsilon_0 \bm{E}$ and $\bm{B} = \mu_0 \bm{H}$ show that
\begin{equation}
\mu_0 \, \bm{\mathcal{F}} = \bm{\star} \bm{F}
		\quad (\mathbb{M} \text{ only}), 
\label{eq:ClosureM}
\end{equation}
so that the homogeneous and inhomogeneous Maxwell equations can be expressed solely in terms of $\bm{F}$:
\begin{equation}
\begin{aligned}
\mathbf{d} \bm{F} &= 0, \\
\mathbf{d}\, {\bm{\star}\bm{F}} &= \mu_0 \, \bm{\mathcal{J}}
\end{aligned}
		\quad (\mathbb{M} \text{ only}).
\label{eq:MaxwellHodgeStar} 
\end{equation}
Moreover, the Hodge star inverse relation of Eq.~(\ref{eq:DoubleHodgeStar}) for 2-forms yields
\begin{equation}
\epsilon_0 \, \bm{F} = - \, {\bm{\star}\bm{\mathcal{F}}}
		\quad (\mathbb{M} \text{ only}),
\label{eq:ClosureInverseM}
\end{equation}
so that the Maxwell equations expressed solely in terms of $\bm{\mathcal{F}}$ as
\begin{equation}
\begin{aligned}
\mathbf{d}\, {\bm{\star}\bm{\mathcal{F}}} &= 0, \\
\mathbf{d} \bm{\mathcal{F}} &= \bm{\mathcal{J}}
\end{aligned}
		\quad (\mathbb{M} \text{ only})
\label{eq:MaxwellHodgeStarInverse}
\end{equation}
contain precisely the same content as Eq.~(\ref{eq:MaxwellHodgeStar}).
The Hodge star operator is not available on $\mathbb{G}$; instead, it will be seen below that the possibilities for Galilei-invariant electrodynamics involve instead the slash-star operator on $\mathbb{G}$ introduced at the end of Sec.~\ref{sec:SpacetimeG}.

Next, on flat manifolds the inverse of $\mathbf{d}^2 = 0$ holds, so that $\mathbf{d} \bm{F} = 0$ implies that the electromagnetic force tensor $\bm{F}$ can be expressed as the exterior derivative of an electromagnetic potential 1-form $\bm{A}$ on both $\mathbb{M}$ and $\mathbb{G}$:
\[
\bm{F} = \mathbf{d} \bm{A}. 
\]
Decomposing $\bm{A}$ as 
\begin{equation}
\bm{A} = 
	\begin{cases}
		- \phi \, \underline{\bm{\chi}} + \bm{\mathscr{a}}
			\quad (\text{on } \mathbb{M}) \\[5pt]
		- \phi \, \bm{\tau} + \bm{\mathscr{a}}
			\quad (\text{on } \mathbb{G}),
	\end{cases}
\label{eq:ElectromagneticPotential}
\end{equation}
where $\phi$ is the scalar potential and the 3-covector potential $\bm{\mathscr{a}}$ is tangent to $\mathbb{S}_t$, the equation $\bm{F} = \mathbf{d} \bm{A}$ corresponds to
\begin{align}
\underline{\bm{E}} &= - \mathrm{\underaccent{\check}{\bm{\nabla}}} \phi 
					- \frac{\partial \bm{\mathscr{a}}}{\partial t}, 
\label{eq:ElectricFieldFromPotential} \\[5pt]
\bm{B} &= \mathrm{\underaccent{\check}{\bm{\nabla}}} \times \overleftarrow{\bm{\mathscr{a}}}
\label{eq:FieldsFromPotential}
\end{align} 
on both $\mathbb{M}$ and $\mathbb{G}$, where $\overleftarrow{\bm{\mathscr{a}}} = \bm{\mathscr{a}} \cdot \overleftrightarrow{\bm{g}} = \bm{\mathscr{a}} \cdot \overleftrightarrow{\bm{\gamma}}$ on $\mathbb{M}$, and $\overleftarrow{\bm{\mathscr{a}}} =  \bm{\mathscr{a}} \cdot \overleftrightarrow{\bm{\gamma}}$ on $\mathbb{G}$.
Note also 
\begin{equation}
\overleftarrow{\bm{A}} = 
	\left\{
	\begin{aligned}
		\bm{A} \cdot \overleftrightarrow{\bm{g}} 
			&= \frac{\phi}{c^2} + \overleftarrow{\bm{\mathscr{a}}}
			& & (\text{on } \mathbb{M}), \\[5pt]
		\bm{A} \cdot \overleftrightarrow{\bm{\gamma}} 
			&= \overleftarrow{\bm{\mathscr{a}}}
			& & (\text{on } \mathbb{G}).
	\end{aligned}
	\right.
\label{eq:ElectromagneticPotentialVector}
\end{equation}
Then on $\mathbb{M}$ the inhomogeneous part of Eq.~(\ref{eq:MaxwellHodgeStar}) can be expressed indifferently in terms of either $\bm{A}$ or $\overleftarrow{\bm{A}}$:
\begin{equation}
\begin{aligned}
\bm{\Box} \bm{A} &= -\mu_0 \, \underline{\bm{J}}, \\[5pt]
\bm{\Box} \overleftarrow{\bm{A}} &= -\mu_0 \, \bm{J}
\end{aligned}
		\quad (\mathbb{M} \text{ only}),
\label{eq:VectorPoissonM}
\end{equation}
in which the Lorenz (not Lorentz! \cite{Gourgoulhon2013Special-Relativ}) gauge characterized by 
\begin{equation}
\bm{\nabla} \cdot \overleftarrow{\bm{A}} = 0 \quad (\mathbb{M} \text{ only})
\label{eq:LorenzGauge}
\end{equation}
has been employed, and
\[
\bm{\Box} = - \frac{1}{c^2} \, \frac{\partial^2}{\partial t^2} 
				+ \bm{\triangle}
		\quad (\mathbb{M} \text{ only}),
\]
where $\bm{\Box} = \bm{\nabla} \cdot \overleftarrow{\bm{\nabla}}$ and $\bm{\triangle} = \mathrm{\underaccent{\check}{\bm{\nabla}}} \cdot \overleftarrow{\mathrm{\underaccent{\check}{\bm{\nabla}}}}$.
The equations 
\[
\bm{\Box}\, \phi = - \frac{\rho}{\epsilon_0}
		\quad (\mathbb{M} \text{ only})
\]
and
\[
\begin{aligned}
\bm{\Box}\, \bm{\mathscr{a}} &= - \mu_0 \, \underline{\bm{j}}, \\[5pt]
\bm{\Box}\, \overleftarrow{\bm{\mathscr{a}}} &= - \mu_0 \, \bm{j}
\end{aligned}
		\quad (\mathbb{M} \text{ only})
\]
are compatible with both the 1-form and vector versions of Eq.~(\ref{eq:VectorPoissonM}).
But on $\mathbb{G}$, noting first that
\[
\bm{\Box} = \bm{\triangle}
		\quad (\mathbb{G} \text{ only})
\]
and comparing Eqs.~(\ref{eq:ElectromagneticPotential}) and (\ref{eq:ElectromagneticPotentialVector}) on $\mathbb{G}$, it is apparent that
the two elements of the $c \rightarrow \infty$ limit of Eq.~(\ref{eq:VectorPoissonM}),
\[
\begin{aligned}
\bm{\triangle} \bm{A} &= -\mu_0 \, \underline{\bm{J}}, \\[5pt]
\bm{\triangle} \overleftarrow{\bm{A}} &= -\mu_0 \, \bm{J}
\end{aligned}
		\quad (\mathbb{G} \text{ only})
\]
contain inequivalent content.
In the 1-form version,
\begin{equation}
\begin{aligned}
\bm{\triangle}\, \phi &= - \frac{\rho}{\epsilon_0}, \\[5pt]
\bm{\triangle}\, \bm{\mathscr{a}} &= - \mu_0 \, \underline{\bm{j}} 
\end{aligned}
		\quad (\mathbb{G} \text{ only, 1-form case}),
\label{eq:PotentialOneFormCase}
\end{equation}
where $J_0 = - \mu_0 \, c^2 \rho = - \rho / \epsilon_0$ (inherited from $\mathbb{M}$, still making sense as $c \rightarrow \infty$ due to the electromagnetic peculiarity $\epsilon_0 \mu_0 = 1 / c^2$) and $\underline{\bm{j}} = \bm{\gamma} \cdot \bm{j}$.
But in the vector version the scalar potential $\phi$ is projected out (and rendered irrelevant) and the charge density $\rho$ is constrained to vanish:
\begin{equation}
\begin{aligned}
0 &= \rho, \\[5pt]
\bm{\triangle}\, \overleftarrow{\bm{\mathscr{a}}} &= - \mu_0 \, \bm{j} 
\end{aligned}
		\quad (\mathbb{G} \text{ only, vector case}).
\label{eq:PotentialVectorCase}
\end{equation}
The existence of two distinct options regarding Galilei-invariant electrodynamics will be further elucidated below.

Finally, reconsider the electromagnetic 4-force and address the energy of the electromagnetic field.
Recognize that in a self-consistent description of an electromagnetic material medium, the electromagnetic force on a test particle in Eq.~(\ref{eq:ElectromagneticForce}) becomes a force density involving the electric current 4-vector $\bm{J}$:
\begin{equation}
n \bm{\Upsilon} = \bm{F}( \, . \, , \bm{J}) = \bm{F} \cdot \bm{J},
\label{eq:ForceDensity}
\end{equation}
where $n$ is the number density of the reference particle type (for instance, baryons) defining the material medium.
With the decompositions of $\bm{F}$ in Eq.~(\ref{eq:ElectromagneticForceTensor31}) and $\bm{J}$ in Eq.~(\ref{eq:ElectricCurrent4Vector31}), the force density counterpart of Eq.~(\ref{eq:ElectromagneticForce31}) is
\begin{equation}
n \bm{\Upsilon} = 
	\begin{cases}
		- \left( \underline{\bm{E}} \cdot \bm{j} \right) \underline{\bm{\chi}}
			\, + \rho \, \underline{\bm{E}}  
				+ \underline{ \bm{j} \times \bm{B} }
					& (\text{on } \mathbb{M}) \\[5pt]
		- \left( \underline{\bm{E}} \cdot \bm{j} \right) \bm{\tau}
					\, + \rho \, \underline{\bm{E}}  
				+ \underline{ \bm{j} \times \bm{B} }
					& (\text{on } \mathbb{G}),
	\end{cases}
\label{eq:ElectromagneticForceDensity31}
\end{equation}
whose time and space parts represent the transfer of energy and 3-momentum respectively from the electromagnetic field to the material medium. 
The energy transfer term $\underline{\bm{E}} \cdot \bm{j}$ appears as a source in the Poynting theorem
\begin{equation}
\frac{\partial}{\partial t} \left( \frac{1}{2} \, \underline{\bm{E}} \cdot \bm{D} 
		+ \frac{1}{2} \, \underline{\bm{H}} \cdot \bm{B} \right)
	+ \mathrm{\underaccent{\check}{\bm{\nabla}}} \cdot \left( \bm{E} \times \bm{H} \right)
= - \underline{\bm{E}} \cdot \bm{j},
\label{eq:PoyntingTheoremM}
\end{equation} 
which follows readily from the Maxwell equations, specifically by contracting the 3-vector relation in Eq.~(\ref{eq:InhomogeneousMaxwell}) with $\underline{\bm{E}}$ and using the 3-vector relation in Eq.~(\ref{eq:HomogeneousMaxwell}).
The Poynting theorem is manifestly a balance equation for the energy of the electromagnetic field.
It is only invariant on $\mathbb{M}$, because the Maxwell equations from which it follows are only invariant on $\mathbb{M}$.

With these preliminaries, an understanding of the possibilities for Galilei-invariant partial electrodynamics follows quickly.
On $\mathbb{G}$ we do not have the Hodge star operator but instead the non-invertible `slash-star' operator, which leads not to two different expressions of the same content, but to two separate options.
It turns out that one of these options requires only that $\bm{B} = \mu_0 \, \bm{H}$ transform properly, and the other requires only that $\bm{D} = \epsilon_0 \, \bm{E}$ transform properly. 
These relaxed requirements on the `constitutive relations' are what enable Galilei invariance.

On the one hand, taking the electromagnetic force tensor $\bm{F}$ as fundamental,
\[
\cancel{\bm{\star}} \bm{F} = \cancel{\bm{\star}}\bm{\mathcal{B}} 
	= \underline{\bm{\chi}} \wedge \underline{\bm{B}}  
		\quad (\mathbb{G} \text{ only, `magnetic'}) 
\]
zeroes out the electric displacement field $\bm{D}$ in the derived electromagnetic source tensor 
\begin{equation}
{\mu_0} \, \bm{\mathcal{F}} = \cancel{\bm{\star}} \bm{F}
	\quad (\mathbb{G} \text{ only, `magnetic'}).
\label{eq:ClosureMagneticG}
\end{equation}
In this `magnetic limit' the spacetime field equations are
\begin{equation}
\begin{aligned}
\mathbf{d} \bm{F} &= 0, \\[5pt]
\mathbf{d}\, {\cancel{\bm{\star}}\bm{F}} &= \mu_0 \, \bm{\mathcal{J}}
\end{aligned}
		\quad (\mathbb{G} \text{ only, `magnetic'}).
\label{eq:MaxwellSlashStarM}
\end{equation} 
The first equation yields
\begin{align*}
 \mathrm{\underaccent{\check}{\mathbf{d}}} \bm{\mathcal{B}} &= 0, \\[5pt]
 \mathrm{\underaccent{\check}{\mathbf{d}}} \underline{\bm{E}} 
			+ \frac{\partial \bm{\mathcal{B}}}{\partial t} &= 0
\end{align*}
as before, but the second now gives only
\[
\begin{aligned}
0 &= \bm{\mathscr{C}}, \\[5pt]
 \mathrm{\underaccent{\check}{\mathbf{d}}} \underline{\bm{B}} 
			 &= \mu_0 \, \bm{\mathscr{J}}.
\end{aligned}
	\quad (\mathbb{G} \text{ only, `magnetic'})
\]
These correspond to the Maxwell equations
\[
\begin{aligned}
\mathrm{\underaccent{\check}{\bm{\nabla}}} \cdot \bm{B} &= 0, \\
\mathrm{\underaccent{\check}{\bm{\nabla}}} \times \bm{E} 
	+ \frac{\partial \bm{B}}{\partial t} &= 0, \\[5pt]
\mathrm{\underaccent{\check}{\bm{\nabla}}} \times \bm{B} &= \mu_0 \, \bm{j}
\end{aligned}
		\quad (\mathbb{G} \text{ only, `magnetic'}), 
\]
with the charge density $\rho$ constrained to vanish (see Eq.~(\ref{eq:ElectricCharge3Form})).
The constraint on the longitudinal part of the electric field has been lost due to the projective character of the slash-star operator, but
\[
\mathrm{\underaccent{\check}{\bm{\nabla}}} \cdot \bm{E} = 0 
	\quad (\mathbb{G} \text{ only, `magnetic'})
\]
may be taken as a minimal and consistent additional assumption.
All in all, in terms of the electromagnetic potential this corresponds to the `vector case' of Eq.~(\ref{eq:PotentialVectorCase}). 
Indeed, noticing that the Lorenz gauge condition of Eq.~(\ref{eq:LorenzGauge}) reduces to
\[
\mathrm{\underaccent{\check}{\bm{\nabla}}} \cdot \overleftarrow{\bm{\mathscr{a}}} = 0 \quad (\mathbb{G} \text{ only, `magnetic'}),
\]
the additional relation $\mathrm{\underaccent{\check}{\bm{\nabla}}} \cdot \bm{E} = 0$ follows from Eq.~(\ref{eq:FieldsFromPotential}) and $\bm{E} = -\partial \overleftarrow{\bm{\mathscr{a}}} / \partial t$ in Eq.~(\ref{eq:ElectricFieldFromPotential}).
Turning to electromagnetic force and energy, thanks to $\rho = 0$ 
the electromagnetic force density of Eq.~(\ref{eq:ElectromagneticForceDensity31}) becomes
\[
n \bm{\Upsilon} = - \left( \underline{\bm{E}} \cdot \bm{j} \right) \bm{\tau}
				+ \underline{ \bm{j} \times \bm{B} }
		\quad (\mathbb{G} \text{ only, `magnetic'}). 
\]
That the electric field term disappears from the Lorentz 3-force leads LeBellac and L\'evy-Leblond \cite{Le-Bellac1973Galilean-electr} to say that the electric field is non-zero but ``does not produce any observable effect'', but it is apparent that the electric field (which is induced by a time-varying magnetic field) is still responsible for energy transfer between the electromagnetic field and the medium.
Moreover the Poynting theorem reads
\[
\frac{\partial}{\partial t} \left( \frac{1}{2} \, \underline{\bm{B}} \cdot \bm{B} \right)
	+ \mathrm{\underaccent{\check}{\bm{\nabla}}} \cdot \left( \bm{E} \times \bm{B} \right)
= - \mu_0 \, \underline{\bm{E}} \cdot \bm{j} \\
	\quad		 (\mathbb{G} \text{ only, `magnetic'}). 
\]
The electric field has also disappeared from the electromagnetic energy density, but is still responsible for an electromagnetic energy flux.
Note however that both $\bm{E}$ and $\bm{B}$ vanish when $\bm{j} = 0$ (assuming vanishing boundary conditions), and there are no electromagnetic waves in vacuum.

On the other hand, taking the electromagnetic source tensor $\bm{\mathcal{F}}$ as fundamental,
\[
\cancel{\bm{\star}} \bm{\mathcal{F}} = \cancel{\bm{\star}}\bm{\mathcal{D}} 
	= \underline{\bm{\chi}} \wedge \underline{\bm{D}}  
		\quad (\mathbb{G} \text{ only, `electric'}) 
\]
zeroes out the magnetic flux density $\bm{B}$ in the derived electromagnetic force tensor 
\begin{equation}
{\epsilon_0} \, \bm{F} = - \cancel{\bm{\star}} \bm{\mathcal{F}}
		\quad (\mathbb{G} \text{ only, `electric'}).
\label{eq:ClosureElectricG}
\end{equation}
In this `electric limit' the spacetime field equations are
\begin{equation}
\begin{aligned}
\mathbf{d} \cancel{\bm{\star}} \bm{\mathcal{F}} &= 0, \\[5pt]
\mathbf{d}\, \bm{\mathcal{F}} &= \bm{\mathcal{J}}
\end{aligned}
		\quad (\mathbb{G} \text{ only, `electric'}).
\label{eq:MaxwellSlashStarE}
\end{equation} 
The first equation gives only
\[
\begin{aligned}
0 &= 0, \\[5pt]
 \mathrm{\underaccent{\check}{\mathbf{d}}} \underline{\bm{D}} 
			 &= 0
\end{aligned}
	\quad (\mathbb{G} \text{ only, `electric'}),
\]
but the second equation yields
\begin{align*}
 \mathrm{\underaccent{\check}{\mathbf{d}}} \bm{\mathcal{D}} &= \bm{\mathscr{C}}, \\[5pt]
 \mathrm{\underaccent{\check}{\mathbf{d}}} \underline{\bm{H}} 
			- \frac{\partial \bm{\mathcal{D}}}{\partial t} &= \bm{\mathscr{J}}
\end{align*}
as before.
These correspond to the Maxwell equations
\[
\begin{aligned}
\mathrm{\underaccent{\check}{\bm{\nabla}}} \times \bm{D} 
	 &= 0, \\[5pt]
\mathrm{\underaccent{\check}{\bm{\nabla}}} \cdot \bm{D} &= \rho, \\
\mathrm{\underaccent{\check}{\bm{\nabla}}} \times \bm{H} - \frac{\partial \bm{D}}{\partial t} 
	&= \bm{j}
\end{aligned}
		\quad (\mathbb{G} \text{ only, `electric'}). 
\]
The constraint on the longitudinal part of the magnetic field has been lost due to the projective character of the slash-star operator, but
\[
\mathrm{\underaccent{\check}{\bm{\nabla}}} \cdot \bm{H} = 0 
	\quad (\mathbb{G} \text{ only, `electric'})
\]
may be taken as a minimal and consistent additional assumption.
All in all, in terms of the electromagnetic potential this corresponds to the `1-form case' of Eq.~(\ref{eq:PotentialOneFormCase}).
Indeed, the additional relation $\mathrm{\underaccent{\check}{\bm{\nabla}}} \cdot \bm{B} = 0$ follows from Eq.~(\ref{eq:FieldsFromPotential}).
Turning to electromagnetic force and energy,
the electromagnetic force density of Eq.~(\ref{eq:ElectromagneticForceDensity31}) becomes
\[
\epsilon_0 \, n \bm{\Upsilon} = - \left( \underline{\bm{D}} \cdot \bm{j} \right) \bm{\tau}
				+ \rho \, \underline{\bm{D}}.
		\quad (\mathbb{G} \text{ only, `electric'}). 
\]
That the magnetic field term disappears from the electromagnetic force leads LeBellac and L\'evy-Leblond \cite{Le-Bellac1973Galilean-electr} to say that the magnetic field is non-zero but ``has no effect at all''.
However, the Poynting theorem reads
\begin{align*}
\frac{\partial}{\partial t} \left( \frac{1}{2} \, \underline{\bm{D}} \cdot \bm{D} \right)
	+ \mathrm{\underaccent{\check}{\bm{\nabla}}} \cdot \left( \bm{D} \times \bm{H} \right)
&= - \underline{\bm{D}} \cdot \bm{j} \\
	&		 (\mathbb{G} \text{ only, `electric'}). 
\end{align*}
The magnetic field has also disappeared from the electromagnetic energy density, but is still responsible for an electromagnetic energy flux.
Note however that both $\bm{D}$ and $\bm{H}$ vanish when $\rho = 0$ and $\bm{j} = 0$ (assuming vanishing boundary conditions), and once again there are no electromagnetic waves in vacuum.

While the invariance on $\mathbb{M}$ and lack of invariance on $\mathbb{G}$ of full electrodynamics were explored with explicit transformations in Sec.~\ref{sec:FullElectrodynamics}, note that the Poincar\'e invariance of full electrodynamics on $\mathbb{M}$ is guaranteed by the spacetime tensor formulation in Eq.~(\ref{eq:MaxwellHodgeStar}), the same content expressed also in Eq.~(\ref{eq:MaxwellHodgeStarInverse}). 
Similarly, the Galilei invariance of the `magnetic' and `electric' versions of partial electrodynamics on $\mathbb{G}$ is guaranteed by the spacetime tensor formulations in Eqs.~(\ref{eq:MaxwellSlashStarM}) and (\ref{eq:MaxwellSlashStarE}) respectively.
A perhaps more compelling way to summarize this is to say that the spacetime invariance of a closed system of electrodynamics equations,
\begin{align*}
\mathbf{d} \bm{F} &= 0 \\
\mathbf{d} \bm{\mathcal{F}} &= \bm{\mathcal{J}},
\end{align*}
is assured when the closure relation can be expressed as a spacetime tensor relation between $\bm{F}$ and $\bm{\mathcal{F}}$.
On $\mathbb{M}$ the closure relation of Eq.~(\ref{eq:ClosureM}) or Eq.~(\ref{eq:ClosureInverseM})---which are merely inverses of one another---is compatible with full electrodynamics as expressed in the familiar Maxwell equations.
On $\mathbb{G}$ the closure relations of Eqs.~(\ref{eq:ClosureMagneticG}) and (\ref{eq:ClosureElectricG})---which are not inverses of one another, but instead two distinct alternatives---yield two different kinds of partial electrodynamics by paring down the Maxwell equations in two different ways.

\section{The extended affine spacetimes $B\mathbb{M}$ and $B\mathbb{G}$}
\label{sec:ExtendedAffineSpacetimes}

Returning to the discussion at the end of Sec.~\ref{sec:MaterialParticleMG}, a means of exhibiting the transformation of kinetic energy while remaining consistent with Poincar\'e and Galilei physics is desired.
This is accomplished by extending the 4-dimensional affine spacetimes $\mathbb{M}$ and $\mathbb{G}$ to the 5-dimensional affine spacetimes $B\mathbb{M}$ and $B\mathbb{G}$.
Remarkably, unlike the relation between $\mathbb{M}$ and $\mathbb{G}$, not only $B\mathbb{M}$ but also $B\mathbb{G}$ is a pseudo-Riemann space, with the Bargmann metric $\bm{G}$ on $B\mathbb{M}$ reducing to that on $B\mathbb{G}$ as $c \rightarrow \infty$.
This metric turns out to be invariant under the groups of Bargmann-Lorentz and homogeneous Bargmann-Galilei transformations designed to exhibit the transformation of kinetic energy. 
As with $\mathbb{M}$ and $\mathbb{G}$, a projection operator to slices of `position space' and a few key vectors and covectors provide for the decomposition of tensors into pieces suitable for the description of observations by fiducial observers.

\subsection{Bargmann spacetime and Bargmann transformations}
\label{sec:BargmannSpacetime}

Work backwards towards Bargmann-Minkowski (or B-Minkowski) spacetime $B \mathbb{M}$ and Bargmann-Galilei (or B-Galilei) spacetime $B \mathbb{G}$ by considering an `inertia-momentum-energy' 5-vector 
\[
\hat{\bm{I}} = M \, \hat{\bm{U}}
\]
that extends the inertia-momentum 4-vector on $\mathbb{M}$ and $\mathbb{G}$.
Relative to a fiducial observer, and with respect to what will be called a Bargmann-Minkowski (or B-Minkowski) or Bargmann-Galilei (or B-Galilei) basis, beyond the time and space components representing inertia and vector 3-momentum respectively, extend Eq.~(\ref{eq:InertiaMomentumFiducial}) to include kinetic energy as a fifth component:
\begin{equation}
\hat{\mathsf{I}} = M \, \hat{\mathsf{U}} = 
	\begin{cases}
		M \begin{bmatrix} \Lambda_\mathsf{v}  \, \phantom{\mathsf{v}} 
				\\ \Lambda_\mathsf{v} \, \mathsf{v}
				\\ c^2 \left( \Lambda_\mathsf{v} - 1 \right)
			\end{bmatrix} 
			& (\text{on } B\mathbb{M}) \\[20pt]
		M \begin{bmatrix} 1
				\\ \mathsf{v}
				\\ \frac{1}{2} \lVert \mathsf{v} \rVert^2
			\end{bmatrix} 
			& (\text{on } B\mathbb{G}),
	\end{cases}
\label{eq:InertiaMomentumEnergyFiducial}
\end{equation}
from which the 5-column $\hat{\mathsf{U}}$ representing the 5-velocity $\hat{\bm{U}}$ can immediately be read.
Note that $\lVert \mathsf{v} \rVert^2 = \mathsf{v}^\mathrm{T} \mathsf{v} = \bm{\gamma} \left( \bm{v}, \bm{v} \right)$ will be appropriate to a B-Minkowski or B-Galilei basis.
As on $\mathbb{M}$ and $\mathbb{G}$, regard the 5-velocity
\begin{equation}
\hat{\bm{U}} = \frac{\mathrm{d} \hat{\mathbf{X}}}{\mathrm{d} \tau} 
		=  \frac{\mathrm{d} t}{\mathrm{d} \tau} \frac{\mathrm{d} \hat{\mathbf{X}}}{\mathrm{d} t}
\label{eq:FiveVelocity}
\end{equation}
as the tangent vector field to a worldline $\left\{ \hat{\mathbf{X}} (\tau) \mid \tau \in \mathbb{R} \right\} \subset B\mathbb{M}, B\mathbb{G}$.
The parameter $\tau$ is to continue to be the proper time governed by Eqs.~(\ref{eq:ProperTimeM}) and (\ref{eq:ProperTimeG}) on $\mathbb{M}$ and $\mathbb{G}$ respectively, with the tensors $\bm{g}$ and $\bm{\tau}$ continuing to be given by Eqs.~(\ref{eq:MetricMinkowski}) and (\ref{eq:TimeForm}) in terms of elements of dual B-Minkowski and B-Galilei bases.

The additional dimension requires an additional coordinate.
With the selection of an origin and a B-Minkowski or B-Galilei basis corresponding to a fiducial observer, a point $\hat{\mathbf{X}} (\tau)$ along the particle worldline is represented by a 5-column
\[
\hat{\mathsf{X}} = \begin{bmatrix} t \\ \mathsf{x} ( t ) \\ \eta ( t ) \end{bmatrix} 
	= \begin{bmatrix} t \\ x^i ( t ) \\ \eta ( t ) \end{bmatrix}
\]
(compare Eq.~(\ref{eq:PointFourColumn})).
Given Eq.~(\ref{eq:FiveVelocity}) and comparing with Eq.~(\ref{eq:InertiaMomentumEnergyFiducial}), it is apparent that the fifth component $\hat{U}^\eta$ of the 5-velocity, associated with the new coordinate $\eta$ along the worldline of a material particle in $B\mathbb{M}$ or $B\mathbb{G}$, must satisfy
\begin{equation}
\hat{U}^\eta
	= \frac{ \mathrm{d} \eta }{ \mathrm{d} \tau } 
	=  \frac{ \mathrm{d} t }{ \mathrm{d} \tau } \frac{ \mathrm{d} \eta }{ \mathrm{d} t }
	=
	\left\{
	\begin{aligned}
		 c^2 \left( \Lambda_\mathsf{v} - 1 \right) & & (\text{on } B\mathbb{M}) \\
	\frac{1}{2} \lVert \mathsf{v} \rVert^2 & & (\text{on } B\mathbb{G}).
	\end{aligned}
	\right. 
\label{eq:FifthVelocityComponent}
\end{equation}
The expressions on the right-hand side might be called the `specific kinetic energy', as they are equivalent to $\epsilon_\mathsf{v} / M$; and as $\epsilon_\mathsf{v} \, \mathrm{d}\tau$ has units of action, $\eta$ might be called the `specific kinetic action coordinate', or `action coordinate' for short.
The action coordinate relation of Eq.~(\ref{eq:FifthVelocityComponent}) will prove crucial to the geometry of $B\mathbb{M}$ and $B\mathbb{G}$.
  
The next step is to determine the $5 \times 5$ B-Lorentz transformation matrices $\hat{\mathsf{P}}_{B\mathbb{M}}^+$ and homogeneous B-Galilei transformation matrices $\hat{\mathsf{P}}_{B\mathbb{G}}^+$ that extend the $4 \times 4$ Lorentz transformation matrices $\mathsf{P}_{\mathbb{M}}^+$ of Eq.~(\ref{eq:LorentzTransformation}) and homogeneous Galilei transformation matrices $\mathsf{P}_{\mathbb{G}}^+$ of Eq.~(\ref{eq:GalileiTransformation}) previously encountered on $\mathbb{M}$ and $\mathbb{G}$ respectively.
The 5-velocity transforms according to
\[
\hat{\mathsf{U}} 
	= \hat{\mathsf{P}}^+ \, \hat{\mathsf{U}}',
\]
which corresponds to either $B\mathbb{M}$ or $B\mathbb{G}$.
Cast this in the (4+1)-dimensional form
\begin{equation}
\begin{bmatrix} \mathsf{U} \\ \hat{U}^\eta \end{bmatrix} 
	= \begin{bmatrix} \mathsf{P}^+ & \mymathsf{0} \\ \sfPhi & 1 \end{bmatrix}
		\begin{bmatrix} \mathsf{U}' \\ \hat{U}'^\eta \end{bmatrix}, 
\label{eq:FiveVelocityTransformation}
\end{equation}
where $\mathsf{U}$, $\mathsf{U}'$, and $\mathsf{P}^+$ correspond to the 4-dimensional spacetimes $\mathbb{M}$ or $\mathbb{G}$.
The 4-column $\mymathsf{0} = \begin{bmatrix} 0^\mu \end{bmatrix}$ in
\begin{equation}
\hat{\mathsf{P}}^+ = \begin{bmatrix} \mathsf{P}^+ & \mymathsf{0} \\ \sfPhi & 1 \end{bmatrix}
\label{eq:BargmannTransformation}
\end{equation}
ensures that the 4-dimensional relation $\mathsf{U} = \mathsf{P}^+ \, \mathsf{U}'$ on $\mathbb{M}$ or $\mathbb{G}$ is preserved when embedded in the 5-dimensional setting of $B\mathbb{M}$ or $B\mathbb{G}$: the fifth component $\hat{U}^\eta$ of $\hat{\mathsf{U}}$ does not `contaminate' the $t$ and $\mathsf{x}$ components.
The 4-column $\mymathsf{0}$ also ensures that the matrix representations of $\bm{g}$ and $\bm{\tau}$ governing causality on $\mathbb{M}$ and $\mathbb{G}$ respectively do not acquire non-vanishing components in the $\eta$ dimension when these are regarded as tensors on  
$B\mathbb{M}$ and $B\mathbb{G}$; this means that $\bm{g} \left( \hat{\bm{U}}, \hat{\bm{U}} \right) =  \bm{g} \left( \bm{U}, \bm{U} \right) = -c^2$ and $\bm{\tau} \left( \hat{\bm{U}} \right) =  \bm{\tau} \left( \bm{U} \right) = 1$, that is, the `timelike 4-velocity' character of $\bm{U}$ on $\mathbb{M}$ or $\mathbb{G}$ is preserved when it is extended to the 5-velocity $\hat{\bm{U}}$ on $B\mathbb{M}$ or $B\mathbb{G}$.

It remains to specify the 4-row $\sfPhi$ in Eq.~(\ref{eq:BargmannTransformation}), which gives the transformation rule for (specific) kinetic energy.
Of course this is already determined by the Lorentz and Galilei transformations on $\mathbb{M}$ and $\mathbb{G}$ respectively.  
In particular, the time component of the transformation rule $\mathsf{U} = \mathsf{P}_\mathbb{M}^+ \, \mathsf{U}'$ on $\mathbb{M}$ allows one to find
\[
\begin{split}
c^2 \left( \Lambda_\mathsf{v} - 1 \right) 
	&=  c^2 \left( \Lambda_\mathsf{u} - 1 \right) \Lambda_{\mathsf{v}'}  
		 	+ \left( \Lambda_\mathsf{u} \,\mathsf{u}^\mathrm{T} \mathsf{R}_\mathbb{S} \right) 
			 \left( \Lambda_{\mathsf{v}'} \, \mathsf{v}' \right)  \\
	& \quad   + \, c^2 \left( \Lambda_{\mathsf{v}'} - 1 \right)
\end{split}
\]
in terms of the boost parameter $\mathsf{u} \in \mathbb{R}^{3 \times 1}$ and rotation $\mathsf{R}_\mathbb{S} \in \mathrm{SO}(3)$.
Moreover the space component of the transformation rule $\mathsf{U} = \mathsf{P}_\mathbb{G}^+ \, \mathsf{U}'$ on $\mathbb{G}$ allows one to find
\[
 \frac{1}{2} \lVert \mathsf{v} \rVert^2 
	=  \frac{1}{2} \lVert \mathsf{u} \rVert^2
		+ \left( \mathsf{u}^\mathrm{T} \mathsf{R}_\mathbb{S} \right) \mathsf{v}'  
		+  \frac{1}{2} \lVert \mathsf{v}' \rVert^2.
\]
From these expressions and use of Eq.~(\ref{eq:InertiaMomentumEnergyFiducial}) in Eq.~(\ref{eq:FiveVelocityTransformation}), the 4-row $\sfPhi$ in Eq.~(\ref{eq:BargmannTransformation}) can be read off:
\begin{equation}
\sfPhi =
	\begin{cases}
		\begin{bmatrix} c^2 \left( \Lambda_\mathsf{u} - 1 \right) 
			& \Lambda_\mathsf{u} \,\mathsf{u}^\mathrm{T} \mathsf{R}_\mathbb{S}\end{bmatrix} 
		& (\text{on } B\mathbb{M}) \\[10pt]
		\begin{bmatrix}    \frac{1}{2} \lVert \mathsf{u} \rVert^2
			& \mathsf{u}^\mathrm{T} \mathsf{R}_\mathbb{S}\end{bmatrix} 
		& (\text{on } B\mathbb{G}).
	\end{cases}
\label{eq:BargmannFourRow}
\end{equation}
The compatibility of these expressions as $c \rightarrow \infty$ is evident.
No new parameters beyond $\mathsf{u} \in \mathbb{R}^{3 \times 1}$ and $\mathsf{R}_\mathbb{S} \in \mathrm{SO}(3)$ already present in a Lorentz transformation $\mathsf{P}_{\mathbb{M}}^+$ or homogeneous Galilei transformation $\mathsf{P}_{\mathbb{G}}^+$ are introduced.
The element $1$ in the last row and column of Eq.~(\ref{eq:BargmannTransformation}) is also confirmed.

This completes specification of the B-Lorentz transformations $\hat{\mathsf{P}}_{B\mathbb{M}}^+$ and the homogeneous B-Galilei transformations $\hat{\mathsf{P}}_{B\mathbb{G}}^+$, which act on the vector spaces $\mathrm{V}_{B\mathbb{M}}$ and $\mathrm{V}_{B\mathbb{G}}$ underlying the extended spacetimes $B\mathbb{M}$ and $B\mathbb{G}$ respectively.

\subsection{Bargmann group and Bargmann metric}
\label{sec:BargmannMetric}

The set of restricted B-Lorentz transformations $\hat{\mathsf{P}}_{B\mathbb{M}}^+$ and the set of restricted homogeneous B-Galilei transformations $\hat{\mathsf{P}}_{B\mathbb{G}}^+$, given by Eqs.~(\ref{eq:BargmannTransformation}) and (\ref{eq:BargmannFourRow}) with $\mathsf{P}^+ = \mathsf{P}_\mathbb{M}^+$ or $\mathsf{P}^+ = \mathsf{P}_\mathbb{G}^+$, are subgroups of $\mathrm{GL}(5)$.
It is evident that these sets of matrices contain the identity ($\mathsf{u} = \mymathsf{0}$ and $\mathsf{R}_\mathbb{S} = \mymathsf{1}$).
To identify the inverse of $\hat{\mathsf{P}}^+$, note again a factorization
\[
\hat{\mathsf{P}}^+ = \hat{\mathsf{L}} \, \hat{\mathsf{R}},
\]
with
\[
\hat{\mathsf{L}} 
	= 	\begin{bmatrix}
			\mathsf{L} & \mymathsf{0} \\
			\sfPhi &  1
		\end{bmatrix}, \ \ \ 
\hat{\mathsf{R}} 
	= 	\begin{bmatrix}
			1 & \mymathsf{0} & 0 \\
			\mymathsf{0} & \mathsf{R}_\mathbb{S} & \mymathsf{0} \\
			0 &
		 	 \mymathsf{0} & 1
		\end{bmatrix},
\]
so that ${{}\hat{\mathsf{P}}^+}^{-1} = \hat{\mathsf{R}}^\mathrm{T} \, \hat{\mathsf{L}}^{-1}$ with $\hat{\mathsf{L}}^{-1}$ obtained from $\hat{\mathsf{L}}$ via $\mathsf{u} \mapsto -\mathsf{u}$.
Closure under matrix multiplication is shown by considering the product
\[
{{}\hat{\mathsf{P}}^+}'' = {{}\hat{\mathsf{P}}^+} \, {{}\hat{\mathsf{P}}^+}',
\]
or in $4+1$ block form,
\[
\begin{bmatrix} {\mathsf{P}^+}'' & \mymathsf{0} \\ \sfPhi'' & 1 \end{bmatrix}
	= \begin{bmatrix} {\mathsf{P}^+} & \mymathsf{0} \\ \sfPhi & 1 \end{bmatrix}
		\begin{bmatrix} {\mathsf{P}^+}' & \mymathsf{0} \\ \sfPhi' & 1 \end{bmatrix} 
	= \begin{bmatrix} {\mathsf{P}^+} \, {\mathsf{P}^+}' && \mymathsf{0} 
		\\ \sfPhi \, {\mathsf{P}^+}' + \sfPhi' && 1 \end{bmatrix}. 
\]
The $4 \times 4$ matrix relation 
\begin{equation}
{\mathsf{P}^+}'' = {\mathsf{P}^+} \, {\mathsf{P}^+}'
\label{eq:LorentzGalileiClosure}
\end{equation}
in the upper-left block is simply the known closure of the restricted Lorentz or restricted homogeneous Galilei group.
The remaining question is whether the 4-row
\[
\sfPhi'' = \sfPhi \, {\mathsf{P}^+}' + \sfPhi' 
\]
is in the form of Eq.~(\ref{eq:BargmannFourRow}), with the relevant expressions involving $\mathsf{u}''$ and $\mathsf{R}''$ determined consistently from Eq.~(\ref{eq:LorentzGalileiClosure}).
Direct computation shows that the answer is yes, completing the demonstration of closure.

The existence of a `Bargmann metric' $\bm{G}$ is suggested by the `action coordinate relation' in Eq.~(\ref{eq:FifthVelocityComponent}) relating coordinate variations along a material particle worldline, and it turns out to be invariant under B-Lorentz or homogeneous B-Galilei transformations, making it a fundamental structure on $B\mathbb{M}$ or $B\mathbb{G}$.
On $B\mathbb{M}$, use $\Lambda_\mathsf{v} = \mathrm{d} t / \mathrm{d} \tau$ and $c^2 \, \mathrm{d} \tau^2 = c^2 \, \mathrm{d} t^2 - \lVert \mathrm{d} \mathsf{x} \rVert^2$ in Eq.~(\ref{eq:FifthVelocityComponent}) to deduce
\[
- 2 \, \mathrm{d} \eta \, \mathrm{d} t + \mathrm{d} x^a \, 1_{a b} \, \mathrm{d} x^b 
	+ \frac{1}{c^2} \, \mathrm{d} \eta^2 = 0 \ \ \ (\mathrm{on \ } B\mathbb{M}).
\]
On $B\mathbb{G}$, use $\mathrm{d} \tau = \mathrm{d} t$ and $\lVert \mathsf{v} \rVert^2 \, \mathrm{dt}^2 =  \lVert \mathrm{d} \mathsf{x} \rVert^2$ to deduce analogously
\[
- 2 \, \mathrm{d} \eta \, \mathrm{d} t + \mathrm{d} x^a \, 1_{a b} \, \mathrm{d} x^b 
	 = 0 \ \ \ (\mathrm{on \ } B\mathbb{G}).
\]
In both cases the left-hand side looks like a line element, suggestive of a Bargmann metric (or B-metric) $\bm{G}$ represented by the B-Minkowski or B-Galilei matrix
\begin{equation}
\mathsf{G}
	=
	\begin{cases}
		\begin{bmatrix} 0 & 0_j & -1 \\ 0_i & 1_{ij} & 0_i \\ -1 & 0_j & \frac{1}{c^2} \end{bmatrix} 
			= \sfeta_{B\mathbb{M}} & (\text{on } B\mathbb{M}) \\[20pt]
		 \begin{bmatrix} 0 & 0_j & -1 \\ 0_i & 1_{ij} & 0_i \\ -1 & 0_j & 0 \end{bmatrix} 
			= \sfeta_{B\mathbb{G}} & (\text{on } B\mathbb{G})
	\end{cases}
\label{eq:BMinkowskiBGalileiMatrices}
\end{equation}
with respect to a B-Minkowski or B-Galilei basis.
(Apologies for the visual similarity of the action coordinate $\eta$, the Minkowski matrix $\sfeta$ related to $\mathbb{M}$, and the B-Minkowski and B-Galilei matrices $\sfeta_{B \mathbb{M}}$ and $\sfeta_{B \mathbb{G}}$. They must not be confused.) 
The Bargmann metric itself is given by
\[
\bm{G} =
	\left\{	
	\begin{aligned}
	& - \, \bm{e}_*^0 \otimes \bm{e}_*^4  \, + \, \bm{e}_*^1 \otimes \bm{e}_*^1
		\, + \, \bm{e}_*^2 \otimes \bm{e}_*^2 \, + \, \bm{e}_*^3 \otimes \bm{e}_*^3  
		 \, - \, \bm{e}_*^4 \otimes \bm{e}_*^0 
		 \, + \, \frac{1}{c^2} \, \bm{e}_*^4 \otimes \bm{e}_*^4
		& & (\text{on } B\mathbb{M})  \\[5pt]
	& - \, \bm{e}_*^0 \otimes \bm{e}_*^4  \, + \, \bm{e}_*^1 \otimes \bm{e}_*^1
		\, + \, \bm{e}_*^2 \otimes \bm{e}_*^2 \, + \, \bm{e}_*^3 \otimes \bm{e}_*^3  
		 \, - \, \bm{e}_*^4 \otimes \bm{e}_*^0 
		& & (\text{on } B\mathbb{G}) 
	\end{aligned}
	\right.
\]
in terms of the elements of a B-Minkowski or B-Galilei dual basis.
The Bargmann metric $\bm{G}$ is a fundamental invariant structure on $B\mathbb{M}$ and $B\mathbb{G}$, in the sense that
\[
\bm{G} \left( \hat{\bm{P}} \left( \bm{a} \right), 
						\hat{\bm{P}} \left( \bm{b} \right) \right) 
		= \bm{G} (\bm{a},\bm{b})
\]
for any $\bm{a}, \bm{b} \in \mathrm{V}_B$.
With respect to a B-Minkowski or B-Galilei basis this condition reads
\begin{align*}
\hat{\mathsf{P}}_{B\mathbb{M}}^\mathrm{T} \,\, \sfeta_{B\mathbb{M}} \,\, \hat{\mathsf{P}}_{B\mathbb{M}}
	&=  \sfeta_{B\mathbb{M}}, \\[5pt]
\hat{\mathsf{P}}_{B\mathbb{G}}^\mathrm{T} \,\, \sfeta_{B\mathbb{G}} \,\, \hat{\mathsf{P}}_{B\mathbb{G}}
	&=  \sfeta_{B\mathbb{G}},
\end{align*}
which are verified by direct computation for both $\hat{\mathsf{P}}_{B\mathbb{M}}^+$ with $\sfeta_{B\mathbb{M}}$ and $\hat{\mathsf{P}}_{B\mathbb{G}}^+$ with $\sfeta_{B\mathbb{G}}$.
Note however that the 6-dimensional Lie groups of restricted B-Lorentz transformations $\hat{\mathsf{P}}_{B\mathbb{M}}^+$ and restricted homogeneous B-Minkowski transformations $\hat{\mathsf{P}}_{B\mathbb{G}}^+$ are only subgroups of the 10-dimensional Lie groups that preserve $\bm{G}$ for $B\mathbb{M}$ and $B\mathbb{G}$ respectively; because of this, invariance of $\bm{G}$ is not sufficient to prove closure, which instead is proved directly.

The above calculation suggesting the existence of $\bm{G}$ also shows that 
\begin{equation}
\bm{G} \left( \hat{\bm{U}}, \hat{\bm{U}} \right) 
	= 	\hat{\mathsf{U}}^\mathrm{T} \, \mathsf{G} \, \hat{\mathsf{U}}
	=	0,
\label{eq:FiveVelocityNorm}
\end{equation}
that is, that $\hat{\bm{U}}$ is null with respect to $\bm{G}$. 
This is so even though $\hat{\bm{U}}$ remains timelike with respect to $\bm{g}$ or $\bm{\tau}$ as appropriate, as noted previously.

The inverse metric $\overleftrightarrow{\bm{G}}$ is represented by
\begin{equation}
\overleftrightarrow{\mathsf{G}}
	= 
	\begin{cases}
		\begin{bmatrix}  -\frac{1}{c^2} & 0^j & -1 \\ 0^i & 1^{ij} & 0^i \\ -1 & 0^j & 0 \end{bmatrix} 
		= \overleftrightarrow{\sfeta}_{\!B\mathbb{M}} & (\text{on } B\mathbb{M}) \\[20pt]
		\begin{bmatrix} 0 & 0^j & -1 \\ 0^i & 1^{ij} & 0^i \\ -1 & 0^j & 0 \end{bmatrix} 
		= \overleftrightarrow{\sfeta}_{\!B\mathbb{G}} & (\text{on } B\mathbb{G})
	\end{cases}
\label{eq:BMinkowskiBGalileiInverseMatrices}
\end{equation}
with respect to a B-Minkowski or B-Galilei basis.
It is given by
\[
\overleftrightarrow{\bm{G}} =
	\left\{	
	\begin{aligned}
	&- \, \frac{1}{c^2} \, \bm{e}_0 \otimes \bm{e}_0 
		\, - \, \bm{e}_0 \otimes \bm{e}_4 \, + \, \bm{e}_1 \otimes \bm{e}_1 
		\, + \, \bm{e}_2 \otimes \bm{e}_2 \, + \, \bm{e}_3 \otimes \bm{e}_3
		\, - \, \bm{e}_4 \otimes \bm{e}_0
		& & (\text{on } B\mathbb{M})  \\[5pt]
	&\phantom{ - \, \frac{1}{c^2} \, \bm{e}_0 \otimes \bm{e}_0 } \ \, 
		\, - \, \bm{e}_0 \otimes \bm{e}_4 \, + \, \bm{e}_1 \otimes \bm{e}_1 
		\, + \, \bm{e}_2 \otimes \bm{e}_2 \, + \, \bm{e}_3 \otimes \bm{e}_3
		\, - \, \bm{e}_4 \otimes \bm{e}_0
		& & (\text{on } B\mathbb{G})
	\end{aligned}
	\right.
\]
in terms of the elements of a B-Minkowski or B-Galilei basis.

Note the remarkable difference in the relationship between $\mathbb{M}$ and $\mathbb{G}$ on the one hand and between $B\mathbb{M}$ and $B\mathbb{G}$ on the other, including startlingly different geometric consequences.
Whereas the spacetime $\mathbb{M}$ is a pseudo-Riemann manifold with metric $\bm{g}$ and inverse $\overleftrightarrow{\bm{g}}$, the spacetime $\mathbb{G}$ obtained as $c \rightarrow \infty$ is not:  instead of a metric and its true inverse one is left with an invariant time form $\bm{\tau}$ and an invariant degenerate inverse `metric' $\overleftrightarrow{\bm{\gamma}}$.
In contrast, both $B\mathbb{M}$ and $B\mathbb{G}$ are pseudo-Riemann manifolds with a metric $\bm{G}$ and inverse $\overleftrightarrow{\bm{G}}$  (of signature $-++++$, and $\mathrm{det} \,\mathsf{G} = -1$ with respect to a B-Minkowski or B-Galilei basis), the versions of both of these on $B\mathbb{M}$ limiting smoothly to those on $B\mathbb{G}$ as $c \rightarrow \infty$, as is evident from the above expressions in terms of B-Minkowski and B-Galilei bases.

With both $B\mathbb{M}$ and $B\mathbb{G}$ as pseudo-Riemann manifolds, henceforth let the underline and overarrow notation denote the raising and lowering of indices with respect to $\bm{G}$.

Exterior differentiation and the (invertible) Hodge star operator---now available on both $B\mathbb{M}$ and $B\mathbb{G}$---will be needed in Sec.~\ref{sec:ElectrodynamicsB}.
Exterior differentiation is the same on the Bargmann spacetimes as on the original spacetimes, because no explicit dependence of tensor fields on the coordinate $\eta$ will be allowed:
\begin{align*}
\hat{\mathbf{d}} 
	&= \bm{e}_*^A \wedge \frac{\partial}{\partial x^A} \\[5pt]
	&= \bm{e}_*^\alpha \wedge \frac{\partial}{\partial x^\alpha} 
		+ \bm{e}_*^4 \wedge \frac{\partial}{\partial x^4}
\end{align*}
reduces to
\begin{equation}
\hat{\mathbf{d}} 
	= \bm{e}_*^\alpha \wedge \frac{\partial}{\partial x^\alpha} 
	= \mathbf{d}
\label{eq:ExteriorDerivativeB}
\end{equation}
(compare Eq.~(\ref{eq:ExteriorDerivativeDecomposition})),
the partial derivative with respect to $x^4 = \eta$ vanishing in all cases.
Note the summation convention, with upper-case Latin indices taking values in $\{0, 1, 2, 3, 4\}$, with letters $A, B, \dots$ near the beginning of the alphabet preferred for dummy indices, and letters $I, J, \dots$ from later in the alphabet preferred for free indices.
An orientation on $B\mathbb{M}$ or $B\mathbb{G}$ is specified with the Levi-Civita tensor $\hat{\bm{\varepsilon}}$ defined such that
\[
\hat{\bm{\varepsilon}}( \bm{e}_0, \bm{e}_1, \bm{e}_2, \bm{e}_3, \bm{e}_4 ) = 1
\]
with components
\[
\hat{\varepsilon}_{ I J K L M } = \left[ I J K L M \right]
\]
for a right-handed B-Minkowski or B-Galilei basis.
With respect to another right-handed but otherwise arbitrary basis $( \bm{e}'_0, \bm{e}'_1, \bm{e}'_2, \bm{e}'_3, \bm{e}'_4 )$, Eq.~(\ref{eq:VolumeFormTransformed}), together with the matrix relation $\mathsf{G}' = \hat{\mathsf{P}}^\mathrm{T} \, \sfeta_B \, \hat{\mathsf{P}}$, show that in the more general basis the components are given by
\[
\hat{\varepsilon}'_{ I J K L M } = \sqrt{ -G' } \, \left[ I J K L M \right],
\]
where $G' = \det \mathsf{G'}$.
Raising all five indices yields
\[
\hat{\varepsilon}'^{ I J K L M } = - \frac{1}{\sqrt{ -G' }} \,\left[ I J K L M \right]
	\quad (\mathbb{M} \text{ only}),
\]
with respect to a general basis, or
\[
\hat{\varepsilon}^{ I J K L M } = - \left[ I J K L M \right]
\]
with respect to a B-Minkowski or B-Galilei basis.
The metric $\bm{G}$ makes the volume form $\hat{\bm{\varepsilon}}$ also a Levi-Civita tensor, and makes available the Hodge star operator $\hat{\bm{\star}}$ that provides a bijection between $p$-forms and $(5-p)$-forms on $B\mathbb{M}$ or $B\mathbb{G}$.
In particular,
\[
\hat{\star} F_{I_1 \dots I_{5-p}} 
	= \frac{1}{p!} \, F^{A_1 \dots A_p} \, \hat{\varepsilon}_{A_1 \dots A_p I_1 \dots I_{5-p}}
\]
gives the components of the $(5-p)$-form $\hat{\bm{\star}}\bm{F}$ dual to the $p$-form $\bm{F}$.

Finally, note that the groups of B-Lorentz and homogeneous B-Galilei transformations discussed here act on the vector spaces $\mathrm{V}_{B\mathbb{M}}$ and $\mathrm{V}_{B\mathbb{G}}$ underlying the extended affine spacetimes $B\mathbb{M}$ and $B\mathbb{G}$ respectively.
The points or events of these extended spacetimes transform by elements of the B-Poincar\'e and B-Galilei groups, which add translations to the B-Lorentz and homogeneous B-Galilei groups, as discussed in the Appendix.

\subsection{Bargmann spacetime foliation and tensor decomposition}
\label{sec:BargmannSpacetimeFoliation}

As was the case with Minkowski spacetime $\mathbb{M}$ and Galilei-Newton spacetime $\mathbb{G}$, it is necessary to decompose the extended spacetimes $B\mathbb{M}$ and $B\mathbb{G}$ and tensor fields thereon in a manner that enables comparison with observations.
Beyond decomposition into `time' and `space', there is now decomposition into `time', `space', and `action', the latter corresponding to the additional coordinate $x^4 = \eta$.

Select an origin $\mathbf{O}$ of $B\mathbb{M}$ or $B\mathbb{G}$, and a fiducial B-Minkowski or B-Galilei basis $( \bm{e}_0, \bm{e}_1, \bm{e}_2, \bm{e}_3, \bm{e}_4 )$ of the underlying vector space $\mathrm{V}_{B\mathbb{M}}$ or $\mathrm{V}_{B\mathbb{G}}$ along with its dual basis $( \bm{e}_*^0, \bm{e}_*^1, \bm{e}_*^2, \bm{e}_*^3, \bm{e}_*^4 )$. 
Regard the affine space $B\mathbb{M}$ or $B\mathbb{G}$ also as a differentiable manifold and think of the above bases as frame fields; then the basis consists of coordinate basis vectors associated with the 5-tuple $( t, x^i, \eta )$ of global B-Minkowski or B-Galilei coordinates, and the dual basis consists of the exterior derivatives or covariant gradients of these coordinate functions.

Consider the 1+3+1 splitting of the extended affine spacetimes $B\mathbb{M}$ and $B\mathbb{G}$ according to a fiducial `inertial observer'. 
As with $\mathbb{M}$ and $\mathbb{G}$ there is a time axis
\[
\mathbb{T} = \{\mathbf{O} +  \bm{e}_0 \, t \mid t \in \mathbb{R} \},
\]
and now also an `action axis'
\[
\mathbb{A} = \{\mathbf{O} +  \bm{e}_4 \, \eta \mid \eta \in \mathbb{R} \}.
\]
Position space as perceived by the fiducial observer at time $t$, for a given value of $\eta$, is the affine 3-plane
\[
\mathbb{S}_{(t,\eta)} = \{ \mathbf{O} + \bm{e}_0 \, t + \bm{e}_i \, x^i + \bm{e}_4 \, \eta \mid ( x^i ) \in \mathbb{R}^3 \}.
\]
The complete collection $\left( \mathbb{S}_{(t,\eta)} \right)_{(t,\eta) \in \mathbb{R}^2}$ is a foliation of $B\mathbb{M}$ or $B\mathbb{G}$ whose leaves are affine 3-planes of codimension 2, instead of hyperplanes of codimension 1 as was the case with $\mathbb{M}$ or $\mathbb{G}$.

In expressing the projection operator $\overleftarrow{\bm{\gamma}}$ used to decompose tensors into pieces along $\mathbb{T}$, tangent to $\mathbb{S}_{(t,\eta)}$, and along $\mathbb{A}$, it will prove convenient to give special labels to the time and action elements of these bases, and in the process to define three special 5-vector fields $\bm{n}$, $\bm{\chi}$, and $\bm{\xi}$.
Similar to $\mathbb{M}$ and $\mathbb{G}$, regard
\[
\bm{n} = \bm{e}_0, \quad \quad 
	\mathsf{n} = \begin{bmatrix} 1 \\ 0^i \\ 0 \end{bmatrix}
\]
as the 5-velocity of the fiducial observer on both $B\mathbb{M}$ and $B\mathbb{G}$.
Its metric dual $\underline{\bm{n}} = \bm{G} \cdot \bm{n}$, represented by $\underline{\mathsf{n}} = \left( \sfeta_B  \mathsf{n} \right)^\mathrm{T}$, is
\[
\underline{\bm{n}} = - \bm{e}_*^4 = - \bm{\nabla}\eta, \quad \quad
	\underline{\mathsf{n}} = \begin{bmatrix} 0 & 0_i & -1 \end{bmatrix}
\]
on both $B\mathbb{M}$ and $B\mathbb{G}$.
On $\mathbb{M}$ but not on $\mathbb{G}$ a dual observer covector was defined by Eq.~(\ref{eq:DualObserverCovector}); similarly a dual observer covector
\[
\underline{\bm{\chi}} = \bm{e}_*^0 = \bm{\nabla}t, \quad \quad
	\underline{\sfchi} = \begin{bmatrix} 1 & 0_i & 0 \end{bmatrix}
\]
satisfying
\[
\underline{\bm{\chi}} \cdot \bm{n} = 1
\]
can now be defined on both $B\mathbb{M}$ and $B\mathbb{G}$.
Note that
\[
\underline{\bm{\chi}} = \bm{\tau} \quad (B\mathbb{G} \mathrm{\ only}),
\]
the linear form $\bm{\tau}$ remaining invariant on $B\mathbb{G}$ as it is on $\mathbb{G}$.
Its metric dual $\bm{\chi} = \underline{\bm{\chi}} \cdot \overleftrightarrow{\bm{G}}$, the dual fiducial observer vector, is
\begin{equation}
\bm{\chi} = 
	\left\{
	\begin{aligned}
		- \frac{1}{c^2} \, \bm{n} - \bm{e}_4 \\ 
		- \bm{e}_4,
	\end{aligned}
	\right.
\quad \quad \quad
\sfchi = 
	\begin{cases}
		\begin{bmatrix} -\frac{1}{c^2} \\[5pt] 0^i \\ -1 \end{bmatrix} 
			& (\text{on } B\mathbb{M}) \\[20pt]
		\begin{bmatrix} 0 \\ 0^i \\ -1 \end{bmatrix} 
			& (\text{on } B\mathbb{G}).
	\end{cases}
\label{eq:DualObserverVectorB}
\end{equation}
Unlike $\mathbb{M}$, on which $\bm{\chi}$ and $\bm{n}$ are collinear according to Eq.~(\ref{eq:DualFiducialObserverVector}), these vectors are linearly independent in the case of $B\mathbb{M}$ or $B\mathbb{G}$.
Finally, it will prove useful to also define the `action vector'
\[
\bm{\xi} =  - \bm{e}_4 =
	\left\{
	\begin{aligned}
		 \frac{1}{c^2} \, \bm{n} + \bm{\chi}  	& & (\text{on } B\mathbb{M}) \\ 
							\bm{\chi}  	& & (\text{on } B\mathbb{G}),
	\end{aligned}
	\right.
\quad \quad \quad
\sfxi = \begin{bmatrix} 0 \\ 0^i \\ -1 \end{bmatrix}.
\]
Note that $\bm{\xi}$ coincides with $\bm{\chi}$ on $B\mathbb{G}$, and is equal to $-\bm{e}_4$ on both $B\mathbb{M}$ and $B\mathbb{G}$.
Its metric dual $\underline{\bm{\xi}} = \bm{G} \cdot \bm{\xi}$ is
\[
\underline{\bm{\xi}} = 
	\left\{
	\begin{aligned}
		\frac{1}{c^2} \, \underline{\bm{n}} + \underline{\bm{\chi}} \\
		\underline{\bm{\chi}}, 
	\end{aligned}
	\right.
\quad \quad \quad
\underline{\sfxi} = 
	\begin{cases}
		\begin{bmatrix} 1 & 0_i & -\frac{1}{c^2} \end{bmatrix} & (\text{on } B\mathbb{M}) \\[10pt]
		\begin{bmatrix} 1 & 0_i & 0 \end{bmatrix}			& (\text{on } B\mathbb{G}).
	\end{cases}
\]
For reference, the norms of these vectors with respect to $\bm{G}$ are
\[
\begin{aligned}
\bm{G}( \bm{n}, \bm{n} ) = \underline{\bm{n}} \cdot \bm{n} 
	&=  0, \\
\bm{G}( \bm{\chi}, \bm{\chi} ) = \underline{\bm{\chi}} \cdot \bm{\chi} 
	&= \begin{cases} 
		- \frac{1}{c^2} & (\text{on } B\mathbb{M})  \\
		\phantom{-} 0 &  (\text{on } B\mathbb{G})
	     \end{cases} \\
\bm{G}( \bm{\xi}, \bm{\xi} ) = \underline{\bm{\xi}} \cdot \bm{\xi} 
	&= \begin{cases}
		\phantom{-} \frac{1}{c^2} & (\text{on } B\mathbb{M}) \\
	   	\phantom{-} 0 	&  (\text{on } B\mathbb{G}). 
	     \end{cases}
\end{aligned}
\]
Their mutual contractions
\begin{align*}
\bm{G}( \bm{\chi}, \bm{n} ) = \underline{\bm{\chi}} \cdot \bm{n} 
	&= 1,  \\
\bm{G}( \bm{n}, \bm{\xi} ) = \underline{\bm{n}} \cdot \bm{\xi} 
	&= 1,  \\
\bm{G}( \bm{\chi}, \bm{\xi} ) = \underline{\bm{\chi}} \cdot \bm{\xi} 
	&= 0
\end{align*}
are the same on $B\mathbb{M}$ and $B\mathbb{G}$.
That $\bm{G}( \bm{n}, \bm{n} ) = 0$ as in Eq.~(\ref{eq:FiveVelocityNorm}) for $\hat{\bm{U}}$, together with $\bm{g}(\bm{n},\bm{n}) = -c^2$ on $B\mathbb{M}$ or $\bm{\tau}(\bm{n}) = 1$ on $B\mathbb{G}$ as is also the case for $\hat{\bm{U}}$, identifies $\bm{n}$ as timelike and suitable as a 5-velocity; indeed the straight line $\mathbb{T}$ to which it is tangent will be regarded as the worldline of the fiducial observer.
In relation to the fiducial vector and covector bases, $\bm{n}$ and $\underline{\bm{n}}$ are equally simple, while the covector $\underline{\bm{\chi}}$ is simpler than $\bm{\chi}$, and the vector $\bm{\xi}$ is simpler than the covector $\underline{\bm{\xi}}$.
This will affect which of these appear in the projection operator $\overleftarrow{\bm{\gamma}}$ and are used in tensor decompositions.

As on $\mathbb{M}$ (but not on $\mathbb{G}$), the projection operator $\overleftarrow{\bm{\gamma}}$ to $\mathbb{S}_{(t,\eta)}$ turns out to be related to the 3-metric $\bm{\gamma}$ by metric duality on both $B\mathbb{M}$ and $B\mathbb{G}$.
The latter can be expressed
\begin{align*}
\bm{\gamma} &= \bm{e}_*^1 \otimes \bm{e}_*^1 + \bm{e}_*^2 \otimes \bm{e}_*^2
				+ \bm{e}_*^3 \otimes \bm{e}_*^3  \\[5pt]
		&= \bm{G} - \underline{\bm{n}} \otimes \underline{\bm{\chi}} 
				- \underline{\bm{\xi}} \otimes \underline{\bm{n}}
\end{align*}
on $B\mathbb{M}$ or $B\mathbb{G}$, provided the appropriate expressions for $\bm{G}$ and $\underline{\bm{\xi}}$ are used.
Raising the first index,
\begin{equation}
\overleftarrow{\bm{\gamma}}	
	= \bm{\delta} - \bm{n} \otimes \underline{\bm{\chi}} 
		- \bm{\xi} \otimes \underline{\bm{n}},
\label{eq:ProjectionFiducialB}
\end{equation}
and one verifies
\begin{align*}
0 &= \overleftarrow{\bm{\gamma}} \cdot \bm{n} 
	= \overleftarrow{\bm{\gamma}} \cdot \bm{\chi}
	= \overleftarrow{\bm{\gamma}} \cdot \bm{\xi}, \\[5pt]
0 &= \underline{\bm{n}} \cdot \overleftarrow{\bm{\gamma}}  
	= \underline{\bm{\chi}} \cdot \overleftarrow{\bm{\gamma}} 
	= \underline{\bm{\xi}} \cdot \overleftarrow{\bm{\gamma}} 
\end{align*}
as desired.
For the decomposition of a vector field on $B\mathbb{M}$ or $B\mathbb{G}$, the time, space, and action components are given by contraction with $\underline{\bm{\chi}}$, $\overleftarrow{\bm{\gamma}}$, and $-\underline{\bm{n}}$ respectively.
For the decomposition of a covector field on $B\mathbb{M}$ or $B\mathbb{G}$, the time, space, and action components are given by contraction with $\bm{n}$, $\overleftarrow{\bm{\gamma}}$, and $-\bm{\xi}$ respectively.

Expressed in terms of the vectors $\bm{n}$, $\bm{\chi}$, and $\bm{\xi}$ and/or their metric duals, the fiducial B-Minkowski or B-Galilei basis and dual basis can be written as a 5-row $\hat{\bm{\mathsf{B}}}$ and 5-column $\hat{\bm{\mathsf{B}}}_*$ respectively: 
\begin{align*}
\hat{\bm{\mathsf{B}}}
	&= \begin{bmatrix} \bm{n} & \bm{e}_i & -\bm{\xi} \end{bmatrix}
		= \begin{bmatrix} \bm{\mathsf{B}} & -\bm{\xi} \end{bmatrix}, \\[5pt]
\hat{\bm{\mathsf{B}}}_*
	&= \begin{bmatrix} \underline{\bm{\chi}} \\[5pt] \bm{e}_*^i \\[5pt] -\underline{\bm{n}} 					\end{bmatrix}
		= \begin{bmatrix} \bm{\mathsf{B}}_* \\[5pt] -\underline{\bm{n}} \end{bmatrix},
\end{align*}
extending the Minkowski or Galilei basis 4-row $\bm{\mathsf{B}}$ and dual basis 4-column $\bm{\mathsf{B}}_*$.
Under a B-Lorentz or homogeneous B-Galilei transformation, the vector basis transforms according to
\[
\hat{\bm{\mathsf{B}}}' = \hat{\bm{\mathsf{B}}} \, \hat{\mathsf{P}} 
	= \begin{bmatrix} \bm{\mathsf{B}} & -\bm{\xi} \end{bmatrix}
		\begin{bmatrix} \mathsf{P} & \mymathsf{0} \\ \sfPhi & 1 \end{bmatrix}
	= \begin{bmatrix} \bm{\mathsf{B}} \, \mathsf{P} - \bm{\xi} \sfPhi  && -\bm{\xi} \end{bmatrix}.
\]
Of note here is that the action vector $\bm{\xi}'$ associated with the new action coordinate $\eta'$ is invariant, that is, 
\[
\bm{\xi}' = \bm{\xi}.
\]
Thus, while the time axis $\mathbb{T}$ and position space 3-planes $\mathbb{S}_{(t,\eta)}$ tilt under B-Lorentz or homogeneous B-Galilei transformations, the action axis $\mathbb{A}$ is invariant.\footnote{
Picture the extended vector space $V_{B\mathbb{M}}$ or $V_{B\mathbb{G}}$ as a collection of conventional spacetime diagrams of $V_{\mathbb{M}}$ or $V_{\mathbb{G}}$ stacked along the action axis $\mathbb{A}$. Under a B-Lorentz or homogeneous B-Galilei transformation, within each subspace (spacetime diagram) $V_{\mathbb{M}}$ or $V_{\mathbb{G}}$ the vectors $(\bm{n}, \bm{e}_i )$  transform as usual according to the term $\bm{\mathsf{B}} \, \mathsf{P}$, but also the entire stack of subspaces (spacetime diagrams) $V_{\mathbb{M}}$ or $V_{\mathbb{G}}$ tilts relative to the invariant action axis $\mathbb{A}$ according to the term $- \bm{\xi} \sfPhi$.}
Meanwhile the dual covector basis transforms according to
\[
\hat{\bm{\mathsf{B}}}_*' =  \hat{\mathsf{P}}^{-1} \, \hat{\bm{\mathsf{B}}}_*
	= \begin{bmatrix} \mathsf{P}^{-1} & \mymathsf{0} \\ \tilde{\sfPhi} & 1 \end{bmatrix}
		\begin{bmatrix} \bm{\mathsf{B}}_* \\ -\underline{\bm{n}} \end{bmatrix}
	= \begin{bmatrix} \mathsf{P}^{-1} \, \bm{\mathsf{B}}_* 
			\\ \tilde{\sfPhi} \, \bm{\mathsf{B}}_* - \underline{\bm{n}} \end{bmatrix}.
\]
Of note here is that the first four dual basis covectors---those that span the dual space of the vector space underlying $\mathbb{M}$ or $\mathbb{G}$---transform under B-Lorentz or homogeneous B-Galilei transformations in the same way they do under Lorentz or homogeneous Galilei transformations:
\[
\bm{\mathsf{B}}_*' =  \mathsf{P}^{-1} \, \bm{\mathsf{B}}_*.
\]
As noted earlier, this means that when $\bm{g}$ on $\mathbb{M}$ given by Eq.~(\ref{eq:MetricMinkowski}), or $\bm{\tau}$ on $\mathbb{G}$ given by Eq.~(\ref{eq:TimeForm}), are regarded as tensors on $B\mathbb{M}$ or $B\mathbb{G}$, the manner in which they govern causality according to Poincar\'e or Galilei physics, by giving a proper time interval $\mathrm{d}\tau$ according to Eq.~(\ref{eq:ProperTimeM}) or (\ref{eq:ProperTimeG}), is preserved in the 5-dimensional Bargmann setting.

The 1+3+1 splittings of the exterior differentiation operator and the volume form will be needed in Sec.~\ref{sec:ElectrodynamicsB}.
Referring to Eqs.~(\ref{eq:ExteriorDerivativeDecomposition}) and (\ref{eq:ExteriorDerivativeB}),
\begin{align*}
\hat{\mathbf{d}} &= \mathbf{d} \\
	&= \underline{\bm{\chi}} \wedge \frac{\partial}{\partial t} 
		+ \mathrm{\underaccent{\check}{\mathbf{d}}}.
\end{align*}
Moreover, because $\bm{n} = \bm{e}_0$ and $\bm{\xi} = -\bm{e}_4$, the contraction
\begin{equation}
\underaccent{\check}{\bm{\varepsilon}} 
	= -\hat{\bm{\varepsilon}}(\bm{n}, \, . \, , \, . \, , \, . \, , \bm{\xi} ) 
		= - \bm{n} \cdot \hat{\bm{\varepsilon}} \cdot \bm{\xi}
\label{eq:VolumeContractionB} 
\end{equation}
(compare Eq.~(\ref{eq:VolumeContraction})) yields the space volume form $ \underaccent{\check}{\bm{\varepsilon}}$ on $\mathbb{S}_{(t,\eta)}$.
Conversely, because $\underline{\bm{\chi}} = \bm{e}_*^0$ and $\underline{\bm{n}} = -\bm{e}_*^4$,
\begin{equation}
\hat{\bm{\varepsilon}} = 
		- \underline{\bm{\chi}} \wedge \underaccent{\check}{\bm{\varepsilon}}
			\wedge \underline{\bm{n}}
\label{eq:VolumeDecompositionB}
\end{equation}
(compare Eq.~(\ref{eq:VolumeDecomposition})) is a useful factorization of the extended spacetime volume form $\hat{\bm{\varepsilon}}$.
These expressions are valid on both $B\mathbb{M}$ and $B\mathbb{G}$.

\section{A material particle on $B\mathbb{M}$ or $B\mathbb{G}$}
\label{sec:MaterialParticleBMBG}

Consider again in passing the 5-velocity $\hat{\bm{U}}$ already described, and note that tensor decompositions with respect to a comoving observer are available.
Consider also a 5-covector version of Newton's second law for a material particle on $B\mathbb{M}$ or $B\mathbb{G}$.

\subsection{Kinematics}
\label{sec:KinematicsB}

The kinematics of a material particle on $B\mathbb{M}$ or $B\mathbb{G}$ has already been given, since the 5-velocity 
\begin{equation}
\hat{\bm{U}} =
	\left\{
	\begin{aligned}
		\Lambda_{\bm{v}} \left( \bm{n} + \bm{v} \right) 
			&- c^2 \left( \Lambda_{\bm{v}} - 1 \right) \bm{\xi} 
			& &(\text{on } B\mathbb{M})  \\
		 \bm{n} + \bm{v}  
			&- \frac{1}{2} \, \bm{\gamma} \left( \bm{v}, \bm{v} \right) \bm{\xi} 
			& &(\text{on } B\mathbb{G}),
	\end{aligned}
	\right. 
\label{eq:FiveVelocityFiducial}
\end{equation}
expressed here decomposed relative to a fiducial inertial observer, was introduced in the course of characterizing these extended spacetimes.

Here it is worth noting that tensors on $B\mathbb{M}$ and $B\mathbb{G}$ can be locally decomposed relative to a comoving observer with 5-velocity $\hat{\bm{U}}$ instead of the fiducial inertial observer with 5-velocity $\bm{n}$.
Key to such a decomposition is the operator
\[
\overleftarrow{\bm{\gamma}_{\hat{\bm{U}}}} 
	= \bm{\delta} - \hat{\bm{U}} \otimes \underline{\bm{\chi}}_{\hat{\bm{U}}}
		- \bm{\xi} \otimes \underline{\hat{\bm{U}}}
\]
that projects vectors to a 3-plane $\mathbb{S}_{(\hat{\bm{U}}(\tau),\eta)}$ constituting position space according a comoving observer.
Comparing with Eq.~(\ref{eq:ProjectionFiducialB}), note that $\bm{n}$ and $\underline{\bm{n}}$ are replaced by $\hat{\bm{U}}$ and $\underline{\hat{\bm{U}}}$, but that $\bm{\xi}$ is unchanged in accord with its invariant status.
The covector $\underline{\hat{\bm{U}}} = \bm{G} \cdot \hat{\bm{U}}$ can be expressed
\[
\underline{\hat{\bm{U}}} =
	\left\{
	\begin{aligned}
		- c^2 \left( \Lambda_{\bm{v}} - 1 \right) \underline{\bm{\chi}}
			+ \Lambda_{\bm{v}} \, \underline{\bm{v}}
			+ \underline{\bm{n}} & & (\text{on } B\mathbb{M}) \\
		- \frac{1}{2} \, \bm{\gamma} \left( \bm{v}, \bm{v} \right) \, \underline{\bm{\chi}}
			+  \underline{\bm{v}}
			+ \underline{\bm{n}} & & (\text{on } B\mathbb{G}). 
	\end{aligned}
	\right.
\]
The covector $\underline{\bm{\chi}}_{\hat{\bm{U}}}$ satisfying $\underline{\bm{\chi}}_{\hat{\bm{U}}} \cdot \hat{\bm{U}} = 1$ can be deduced from its counterpart $\underline{\bm{\chi}}_{\bm{U}} = -\bm{g} \cdot \bm{U} / c^2 = - ( \bm{\gamma} + \underline{\bm{n}} \otimes \underline{\bm{\chi}} ) \cdot \bm{U} / c^2$ on $\mathbb{M}$ and  $\underline{\bm{\chi}}_{\bm{U}} \rightarrow \bm{\tau}$ on $\mathbb{G}$ with the result
\[
\underline{\bm{\chi}}_{\hat{\bm{U}}} =
	\left\{
	\begin{aligned}
		& \Lambda_{\bm{v}} \left( \underline{\bm{\chi}} - \frac{1}{c^2} \, \underline{\bm{v}} \right)
		& & (\text{on } B\mathbb{M}) \\
		 & \underline{\bm{\chi}} & & (\text{on } B\mathbb{G}).
	\end{aligned}
	\right.
\]
Analogous to relations involving $\bm{n}$ and $\underline{\bm{\chi}}$ in the case of the fiducial observer, one finds the relation
\[
\bm{\xi} = 
	\left\{
	\begin{aligned}
		 \frac{1}{c^2} \, \hat{\bm{U}} + \bm{\chi}_{\hat{\bm{U}}}  & & (\text{on } B\mathbb{M}) \\ 
							\bm{\chi}_{\hat{\bm{U}}}  	& & (\text{on } B\mathbb{G}),
	\end{aligned}
	\right.
\]
the norms
\begin{align*}
\bm{G}( \hat{\bm{U}}, \hat{\bm{U}} ) = \underline{\hat{\bm{U}}} \cdot \hat{\bm{U}} 
	&= 0, \\
\bm{G}( \bm{\chi}_{\hat{\bm{U}}}, \bm{\chi}_{\hat{\bm{U}}} ) 
		= \underline{\bm{\chi}}_{\hat{\bm{U}}} \cdot \bm{\chi}_{\hat{\bm{U}}} 
	&= \begin{cases}
		 - \frac{1}{c^2} \quad (\text{on } B\mathbb{M}) \\
		\phantom{ - } 0 \quad \ \  (\text{on } B\mathbb{G}),
	     \end{cases}
\end{align*}
the mutual contractions
\begin{align*}
\bm{G}( \bm{\chi}_{\hat{\bm{U}}}, \hat{\bm{U}} ) 
		= \underline{\bm{\chi}}_{\hat{\bm{U}}} \cdot \hat{\bm{U}} 
	&= 1, \\
\bm{G}( \hat{\bm{U}}, \bm{\xi} ) = \underline{\bm{U}} \cdot \bm{\xi} 
	&= 1, \\
\bm{G}( \bm{\chi}_{\hat{\bm{U}}}, \bm{\xi} ) = \underline{\bm{\chi}}_{\hat{\bm{U}}} \cdot \bm{\xi} 
	&= 0,
\end{align*}
and the vanishing projections
\begin{align*}
0 &= \overleftarrow{\bm{\gamma}_{\hat{\bm{U}}}} \cdot \hat{\bm{U}} 
	= \overleftarrow{\bm{\gamma}_{\hat{\bm{U}}}} \cdot \bm{\chi}_{\hat{\bm{U}}}
	= \overleftarrow{\bm{\gamma}_{\hat{\bm{U}}}} \cdot \bm{\xi}, \\[5pt]
0 &= \underline{\hat{\bm{U}}} \cdot \overleftarrow{\bm{\gamma}_{\hat{\bm{U}}}}  
	= \underline{\bm{\chi}}_{\hat{\bm{U}}} \cdot \overleftarrow{\bm{\gamma}_{\hat{\bm{U}}}} 
	= \underline{\bm{\xi}} \cdot \overleftarrow{\bm{\gamma}_{\hat{\bm{U}}}}.
\end{align*}
Taken together these show that $\hat{\bm{U}}$, $\underline{\hat{\bm{U}}}$, $\underline{\bm{\chi}}_{\hat{\bm{U}}}$, $\bm{\xi}$, and $\overleftarrow{\bm{\gamma}_{\hat{\bm{U}}}}$ provide for the decomposition of tensors according to a comoving observer.

\subsection{Dynamics}
\label{sec:DynamicsB}

Since the extended spacetime B-metric $\bm{G}$ is available on both $B\mathbb{M}$ and $B\mathbb{G}$, consider the 5-covector version of Newton's second law:
\begin{equation}
\frac{\mathrm{d}}{\mathrm{d}\tau} \, \underline{\hat{\bm{I}}} = \hat{\bm{\Upsilon}},
\label{eq:NewtonSecondLawB}
\end{equation}
where $\underline{\hat{\bm{I}}} = \bm{G} \cdot \hat{\bm{I}} = M \, \hat{\underline{\bm{U}}}$.

With respect to a B-Minkowski or B-Galilei basis,
$\underline{\hat{\bm{I}}}$ is represented by the 5-row
\begin{equation}
\underline{\hat{\mathsf{I}}} = M \, \underline{\hat{\mathsf{U}}} = 
	\begin{cases}
		\begin{bmatrix} - M c^2 \left( \Lambda_\mathsf{v} - 1 \right) 
				& M \Lambda_\mathsf{v} \, \underline{\mathsf{v}}
				& -M
			\end{bmatrix} 
			& (\text{on } B\mathbb{M}) \\[10pt]
		\begin{bmatrix} - \frac{1}{2} M \lVert \mathsf{v} \rVert^2
				& M \underline{\mathsf{v}}
				& -M
			\end{bmatrix} 
			& (\text{on } B\mathbb{G}).
	\end{cases}
\label{eq:EnergyMomentumMassFiducial}
\end{equation}
Compare the relationship between Eqs.~(\ref{eq:InertiaMomentumFiducial}) and (\ref{eq:EnergyMomentumFiducial}) on the one hand, and between Eqs.~(\ref{eq:InertiaMomentumEnergyFiducial}) and (\ref{eq:EnergyMomentumMassFiducial}) on the other, in order to appreciate the different impacts of index lowering via a metric on the 4D spacetimes $\mathbb{M}$ and $\mathbb{G}$ vs. the 5D Bargmann extended spacetimes $B\mathbb{M}$ and $B\mathbb{G}$.
The inertia-momentum $\bm{I}$ exists on both $\mathbb{M}$ and $\mathbb{G}$, even though 
in terms of bulk motion the inertia $M \Lambda_\mathsf{v}$ is dynamic on $\mathbb{M}$ while it is fixed to the rest mass $M$ on $\mathbb{G}$.
Because the Minkowski metric $\bm{g}$ exists on $\mathbb{M}$ but not on $\mathbb{G}$, the total-energy--momentum $\underline{\bm{I}}$ exists on $\mathbb{M}$ but not on $\mathbb{G}$. 
Lowering the index converts dynamic inertia $M \Lambda_\mathsf{v}$ to (the negative of) total energy $M c^2 \Lambda_\mathsf{v}$ in the time component.
In contrast, on both $B\mathbb{M}$ and $B\mathbb{G}$ index lowering with the B-metric $\bm{G}$ converts the inertia--momentum--kinetic-energy $\hat{\bm{I}}$ to the kinetic-energy--momentum--mass $\underline{\hat{\bm{I}}}$:
the off-diagonal time/action components in the B-Minkowski and B-Galilei matrices $\sfeta_{B\mathbb{M}}$ and $\sfeta_{B\mathbb{G}}$ of Eq.~(\ref{eq:BMinkowskiBGalileiMatrices}) swap the places (and change the signs) of inertia and kinetic energy;
and the diagonal action-action component $1 / c^2$ in $\sfeta_{B\mathbb{M}}$ has the effect of converting the dynamic inertia $M \Lambda_\mathsf{v}$ into the rest mass $M$ on $B\mathbb{M}$, resulting in the same rest mass $M$ that constitutes inertia on $B\mathbb{G}$.

With the definitions of 3-momentum $\bm{p}$ and kinetic energy $\epsilon_{\bm{p}}$ in Eqs.~(\ref{eq:ThreeMomentum}) and (\ref{eq:KineticEnergy}) respectively, the kinetic-energy--momentum--mass 5-covector can be written
\begin{equation}
\hat{\bm{\Pi}} = \underline{\hat{\bm{I}}} = - \epsilon_{\bm{p}} \, \underline{\bm{\chi}} + \bm{p} + M \, \underline{\bm{n}}
\label{eq:EnergyMomentumMass}
\end{equation}
on both $B\mathbb{M}$ and $B\mathbb{G}$, so that Eq.~(\ref{eq:NewtonSecondLawB}) reads
\begin{equation}
\frac{\mathrm{d}}{\mathrm{d}\tau} \, \hat{\bm{\Pi}} = \hat{\bm{\Upsilon}}.
\label{eq:NewtonSecondLawB2}
\end{equation}
The denomination $\hat{\bm{\Pi}} = \underline{\hat{\bm{I}}}$ is motivated by the fact that its time and space pieces are precisely the relative energy-momentum $\bm{\Pi}$ of Eq.~(\ref{eq:RelativeEnergyMomentumExplicit}) on $\mathbb{M}$ and $\mathbb{G}$. 
Just as $\hat{\bm{U}}$ extends $\bm{U}$ with kinetic energy as a fifth component, in a similar manner $\hat{\bm{\Pi}}$ extends $\bm{\Pi}$ with rest mass as a fifth component.
(Note that while the notation $\underline{\hat{\bm{U}}} = \bm{G} \cdot \bm{U}$ and $\underline{\hat{\bm{I}}} = \bm{G} \cdot \bm{I}$ have been used here, these 5-covectors do not extend the 4-covectors $\underline{\bm{U}}$ and $\underline{\bm{I}}$ with an additional component; instead, $\underline{\hat{\bm{U}}}$ and $\underline{\hat{\bm{I}}}$ are simply the metric duals with respect to $\bm{G}$ of $\hat{\bm{U}}$ and $\hat{\bm{I}}$, which do extend the 4-vectors $\bm{U}$ and $\bm{I}$.)

Turn to the 5-force covector $\hat{\bm{\Upsilon}}$ and find its decomposition relative to the fiducial observer.
(Note that the notation has been arranged in such a way that Eq.~(\ref{eq:NewtonSecondLawB2}) extends to Bargmann spacetimes Eq.~(\ref{eq:CovectorNewtonSecondRelative}) rather than Eq.~(\ref{eq:CovectorNewtonSecond}); in particular, $\hat{\bm{\Upsilon}}$ extends $\bm{\Upsilon}$ of Eq.~(\ref{eq:RelativeFourForce}) rather than $\bm{\Upsilon}_{\bm{\!I}}$.)
Using Eq.~(\ref{eq:EnergyMomentumMass}) in Eq.~(\ref{eq:NewtonSecondLawB2}) yields
\begin{equation}
-\frac{\mathrm{d} \epsilon_{\bm{p}}}{\mathrm{d}\tau} \, \underline{\bm{\chi}} 
	+ \frac{\mathrm{d}\bm{p}}{\mathrm{d}\tau} 
	+ \frac{\mathrm{d}M}{\mathrm{d}\tau} \, \underline{\bm{n}}
= \hat{\bm{\Upsilon}}.
\label{eq:FiveForceB}
\end{equation}
Following the definitions 
\[
\hat{\bm{\Upsilon}} \cdot \bm{\xi} 
	= \frac{\mathrm{d}M}{\mathrm{d}\tau} = 	
	\left\{
	\begin{aligned}
		\frac{\theta}{c^2} & & (\text{on } B\mathbb{M}) \\[5pt]
		0 & & (\text{on } B\mathbb{G})
	\end{aligned}
	\right.	
\]
and
\[
\hat{\bm{\Upsilon}} \cdot \overleftarrow{\bm{\gamma}}
	= \frac{\mathrm{d}\bm{p}}{\mathrm{d}\tau} =
	\left\{
	\begin{aligned}
		\Lambda_{\bm{v}} \, \frac{\mathrm{d}\bm{p}}{\mathrm{d}t} 
			&= \Lambda_{\bm{v}} \, \bm{\mathscr{F}}  & & (\text{on } B\mathbb{M}) \\[5pt]
		\frac{\mathrm{d}\bm{p}}{\mathrm{d}t} 
			&= \bm{\mathscr{F}}  & & (\text{on } B\mathbb{G}) 
	\end{aligned}
	\right.	
\]
of heating rate $\theta$ and 3-force covector $\bm{\mathscr{F}}$ utilized on $\mathbb{M}$ and $\mathbb{G}$, consider whether
\[
-\hat{\bm{\Upsilon}} \cdot \bm{n} 
	= \frac{\mathrm{d}\epsilon_{\bm{p}}}{\mathrm{d}\tau}
\]
agrees with the result found on $\mathbb{M}$ and $\mathbb{G}$.
In order to determine this, note that since
\[
\frac{\mathrm{d}\hat{\bm{\Pi}}}{\mathrm{d}\tau}
	= \frac{\mathrm{d}M}{\mathrm{d}\tau} \, \underline{\hat{\bm{U}}} 
		+ M \,\frac{\mathrm{d}}{\mathrm{d}\tau} \, \underline{\hat{\bm{U}}}
\]
and $\bm{G}\left( \hat{\bm{U}}, \hat{\bm{U}} \right) = \underline{\hat{\bm{U}}} \cdot \hat{\bm{U}} = 0$, it holds that
\[
\frac{\mathrm{d}\hat{\bm{\Pi}}}{\mathrm{d}\tau} \cdot \hat{\bm{U}} = 0.
\]
Therefore use of Eqs.~(\ref{eq:FiveVelocityFiducial}) and (\ref{eq:FiveForceB}) together with the relations immediately above results in
\begin{align*}
0 	&= \hat{\bm{\Upsilon}} \cdot \hat{\bm{U}} \\[5pt]
 	&= 
	\left\{
	\begin{aligned}
		&\Lambda_{\bm{v}}^2 \left( 
			-\frac{\mathrm{d}\epsilon_{\bm{p}}}{\mathrm{d}t} 
				+ \bm{\mathscr{F}} \cdot \bm{v} \right) 
			- \left( \Lambda_{\bm{v}} - 1 \right)  \theta  & & (\text{on } B\mathbb{M}) \\[5pt]
		& \quad \ \   -\frac{\mathrm{d}\epsilon_{\bm{p}}}{\mathrm{d}t} 
				+ \bm{\mathscr{F}} \cdot \bm{v} & & (\text{on } B\mathbb{G}),
	\end{aligned}
	\right.
\end{align*}
which yields
\[
\frac{\mathrm{d} \epsilon_{\bm{p}}  }{\mathrm{d} t} =
	\left\{
	\begin{aligned}  
		 & \bm{\mathscr{F}} \cdot \bm{v} 
			- \frac{ \left( \Lambda_{\bm{v}} - 1 \right) }{ \Lambda_{\bm{v}}^2 } \, \theta
		& & (\mathrm{on \ } B\mathbb{M})  \\[5pt]
		 & \bm{\mathscr{F}} \cdot \bm{v} 
		& & (\mathrm{on \ } B\mathbb{G})
	\end{aligned}
	\right.
\]
in agreement with Eq.~(\ref{eq:WorkEnergyTheorem}).

\section{Electrodynamics on $B\mathbb{M}$ and $B\mathbb{G}$}
\label{sec:ElectrodynamicsB}

Each of the spacetime formulations of electrodynamics on $\mathbb{M}$ or $\mathbb{G}$ given in Sec.~\ref{sec:Electrodynamics} consists of two sets of field equations, a set of closure relations, and a force law describing the interaction of the electromagnetic field with a charged material particle or a material medium possessing an electromagnetic current.
These formulations can be placed directly in the extended setting of the Bargmann spacetimes $B\mathbb{M}$ and $B\mathbb{G}$, without altering the physics, thanks to three facts noted at the end of Sec.~\ref{sec:BargmannSpacetimeFoliation}: 
first, $\hat{\mathbf{d}} = \mathbf{d}$, that is, no dependence of fields on coordinate $x^4 = \eta$ is allowed; 
second, the first four dual basis vectors $(\bm{e}_*^0, \bm{e}_*^i ) = (\underline{\bm{\chi}}, \bm{e}_*^i )$ on $B\mathbb{M}$ or $B\mathbb{G}$ transform under B-Lorentz or homogeneous B-Galilei transformations just as they do under Lorentz or homogeneous Galilei transformations on $\mathbb{M}$ or $\mathbb{G}$, without admixture of the fifth dual basis vector $\bm{e}_*^4 = - \underline{\bm{n}}$;
and third, the fifth basis vector $\bm{e}_4 = -\bm{\xi}$ on $B\mathbb{M}$ or $B\mathbb{G}$ is invariant under B-Lorentz or homogeneous B-Galilei transformations, without admixture of the first four basis vectors $(\bm{e}_0, \bm{e}_i ) = (\bm{n}, \bm{e}_i )$.

The first two facts recalled above from the end of Sec.~\ref{sec:BargmannSpacetimeFoliation} are consequential for the field equations.
Writing them as
\begin{equation}
\begin{aligned}
\hat{\mathbf{d}} \hat{\bm{F}} &= 0, \\
\hat{\mathbf{d}} \hat{\bm{\mathcal{F}}} &= \hat{\bm{\mathcal{J}}}
\end{aligned}
\label{eq:ElectrodynamicsEquationsB}
\end{equation}
on $B\mathbb{M}$ or $B\mathbb{G}$, they have exactly the same content as
\begin{align*}
\mathbf{d} \bm{F} &= 0 \\
\mathbf{d} \bm{\mathcal{F}} &= \bm{\mathcal{J}},
\end{align*}
on $\mathbb{M}$ or $\mathbb{G}$, provided one simply takes
\begin{equation}
\hat{\bm{F}} = \bm{F}, 
	\quad \hat{\bm{\mathcal{F}}} = \bm{\mathcal{F}}, 
	\quad \hat{\bm{\mathcal{J}}} = \bm{\mathcal{J}}
\label{eq:ElectromagneticFormsB}
\end{equation}
for the 2-form $\hat{\bm{F}}$, the 2-form $\hat{\bm{\mathcal{F}}}$, and the 3-form $\bm{\mathcal{J}}$ on $B\mathbb{M}$ or $B\mathbb{G}$.
What were 1+3 decompositions on $\mathbb{M}$ or $\mathbb{G}$,
\begin{align*}
\bm{F} &= -\underline{\bm{\chi}} \wedge \underline{\bm{E}} + \bm{\mathcal{B}}, \\
\bm{\mathcal{F}} &= \phantom{-}\underline{\bm{\chi}} \wedge \underline{\bm{H}} + \bm{\mathcal{D}}, \\
\bm{\mathcal{J}} &= -\underline{\bm{\chi}} \wedge \bm{\mathscr{J}} + \bm{\mathscr{C}}, 
\end{align*}
are now 1+3+1 decompositions $B\mathbb{M}$ or $B\mathbb{G}$---that is, $\underline{\bm{E}}$ and $\bm{\mathcal{B}}$, $\underline{\bm{H}}$ and $\bm{\mathcal{D}}$, and $\bm{\mathscr{J}}$ and $\bm{\mathscr{C}}$ are all tangent to the 3-space slices $\mathbb{S}_{(t,\eta)}$ (recall also that $\bm{\mathcal{B}} = \bm{B} \cdot \underaccent{\check}{\bm{\varepsilon}}$, and $\bm{\mathcal{D}} = \bm{D} \cdot \underaccent{\check}{\bm{\varepsilon}}$, and $\bm{\mathscr{J}} = \bm{j} \cdot \underaccent{\check}{\bm{\varepsilon}}$, and $\bm{\mathscr{C}} = \rho \, \underaccent{\check}{\bm{\varepsilon}}$).
That $\hat{\mathbf{d}} = \mathbf{d}$ was the first fact recalled above from Sec.~\ref{sec:BargmannSpacetimeFoliation}.
The invariance of the field equations of Eq.~(\ref{eq:ElectrodynamicsEquationsB}) follows from the second fact recalled above from Sec.~\ref{sec:BargmannSpacetimeFoliation}. 
This implies that if covariant (that is, type $(0,p)$) tensors---here $\hat{\bm{F}}$, $\hat{\bm{\mathcal{F}}}$, and $\hat{\bm{\mathcal{J}}}$---have vanishing `action' (fifth-dimension) components with respect to one B-Minkowski or B-Galilei basis, it is so with respect to all such bases.
Moreover the time/space components transform just as they do on $\mathbb{M}$ or $\mathbb{G}$.

The third fact recalled above from Sec.~\ref{sec:BargmannSpacetimeFoliation}---the invariance of the action vector $\bm{\xi}$---is consequential for the closure relations connecting $\hat{\bm{F}}$ and $\hat{\bm{\mathcal{F}}}$, and for the electromagnetic force law.

On the one hand, consider the 3-form $\hat{\bm{\star}} \hat{\bm{F}}$, the Hodge dual of the 2-form $\hat{\bm{F}}$ on $B\mathbb{M}$ or $B\mathbb{G}$:
\begin{equation}
\hat{\bm{\star}} \hat{\bm{F}} = 
	\left\{
	\begin{aligned}
		&- \underline{\bm{\chi}} \wedge \bm{\mathcal{E}} 
			- \underline{\bm{\chi}} \wedge \underline{\bm{B}} \wedge \underline{\bm{n}} 
			- \frac{1}{c^2} \, \bm{\mathcal{E}} \wedge \underline{\bm{n}} 
			 & (\text{on } B\mathbb{M}) \\[5pt]
		&- \underline{\bm{\chi}} \wedge \bm{\mathcal{E}} 
			- \underline{\bm{\chi}} \wedge \underline{\bm{B}} \wedge \underline{\bm{n}} 
			 & (\text{on } B\mathbb{G}), \\[5pt]
	\end{aligned}
	\right.
\label{eq:DualForceFormB}
\end{equation}
where $\bm{\mathcal{E}} = \bm{E} \cdot \underaccent{\check}{\bm{\varepsilon}}$.
(The difference between the results on $B\mathbb{M}$ and $B\mathbb{G}$ arises from the index raising of $\underline{\bm{\chi}}$ to $\bm{\chi}$ in taking the Hodge dual: $\underline{\bm{\chi}}$ is the same on $B\mathbb{M}$ and $B\mathbb{G}$, but $\bm{\chi}$ differs according to Eq.~(\ref{eq:DualObserverVectorB}).)
An immediate consequence is that contraction $\hat{\bm{\star}} \hat{\bm{F}} ( \, . \, , \, . \, , \bm{\xi} ) = \hat{\bm{\star}} \hat{\bm{F}} \cdot \bm{\xi}$ yields the 2-form
\[
\hat{\bm{\star}} \hat{\bm{F}} \cdot \bm{\xi} = 
	\left\{
	\begin{aligned}
		&- \underline{\bm{\chi}} \wedge \underline{\bm{B}} 
			- \frac{1}{c^2} \, \bm{\mathcal{E}} 
			 & (\text{on } B\mathbb{M}) \\[5pt]
		&- \underline{\bm{\chi}} \wedge \underline{\bm{B}} 
			 & (\text{on } B\mathbb{G}). \\[5pt]
	\end{aligned}
	\right.
\]
This is amenable to the closure relation
\[
\mu_0 \, \hat{\bm{\mathcal{F}}} = - \, \hat{\bm{\star}}{\hat{\bm{F}}} \cdot \bm{\xi},
\]
which is invariant because $\bm{\xi}$ is invariant.
On $B\mathbb{M}$ this closure relation is precisely that of Eq.~(\ref{eq:ClosureM}) on $\mathbb{M}$, and the field equations of Eq.~(\ref{eq:ElectrodynamicsEquationsB}) give the full Maxwell equations.
But on $B\mathbb{G}$ this closure relation is precisely that of Eq.~(\ref{eq:ClosureMagneticG}) on $\mathbb{G}$, with the electric field disappearing from the electromagnetic source tensor $\hat{\bm{\mathcal{F}}}$, and the field equations of Eq.~(\ref{eq:ElectrodynamicsEquationsB}) giving the truncated Maxwell equations of the Galilei magnetic limit, including the vanishing charge density constraint $\rho = 0$.

On the other hand, consider the 3-form $\hat{\bm{\star}} \hat{\bm{\mathcal{F}}}$, the Hodge dual of the 2-form $\hat{\bm{\mathcal{F}}}$ on $B\mathbb{M}$ or $B\mathbb{G}$:
\begin{equation}
\hat{\bm{\star}} \hat{\bm{\mathcal{F}}} = 
	\left\{
	\begin{aligned}
		& \underline{\bm{\chi}} \wedge \bm{\mathcal{H}} 
			- \underline{\bm{\chi}} \wedge \underline{\bm{D}} \wedge \underline{\bm{n}} 
			+ \frac{1}{c^2} \, \bm{\mathcal{H}} \wedge \underline{\bm{n}} 
			 & (\text{on } B\mathbb{M}) \\[5pt]
		& \underline{\bm{\chi}} \wedge \bm{\mathcal{H}} 
			- \underline{\bm{\chi}} \wedge \underline{\bm{D}} \wedge \underline{\bm{n}} 
			 & (\text{on } B\mathbb{G}), \\[5pt]
	\end{aligned}
	\right.
\label{eq:DualSourceFormB}
\end{equation}
where $\bm{\mathcal{H}} = \bm{H} \cdot \underaccent{\check}{\bm{\varepsilon}}$.
An immediate consequence is that contraction $\hat{\bm{\star}} \hat{\bm{\mathcal{F}}} ( \, . \, , \, . \, , \bm{\xi} ) = \hat{\bm{\star}} \hat{\bm{\mathcal{F}}} \cdot \bm{\xi}$ yields the 2-form
\[
\hat{\bm{\star}} \hat{\bm{\mathcal{F}}} \cdot \bm{\xi} = 
	\left\{
	\begin{aligned}
		& - \underline{\bm{\chi}} \wedge \underline{\bm{D}} 
			+ \frac{1}{c^2} \, \bm{\mathcal{H}} 
			 & (\text{on } B\mathbb{M}) \\[5pt]
		& - \underline{\bm{\chi}} \wedge \underline{\bm{D}} 
			 & (\text{on } B\mathbb{G}). \\[5pt]
	\end{aligned}
	\right.
\]
This is amenable to the closure relation
\[
\epsilon_0 \, \hat{\bm{F}} = \hat{\bm{\star}}{\hat{\bm{\mathcal{F}}}} \cdot \bm{\xi},
\]
which is invariant because $\bm{\xi}$ is invariant.
On $B\mathbb{M}$ this closure relation is precisely that of Eq.~(\ref{eq:ClosureInverseM}) on $\mathbb{M}$, and once again the field equations of Eq.~(\ref{eq:ElectrodynamicsEquationsB}) give the full Maxwell equations.
But on $B\mathbb{G}$ this closure relation is precisely that of Eq.~(\ref{eq:ClosureElectricG}) on $\mathbb{G}$, with the magnetic field disappearing from the electromagnetic force tensor $\hat{\bm{F}}$, and the field equations of Eq.~(\ref{eq:ElectrodynamicsEquationsB}) giving the truncated Maxwell equations of the Galilei electric limit.

Turn finally to the electromagnetic force law.
On $\mathbb{M}$ or $\mathbb{G}$, the force density on a material medium with an electric current is given by Eq.~(\ref{eq:ForceDensity}).
Consider an extended version of this equation on $B\mathbb{M}$ or $B\mathbb{G}$:
\begin{equation}
n \hat{\bm{\Upsilon}} = \hat{\bm{F}}( \, . \, , \hat{\bm{J}}) = \hat{\bm{F}} \cdot \hat{\bm{J}}.
\label{eq:ForceDensityB}
\end{equation}
On the left-hand side, Eq.~(\ref{eq:FiveForceB}) gives
\[
\hat{\bm{\Upsilon}} = \bm{\Upsilon} + \frac{\mathrm{d}M}{\mathrm{d}\tau} \, \underline{\bm{n}}
	= \bm{\Upsilon}
\]
because $\mathrm{d}M / \mathrm{d}\tau = 0$ for the electromagnetic force.
This absence of an action component $\hat{\Upsilon}_\eta$ is already guaranteed by the right-hand side of Eq.~(\ref{eq:ForceDensityB}), where as discussed above $\hat{\bm{F}} = \bm{F}$ has vanishing action components in any B-Minkowski or B-Galilei basis.
The properties of the extended electric current vector 
\[
\hat{\bm{J}} = \bm{J} - \hat{J}^\eta \, \bm{\xi}
\]
are consistent with the discussion of $\hat{\bm{U}}$ in Sec.~\ref{sec:BargmannSpacetime}.
In particular, the time and space components, that is, the components of $\bm{J}$, transform as they do on $\mathbb{M}$ and $\mathbb{G}$, without admixture of the action component $\hat{J}^\eta$.
Because $\hat{\bm{J}}$ is a contravariant vector, unlike the covariant tensors in Eq.~(\ref{eq:ElectromagneticFormsB})---to which list, by the way, an electromagnetic 1-form $\hat{\bm{A}} = \bm{A}$ could be added---it is not possible to assert that $\hat{J}^\eta$ vanishes with respect to all B-Minkowski or B-Galilei bases.
However, the component $\hat{J}^\eta$ plays no physical role in the electromagnetic force, because Eq.~(\ref{eq:ForceDensityB}) actually reads
\[
n \hat{\bm{\Upsilon}} = \hat{\bm{F}} \cdot \hat{\bm{J}} 
	= \bm{F} \cdot \hat{\bm{J}}
	= \bm{F} \cdot \bm{J}
	= n \bm{\Upsilon}
\]
thanks to the projective nature of $\hat{\bm{F}} = \bm{F}$.
Moreover using $\bm{J} = \rho \, \bm{n} + \bm{j}$ one can show that
\begin{align*}
\hat{\bm{J}} \cdot \hat{\bm{\varepsilon}}
	&= \hat{\bm{\varepsilon}} \left( \hat{\bm{J}}, \, . \, , \, . \, , \, . \, , \, . \, \right) \\
	&= \underline{\bm{\chi}} \wedge \hat{J}^\eta \, \mathrm{\underaccent{\check}{\bm{\varepsilon}}}
		- \bm{\mathcal{J}} \wedge \underline{\bm{n}},
\end{align*}
so that
\begin{align*}
\bm{\mathcal{J}} 
	&= - \hat{\bm{\varepsilon}} \left( \hat{\bm{J}}, \, . \, , \, . \, , \, . \, , \bm{\xi} \right) \\
	&= - \hat{\bm{J}} \cdot  \hat{\bm{\varepsilon}} \cdot \bm{\xi}
\end{align*}
relates the current 5-vector in the electromagnetic force to the current 3-form appearing in the field equations in an invariant manner, thanks to the invariance of $\bm{\xi}$.

\section{Conclusion}
\label{sec:Conclusion}

This work begins by suggesting a semantic shift in the way physicists use the terms `special relativity' and `general relativity'. 
The suggestion is that these terms be used to refer to physics on affine (flat) spacetimes on the one hand, or spacetimes with curvature on the other, regardless of whether the physics is governed by the Poincar\'e group or by the Galilei group.
In this perspective these spacetime symmetry groups apply globally in `special relativity' but only locally in `general relativity'. 
This semantic shift leads to a conceptual shift to a more unified perspective on Poincar\'e and Galilei physics.
This paper focuses on special relativity---Poincar\'e physics and Galilei physics on affine spacetimes---and the sequel will address general relativity.

The 4-dimensional affine spacetimes governed by the Poincar\'e and Galilei groups respectively---Minkowski spacetime $\mathbb{M}$, and Galilei-Newton spacetime $\mathbb{G}$---have important differences and similarities.
Causality is governed by the null cone on $\mathbb{M}$, embodied by the spacetime metric $\bm{g}$, whose inverse is $\overleftrightarrow{\bm{g}}$.
A fulness of tensor algebra and tensor calculus is available on $\mathbb{M}$, including metric duality (raising and lowering of tensor indices), a Levi-Civita connection and Levi-Civita volume form, and Hodge duality.
This technology is more limited on $\mathbb{G}$ due to the lack of a non-degenerate spacetime metric.
The asymptotic behavior of $\bm{g}$ as $c \rightarrow \infty$ leads to a 1-form $\bm{\tau}$ embodying absolute time on $\mathbb{G}$.
The limit of $\overleftrightarrow{\bm{g}}$ as $c \rightarrow \infty$ is a $(2,0)$ tensor $\overleftrightarrow{\bm{\gamma}}$ tangent to the leaves $\mathbb{S}_t$ (position space 3-slices) of the given foliation implied by the absolute time 1-form $\bm{\tau}$.
Regarded as a tensor on spacetime $\mathbb{G}$, this tangency to $\mathbb{S}_t$ renders $\overleftrightarrow{\bm{\gamma}}$ degenerate, in that $\bm{\tau} \cdot \overleftrightarrow{\bm{\gamma}} = 0$ in contrast to  $\bm{g} \cdot \overleftrightarrow{\bm{g}} = 1$ on $\mathbb{M}$. 
As for the characteristic groups of Poincar\'e and Galilei physics, as $c \rightarrow \infty$ the Lorentz transformations that preserve preferred representations of the fundamental structures $\bm{g}$ and $\overleftrightarrow{\bm{g}}$ on $\mathbb{M}$ limit smoothly to the homogeneous Galilei transformations that preserve preferred representations of the corresponding fundamental structures $\bm{\tau}$ and $\overleftrightarrow{\bm{\gamma}}$ on $\mathbb{G}$. 
While transformations of 1+3 (time/space) foliations of spacetime according to different inertial observers are geometrically different for $\mathbb{M}$ and $\mathbb{G}$---pseudo-rotations of time axis and 3-space slices on $\mathbb{M}$, vs. a `beveling' of absolute 3-space slices according to a tilted time axis on $\mathbb{G}$---for a single inertial observer the splitting of spacetime into space and time is formally similar.
Associated tensor decompositions into time/space pieces relatable to human observation and measurement are crucial; the projection operator $\overleftarrow{\bm{\gamma}}$ to 3-slices $\mathbb{S}_t$, fiducial observer 4-vector $\bm{n}$, and dual observer covector $\underline{\bm{\chi}}$ (for $\mathbb{M}$) or absolute time form $\bm{\tau}$ (for $\mathbb{G}$) are indispensable tools for effecting such decompositions.

Classical physics on $\mathbb{M}$ and $\mathbb{G}$ begins with consideration of a material particle. 
Kinematics---a description of where a particle is (its position along a worldline) and how fast it is moving (via the 4-velocity $\bm{U}$, tangent to the worldline)---is unproblematic on both $\mathbb{M}$ and $\mathbb{G}$.
But a unified perspective on dynamics on $\mathbb{M}$ and $\mathbb{G}$---a prescription of what determines the shape of the worldline---is more problematic because of the absence of a spacetime metric on $\mathbb{G}$.
The Poincar\'e and Galilei groups naturally address the the transformation of inertia and 3-momentum, combined in the inertia-momentum 4-vector.
On $\mathbb{M}$ this also includes energy thanks to the equivalence of inertia and total energy modulo the factor $c^2$;
but the geometry of $\mathbb{G}$ enforces the invariance of inertia and its strict separation from kinetic energy, precluding a Galilei tensor formalism on 4-dimensional spacetime that explicitly exhibits the transformation of energy.

More on this shortly; but first, no treatment of `special relativity' (including as redefined here) would be adequate without a discussion of electrodynamics.
This paper presents a fresh---indeed, taken as a whole, apparently novel---spacetime approach to this subject well suited to a unified perspective on $\mathbb{M}$ and $\mathbb{G}$: due to the absence of a spacetime metric on $\mathbb{G}$, the fundamental equations are given only in terms of the spacetime exterior derivative operator $\mathbf{d}$, acting on a 2-form $\bm{F}$ (the `electric force tensor' encoding the electric field strength $\bm{E}$ and magnetic flux density $\bm{B}$) and a separate 2-form $\bm{\mathcal{F}}$ (the `electric source tensor' encoding the electric displacement field $\bm{D}$ and magnetic field strength $\bm{H}$).
The manifestly Poincar\'e- or Galilei-invariant theories in vacuum then follow from closing the system with `constitutive relations' given by spacetime tensor relations between $\bm{F}$ and $\bm{\mathcal{F}}$.
On $\mathbb{M}$, the natural home of full electrodynamics, this is an invertible relationship of Hodge duality, consistent with the Poincar\'e invariance.
On $\mathbb{G}$ the degeneracy of the inverse `metric' $\overleftrightarrow{\bm{\gamma}}$ now rears its head: there is no Hodge star operator, but instead a non-invertible `slash-star' operator whose use in closure relations results in partly truncated versions of electrodynamics.
Depending on whether one takes $\bm{F}$ or $\bm{\mathcal{F}}$ as fundamental, one obtains either the `magnetic limit' or the `electric limit' originally found and discussed by LeBellac and L\'evy-Leblond \cite{Le-Bellac1973Galilean-electr} without the benefit of a spacetime perspective.
These authors say that the electric field (in the magnetic limit) or the magnetic field (in the electric limit) exists but has no physical effect in these respective limits, but the presentation here shows that this conclusion is too hasty: while it is true that these `opposite' fields disappear from the Lorentz force and the electromagnetic energy density, consideration of the Poynting theorem in these limits shows that electric field (in the magnetic limit) or magnetic field (in the electric limit) still plays a role in the electromagnetic transport of energy. 

Returning to the question of explicit accommodation of the transformation of kinetic energy in a tensor formalism, this can be accomplished for both Poincar\'e and Galilei physics by moving to a 5-dimensional setting, leading to the extended spacetimes $B\mathbb{M}$ and $B\mathbb{G}$.
The fifth coordinate, $\eta$ (not to be confused with preferred matrix representations $\sfeta$ of metric tensors), has units of action/mass and is called the `action coordinate' for short.
The `action coordinate relation' of Eq.~(\ref{eq:FifthVelocityComponent}) is crucial to the geometry of $B\mathbb{M}$ and $B\mathbb{G}$, for in both cases it leads to a non-degenerate metric tensor labeled $\bm{G}$, with preferred matrix representations $\mathsf{G}$ given by Eq.~(\ref{eq:BMinkowskiBGalileiMatrices}).
Unlike the relationship between 4-dimensional $\mathbb{M}$ and $\mathbb{G}$, in this case the expressions for both $\mathsf{G}$ and its inverse $\overleftrightarrow{\mathsf{G}}$ on $B\mathbb{M}$ limit smoothly to the corresponding expressions on $B\mathbb{G}$.
And unlike 4-dimensional $\mathbb{G}$, 5-dimensional $B\mathbb{G}$ is a pseudo-Riemann manifold, making available the corresponding full technology of tensor algebra and tensor calculus. 
This allows an even more deeply unified perspective on Poincar\'e and Galilei physics, via their more parallel treatment on the extended spacetimes $B\mathbb{M}$ and $B\mathbb{G}$.
Similar to the 1+3 (time/space) splitting of 4-dimensional spacetimes and tensors thereon, a 1+3+1 (time/space/action) splitting of $B\mathbb{M}$ and $B\mathbb{G}$ and associated tensor decompositions into time/space/action pieces relatable to human observation and measurement are crucial; the projection operator $\overleftarrow{\bm{\gamma}}$ to 3-slices $\mathbb{S}_{(t,\eta)}$, fiducial observer 5-vector $\bm{n}$, action 5-vector $\bm{\xi}$, and 5-covectors $\underline{\bm{\chi}}$ and $\underline{\bm{n}}$ are indispensable tools for effecting such decompositions.

The B-Lorentz and homogeneous B-Galilei groups, which act on the vector spaces underlying the extended spacetimes $B\mathbb{M}$ and $B\mathbb{G}$ respectively, have some notable properties.\footnote{The Lie theory of these groups, including their Lie algebra cohomology and the status of the B-Poincar\'e and B-Galilei groups as central extensions of the Poincar\'e and Galilei groups, will be left for separate exposition.}
These transformations are represented by the matrices given by Eqs.~(\ref{eq:BargmannTransformation}) and (\ref{eq:BargmannFourRow}) with respect to a B-Minkowski or B-Galilei basis.
They respectively preserve the versions of the metric $\bm{G}$ on $B\mathbb{M}$ and $B\mathbb{G}$.
The signature of $\bm{G}$ is $-++++$ in both cases;
thus the B-Lorentz and homogeneous B-Galilei groups can be understood as subgroups of $\mathrm{SO}(1,4)$ (which itself is a subgroup of $\mathrm{GL}(5)$) that satisfy additional properties.
One additional property is that the first four B-Minkowski or B-Galilei dual basis covectors $\left( \bm{e}_*^\mu \right) = \left( \underline{\bm{\chi}}, \bm{e}_*^i \right)$ transform just as they do under Lorentz or homogeneous Galilei transformations, without admixture of the last B-Minkowski or B-Galilei dual basis covector $\bm{e}_*^4 = -\underline{\bm{n}}$.
Another property is that the last B-Minkowski or B-Galilei basis vector, $\bm{e}_4 = -\bm{\xi}$, remains invariant under these transformations.

As a consequence of these properties, the Lorentz- and homogeneous Galilei-invariant physics on $\mathbb{M}$ and $\mathbb{G}$ discussed in this paper translate into manifestly B-Lorentz- and B-Galilei-invariant physics on $B\mathbb{M}$ and $B\mathbb{G}$.
In the case of material particles, the metric $\bm{g}$ on $\mathbb{M}$ and the time covector $\bm{\tau}$ on $\mathbb{G}$ governing causality are both covariant tensors (that is, of type $(0,p))$ that play the same role on $B\mathbb{M}$ and $B\mathbb{G}$, and are uncontaminated by $\bm{e}_*^4 = -\underline{\bm{n}}$ under B-Lorentz or B-Galilei transformations.
The same is true of the 2-forms $\bm{F}$ and $\bm{\mathcal{F}}$ in the formulation of electrodynamics on $\mathbb{M}$ and $\mathbb{G}$ presented here, which, together with the Hodge star operator now available on both $B\mathbb{M}$ and $B\mathbb{G}$ and the invariance of $\bm{e}_4 = -\bm{\xi}$ used in closure relations, provide for a straightforward invariant translation of Poincar\'e and Galilei electrodynamics into the 5-dimensional setting.

In the case of a material particle no fundamentally new physics emerges in the 5-dimensional setting of $B\mathbb{M}$ and $B\mathbb{G}$ relative to the 4-dimensional setting of $\mathbb{M}$ and $\mathbb{G}$, but things are rearranged in such a way that Poincar\'e and Galilei versions can be treated in parallel.
In a sense, Poincar\'e physics gives up a bit for the benefit of Galilei physics: the Poincar\'e union of mass and kinetic energy is less apparent, but explicitly separating kinetic energy allows Galilei physics to also handle energy in a tensor formalism.  
In the inertia--momentum--kinetic-energy 5-vector $\hat{\bm{I}} = M \hat{\bm{U}}$, with rest mass $M$ and 5-velocity $\hat{\bm{U}}$ tangent to the worldline in extended spacetime, the usual inertia-momentum 4-vector is extended to include a fifth component, the kinetic energy; and as on $\mathbb{M}$ and $\mathbb{G}$, the first component is inertia, the dynamic $M \Lambda_{\bm{v}}$ in the case of $B\mathbb{M}$ and the invariant $M$ in the case of $B\mathbb{G}$.
In the kinetic-energy--momentum--mass 5-covector $\hat{\bm{\Pi}} = \bm{G} \cdot \hat{\bm{I}}$ obtained by metric duality, the first component is the (negative of) kinetic energy, and the last component is invariant, the (negative of) rest mass $M$ on both $B\mathbb{M}$ and $B\mathbb{G}$. 
Newton's second law is most naturally handled in its covector version and the work-energy theorem is directly present in the tensor formalism.

As for electrodynamics, no fundamentally different physics arises in the 5-dimensional setting of $B\mathbb{M}$ and $B\mathbb{G}$ either, at least in the straightforward translation to the 5-dimensional setting presented here.
In this paper the two different Galilei-invariant theories on $\mathbb{G}$ (the so-called magnetic and electric limits) arise because of the non-invertible nature of the `slash-star' operator (the Galilei-invariant counterpart of the invertible Hodge star operator on $\mathbb{M}$).
One might have wondered whether the availability of a true invertible Hodge star operator on $B\mathbb{G}$ changes things, but this turns out not to be the case.
The reason has to do with the way the inverse metric, which appears in the Hodge star and slash-star operators, manifests in the 5-dimensional setting.
The metric $\bm{g}$ on $\mathbb{M}$ is necessarily `scrambled' in going over to the metric $\bm{G}$ on $B\mathbb{M}$; if this were not so, $\bm{G}$ on $B\mathbb{M}$ could not limit sensibly to a metric $\bm{G}$ on $B\mathbb{G}$.
In contrast, the inverse metrics $\overleftrightarrow{\bm{g}}$ on $\mathbb{M}$ and $\overleftrightarrow{\bm{\gamma}}$ on $\mathbb{G}$ are directly extended into the inverse metric $\overleftrightarrow{\bm{G}}$ on $B\mathbb{M}$ and $B\mathbb{G}$; see the upper-left $4 \times 4$ blocks in Eq.~(\ref{eq:BMinkowskiBGalileiInverseMatrices}).
Even though the metric is now invertible and a true Hodge star operator exists on $B\mathbb{G}$, the vanishing time-time component $-1/c^2 \rightarrow 0$ in these expressions ends up projecting out the electric field (in the magnetic limit) or the magnetic field (in the electric limit) when taking the Hodge dual.
 
The question to be addressed in the sequel to this paper is what may exist in terms of a more complete `Galilei general relativity' in the 5-dimensional setting.
The Newton-Cartan spacetime $\mathcal{N}$ in Fig.~\ref{fig:SpacetimeRoadmap} has flat position space 3-slices, with the presence of space\textit{time} curvature encoding the Newton gravitational potential.
In this theory the mass density is the only source of spacetime curvature, with bulk kinetic energy density, internal energy density, and stresses apparently disappearing as sources in comparison with mass density due to their `inertia' being given by multiplication by $1/c^2$.
Moreover, in studying 4-dimensional curved spacetime with local Galilei symmetry related to the usual Einstein spacetime $\mathcal{E}$ with Poincar\'e symmetry (traditionally known as `general relativity') it is common to reduce the number of degrees of freedom with additional constraints that result in flat position space 3-slices (e.g. \cite{Dixon1975On-the-uniquene,Dautcourt1997Post-Newtonian-,Andringa2011Newtonian-gravi}). And at least some translations of $\mathcal{N}$ into the 5-dimensional setting preserve this spatial flatness (e.g. \cite{Duval1985Bargmann-struct,de-Saxce2016Galilean-Mechan,de-Saxce20175-Dimensional-T}); this is labeled $B\mathcal{N}$ in Fig.~\ref{fig:SpacetimeRoadmap}.

But the possibility of spatial curvature---and indeed strong spacetime curvature---consistent with Galilei physics may be open and more accessible in a 5-dimensional setting.
Consider in particular a formulation---apparently not suggested before, apart from \cite{Cardall2023Towards-full-Ga}---motivated by the `action coordinate relation' of Eq.~(\ref{eq:FifthVelocityComponent}) key to the present exposition.
In the 1+3 formulation (traditionally called `3+1', e.g. \cite{Gourgoulhon201231-Formalism-in}) of Einstein spacetime $\mathcal{E}$ (traditionally known as `general relativity') in terms of the lapse function $\alpha$, shift 3-vector $\bm{\beta}$, and 3-metric $\bm{\gamma}$, proper time intervals are given by 
\[
c^2 \, \mathrm{d}\tau^2 = c^2 \alpha^2 \, \mathrm{d}t^2 - \bm{\gamma} \left( \mathrm{d} \bm{x} + \bm{\beta} \, \mathrm{d}t, \mathrm{d} \bm{x} + \bm{\beta} \, \mathrm{d}t \right)
\quad (\mathrm{on \ } \mathcal{E})
\]
and the Lorentz factor of a material particle is $\Lambda = \alpha \, \mathrm{d}t / \mathrm{d}\tau$.
Use of these expressions in the action coordinate relation of Eq.~(\ref{eq:FifthVelocityComponent}) yields
\[
0 = \beta_a \beta^a \, \mathrm{d}t^2 
		- 2 \, \mathrm{d}t \, \beta_a \mathrm{d}x^a  
		+ \mathrm{d} x^a \, \gamma_{a b} \, \mathrm{d} x^b
		 - 2 \, \alpha \, 	\mathrm{d} \eta \, \mathrm{d} t + \frac{1}{c^2} \, \mathrm{d} \eta^2
	\quad (\mathrm{on \ } B\mathcal{E}).
\]
This is suggestive of a 5D Bargmann-Einstein spacetime $B\mathcal{E}$ and its $c\rightarrow \infty$ limit $B\mathcal{G}$ compatible with Galilei physics, with metric $\bm{G}$ represented by
\[
\mathsf{G} =
	\begin{cases}
		 \begin{bmatrix} 
		 	\beta_a \beta^a && \beta_j && -\alpha \\[5pt]
			 \beta_i && \gamma_{ij} && 0_i \\[5pt]
			  -\alpha && 0_j && \frac{1}{c^2}
		\end{bmatrix} 
			& (\mathrm{on \ } B\mathcal{E}) \\[20pt]
		 \begin{bmatrix} \beta_a \beta^a && \beta_j && -\alpha \\[5pt]
			 \beta_i && \gamma_{ij} && 0_i \\[5pt]
			  -\alpha && 0_j && 0
		\end{bmatrix} 
			& (\mathrm{on \ } B\mathcal{G})
	\end{cases}
\]
and inverse metric $\overleftrightarrow{\bm{G}}$ represented by
\[
\overleftrightarrow{\mathsf{G}} =
	\begin{cases}
		\begin{bmatrix} 
			-\frac{1}{c^2  \alpha^2} && \frac{1}{c^2  \alpha^2} \, \beta^j  
				&& -\frac{1}{\alpha} \\[5pt]
			\frac{1}{c^2  \alpha^2} \, \beta^i 
				&& \gamma^{ij} -  \frac{1}{c^2  \alpha^2} \, \beta^i \beta^j
				&& \frac{1}{\alpha} \, \beta^i \\[5pt]
			-\frac{1}{\alpha} && \frac{1}{\alpha} \, \beta^j  && 0 
		\end{bmatrix} 
			& (\mathrm{on \ } B\mathcal{E}) \\[20pt]
		\begin{bmatrix} 
			0 && 0^j && -\frac{1}{\alpha} \\[5pt]
			0^i && \gamma^{ij} && \frac{1}{\alpha} \, \beta^i \\[5pt]
			-\frac{1}{\alpha} && \frac{1}{\alpha} \, \beta^j  && 0 
		\end{bmatrix} 
			& (\mathrm{on \ } B\mathcal{G}).
	\end{cases}
\]
(These reduce to Eqs.~(\ref{eq:BMinkowskiBGalileiMatrices}) and (\ref{eq:BMinkowskiBGalileiInverseMatrices}) on affine spacetimes $B\mathbb{M}$ and $B\mathbb{G}$ as $\alpha \rightarrow 1$ and $\bm{\beta} \rightarrow 0$ and $\gamma_{ij} \rightarrow 1_{ij}$.)
Thus there is a reasonable prospect that recasting the $1+3$ (time/space) formulation of the Einstein equations on $\mathcal{E}$ as a $1+3+1$ (time/space/action) formulation on $B\mathcal{E}$ and taking the $c \rightarrow \infty$ limit could yield a Galilei gravitation of enhanced strength on spacetime $B\mathcal{G}$ in which energy density and stress contribute as sources and give rise to position space 3-slice curvature as well as spacetime curvature, beyond the flat position space 3-slices and spacetime curvature determined by mass density alone on $\mathcal{N}$ and $B\mathcal{N}$.
This is the reason for suspecting that there exists a $B\mathcal{G}$ distinct from $B\mathcal{N}$ if Fig.~\ref{fig:SpacetimeRoadmap}.
This suspicion is heightened by the fact that the metric given above for $B\mathcal{G}$ seems incommensurate with that on $B\mathcal{N}$ given by de Saxc\'e \cite{de-Saxce2016Galilean-Mechan,de-Saxce20175-Dimensional-T}: the latter contains $-1$ in the off-diagonal time-action components rather than $-\alpha$, instead locating the Newton gravitational potential in the time-time component, and exhibits manifestly flat position space 3-slices.

That a large-$c$ limit is not necessarily a weak-field (small curvature) limit has become apparent in recent work allowing for torsion on 4-dimensional spacetime (e.g. \cite{Geracie2015Curved-non-rela,Van-den-Bleeken2017Torsional-Newto,Hansen2019Action-Principl,Cariglia2018General-theory-,Hansen2020Non-relativisti}).
In Newton-Cartan spacetime $\mathcal{N}$, flat position space 3-slices go hand-in-hand with vanishing torsion through the `absolute time' condition that the time form be closed ($\mathbf{d} \bm{\tau} = 0$), but the generalizations in the above-cited works consider a weaker `twistless torsion' condition ($\bm{\tau} \wedge \mathbf{d} \bm{\tau} = 0$) requiring only a foliation of spacetime according to a global time coordinate, with proper time between leaves of the foliation governed by a lapse function as in the usual Poincar\'e-Einstein case.
These works typically glance at or even make partial use of a 5-dimensional setting, but ultimately boil down to consideration of curved 4-dimensional spacetimes consistent with Galilei physics. 
The approach advocated in this paper is somewhat different: the idea is to express standard Poincar\'e physics also in a 5-dimensional setting where Galilei physics can most naturally breathe, and use it as a guide to obtaining Galilei physics without ever subjecting the latter to a `reduction' to a 4-dimensional setting.
Nevertheless the exhibition of both weak-field and strong-field versions of Galilei-compatible $c \rightarrow \infty$ Schwarzschild geometry presented by Van den Bleeken \cite{Van-den-Bleeken2017Torsional-Newto} is particularly striking, and may correspond to the fundamental distinction between $B\mathcal{N}$ and $B\mathcal{G}$ conjectured in the previous paragraph.
Indeed if the strong-field 5-dimensional $B\mathcal{G}$ distinct from the weak-field $B\mathcal{N}$ materializes as conjectured above, exploration of a possible relationship between such a formulation and the recently discovered twistless-torsional generalizations of 4-dimensional Newton-Cartan spacetime will be of keen interest.

Strong-field gravitation consistent with Galilei physics would be a useful---and conceptually and mathematically sound---approximation in astrophysical scenarios such as core-collapse supernovae, in which the energy density and pressure of the nascent neutron star contribute to enhanced gravity at the 10-20\% level, but for which the computationally/numerically fraught phenomena of `Minkowski' bulk fluid flow and back-reaction of gravitational radiation are much less significant.
The most commonly used procedure \cite{Marek-A.-and-Dimmelmeier-H.-and-Janka-H.-T.-and-Muller-E.-and-Buras-R.2006Exploring-the-r} for approximating strong gravity in core-collapse supernova simulations at present---keeping higher-order Newton multipole moments while swapping the Newton monopole for a Poincar\'e-Einstein (traditionally, `general relativistic') one---is physically motivated but uncontrolled mathematically, precluding any handle on global conservation properties.

\vspace{6pt} 

\funding{This work was supported by the U.S. Department of Energy, Office of Science, Office of Nuclear Physics under contract number DE-AC05-00OR22725.}

\acknowledgments{Thanks are due to G\'ery de Saxc\'e for pointing out that preservation of the B-metric $\bm{G}$ is not sufficient to prove closure of the B-Lorentz transformations $\hat{\mathsf{P}}_{B\mathbb{M}}^+$, but that closure directly follows instead from relations obtained from closure of the Lorentz group.
A review with helpful comments by Vassilios Mewes is also appreciated.}

\conflictsofinterest{The author declares no conflict of interest. The funders had no role in the design of the study; in the collection, analyses, or interpretation of data; in the writing of the manuscript, or in the decision to publish the results.}



%

\appendixtitles{yes} 
\appendixstart
\appendix
%
%

\section{Affine spaces and linear tensors}

Begin by establishing a unified conceptual framework for the flat 4-dimensional Minkowski and Galilei-Newton spacetimes and their 5-dimensional Bargmann extensions treated in this work.
This Appendix includes descriptions of a generic affine space and of linear tensors on the vector space underlying an affine space, along with a discussion of treating an affine space as a differentiable manifold with connection even in the absence of a metric.
To maintain the flavor of coordinate-free formulations while referring to specific bases, a matrix formalism is introduced to reduce, where feasible, the index clutter associated with tensor component expressions.
Books by Gourgoulhon \cite{Gourgoulhon2013Special-Relativ,Gourgoulhon201231-Formalism-in}, by de Saxc\'e \cite{de-Saxce2016Galilean-Mechan}, and by Frankel \cite{Frankel2012The-Geometry-of}, written for the perspective of physicists, may be helpful for understanding the geometric approach, mathematical tools, and (to some extent) notation employed here.
 
\subsection{Affine spaces}
\label{sec:AffineSpaces}

Informally, a real affine space $\mathbb{A}$ of dimension $n$ is essentially a real vector space $\mathrm{V}$ of dimension $n$ for which one `forgets' the origin (zero vector), so that translations become symmetries of the space.
Indeed vectors in a vector space $\mathrm{V}$ underlying an affine space $\mathbb{A}$ act on points of $\mathbb{A}$ according to an `addition' mapping
\begin{equation}
\begin{array}{rclcl} 
 + & : & \mathbb{A} \times \mathrm{V} & \rightarrow & \mathbb{A} \\
   &   & (\mathbf{x}, \overrightarrow{\mathbf{x y}}) & \mapsto & \mathbf{y} = \mathbf{x} + \overrightarrow{\mathbf{x y}}.
\end{array}
\label{eq:AffinePoints}
\end{equation}
For points $\mathbf{x}, \mathbf{y} \in \mathbb{A}$, this says that $\overrightarrow{\mathbf{x y}} \in \mathrm{V}$ is the unique vector that `points from $\mathbf{x}$ to $\mathbf{y}$' or `translates $\mathbf{x}$ to $\mathbf{y}$', with $\overrightarrow{\mathbf{x y}} = \bm{0}$ (the zero vector) if $\mathbf{y} = \mathbf{x}$.
This action of $\mathrm{V}$ on $\mathbb{A}$ is taken to be compatible with the vector addition of $\mathrm{V}$, as follows. 
Consider a third point $\mathbf{z} \in \mathbb{A}$, which can be written in two ways:
\[
\mathbf{z} = \mathbf{x} + \overrightarrow{\mathbf{x z}},
\]
or
\begin{align*}
\mathbf{z} &= \mathbf{y} + \overrightarrow{\mathbf{y z}} \\
                 &= ( \mathbf{x} + \overrightarrow{\mathbf{x y}}  ) + \overrightarrow{\mathbf{y z}} \\
                 &= \mathbf{x} + ( \overrightarrow{\mathbf{x y}}  + \overrightarrow{\mathbf{y z}} ). 
\end{align*}
Equating these two different ways of writing $\mathbf{z}$ implies
\[
\overrightarrow{\mathbf{x z}} = \overrightarrow{\mathbf{x y}}  + \overrightarrow{\mathbf{y z}},
\] 
which is nothing but the addition operation of the underlying vector space $\mathrm{V}$.

A symmetry of a space---here an affine space $\mathbb{A}$ or its underlying vector space $\mathrm{V}$, or specializations of these---is an automorphism: a one-to-one and onto mapping of the space to itself, a transformation that leaves it `unchanged', that is, indistinguishable from its previous state, while preserving any (possibly additional) mathematical structure with which it is endowed.
The set of symmetry transformations of a space forms a group: 
the identity transformation, in which each element of the space is mapped to itself, is an obvious symmetry;
if one transformation leaves the space unchanged, a succession of two of them also leaves the space unchanged; 
and if a transformation leaves the space unchanged, there is an inverse transformation which maps each element of the transformed space back to the one from which it was mapped.

The symmetries of $\mathbb{A}$---translations of $\mathbb{A}$, along with invertible linear transformations of its underlying vector space $\mathrm{V}$---can begin to be described concretely with the selection of a point $\mathbf{O} \in \mathbb{A}$ as origin and a basis $( \bm{e}_1, \dots, \bm{e}_n )$ of $\mathrm{V}$.
Then via Eq.~(\ref{eq:AffinePoints}), the points $\mathbf{x} \in \mathbb{A}$ are put into one-to-one correspondence with the elements of $\mathbb{R}^n$, the set of all ordered $n$-tuples of real numbers, according to
\begin{equation}
\begin{aligned}
\mathbf{x} 
	&= \mathbf{O} + \overrightarrow{\mathbf{O}\mathbf{x}} \\
	&= \mathbf{O} + \bm{e}_1 \, x^1 + \dots + \bm{e}_n \, x^n \\
	&= \mathbf{O} + \bm{e}_a \, x^a,
\end{aligned}
\label{eq:PointExpansion}
\end{equation}
where $(x^i ) = ( x^1, \dots, x^n) \in \mathbb{R}^n$ collects the components of $\overrightarrow{\mathbf{O}\mathbf{x}} \in \mathrm{V}$ with respect to the basis $( \bm{e}_1, \dots, \bm{e}_n )$.
Note the summation convention on repeated dummy index $a$, one a superscript and one a subscript; the fixed integer $n$ denoting the dimension is not a dummy index.

Two perspectives on symmetry transformations of $\mathbb{A}$ are available.
In active transformations the origin of $\mathbb{A}$ and the basis of $\mathrm{V}$ are fixed, while the points of $\mathbb{A}$ are moved according to $\mathbf{x} \mapsto \mathbf{x}' = \mathbf{x} + \bm{C}$, where $\bm{C} \in \mathrm{V}$ is a translation vector, along with $\overrightarrow{\mathbf{O}\mathbf{x}} \mapsto \overrightarrow{\mathbf{O}\mathbf{x}'} = \bm{P}\left( \overrightarrow{\mathbf{O}\mathbf{x}} \right)$, where $\bm{P} \in \mathrm{GL}(\mathrm{V})$ is an invertible linear transformation of $\mathrm{V}$.
Here passive transformations are adopted instead, in which the points of $\mathbb{A}$ are fixed while the origin is translated and the basis of $\mathrm{V}$ is transformed.
In terms of the new origin and basis,
\begin{equation}
\begin{aligned}
\mathbf{x} 
	&= \mathbf{O}' + \overrightarrow{\mathbf{O}'\mathbf{x}'} \\
	&= \mathbf{O} + \bm{C} + \bm{e}'_a x'^a,
\end{aligned} 
\label{eq:PointExpansionNew}
\end{equation}
where $\bm{C} = \overrightarrow{\mathbf{O}\mathbf{O}'} \in \mathrm{V}$ is the translation of the origin, and $( \bm{e}'_1, \dots, \bm{e}'_n ) = \left( \bm{P}(\bm{e}_1), \dots, \bm{P}(\bm{e}_n) \right)$ with $\bm{P} \in \mathrm{GL(V)}$ is the transformed basis.

It is convenient to introduce matrix representations.
Write the original basis of $\mathrm{V}$, the $n$-tuple $( \bm{e}_1, \dots, \bm{e}_n )$, as the $n$-row (that is, $1 \times n$ matrix) 
\[
\bm{\mathsf{B}} = \begin{bmatrix} \bm{e}_1 & \dots & \bm{e}_n \end{bmatrix}.
\]
Let $\mathbb{R}^{n\times 1}$ denote the vector space of $n$-columns (that is, $n \times 1$ matrices) of real numbers, naturally isomorphic to the vector space $\mathbb{R}^n$ of $n$-tuples of real numbers.
Moreover take $\mathsf{x} \in \mathbb{R}^{n\times 1}$ to be
\[
\mathsf{x} = \begin{bmatrix} x^1 \\ \vdots \\ x^n \end{bmatrix}
\]
so that $\overrightarrow{\mathbf{O x}} = \bm{e}_a \, x^a = \bm{\mathsf{B}}\, \mathsf{x}$ is given by matrix multiplication.
Expand the transformed basis elements as 
\begin{equation}
\bm{e}'_j = \bm{P}(\bm{e}_j) = \bm{e}_i \, {P^i}_j, 
\label{eq:TransformedBasisExpansion}
\end{equation}
and collect the expansion coefficients in the matrix $[ {P^i}_j ] = \mathsf{P}  \in \mathrm{GL}(n)$, where $\mathrm{GL}(n)$ is the group of invertible $n \times n$ real matrices.
Then the transformed basis is given by matrix multiplication as
\[
\bm{\mathsf{B}}' = \begin{bmatrix} \bm{e}'_1 & \dots & \bm{e}'_n \end{bmatrix} = \begin{bmatrix} \bm{e}_1 & \dots & \bm{e}_n \end{bmatrix} \mathsf{P} = \bm{\mathsf{B}}\, \mathsf{P},
\]
so that $\overrightarrow{\mathbf{O}'\mathbf{x}'} = \bm{e}'_a \, x'^a = \bm{\mathsf{B}}'\, \mathsf{x}' = \bm{\mathsf{B}}\, \mathsf{P} \, \mathsf{x}'$.
Finally, expand $\bm{C} = \overrightarrow{\mathbf{O}\mathbf{O}'} = \bm{\mathsf{B}}\,\mathsf{C}$ in the original basis.
Then equating Eqs.~(\ref{eq:PointExpansion}) and (\ref{eq:PointExpansionNew}) yields the transformation rule
\begin{equation}
\mathsf{x} = \mathsf{P}\, \mathsf{x}' + \mathsf{C}
\label{eq:AffineTransformation}
\end{equation}
relating the two $n$-column representations of a point $\mathbf{x} \in \mathbb{A}$.

Thus the symmetry group $\mathrm{Aff}(\mathbb{A})$ of an affine space $\mathbb{A}$ of dimension $n$ with underlying vector space $\mathrm{V}$ comprises 
the combined actions of translations of $\mathbb{A}$ and invertible linear transformations of $\mathrm{V}$. (In mathematical terms this `combination' is the semidirect product of $\mathrm{V}$, understood as an abelian group under addition, and $\mathrm{GL}(\mathrm{V})$, a group under composition.)
In the $n$-column representation of points of $\mathbb{A}$, elements of the group $\mathrm{Aff}(n)$ are pairs $(\mathsf{C},\mathsf{P})$ which act according to Eq.~(\ref{eq:AffineTransformation}).
It is readily seen that the group multiplication law is given by $(\mathsf{C}_{2}, \mathsf{P}_{2}) (\mathsf{C}_{1}, \mathsf{P}_{1}) = (\mathsf{C}_{21}, \mathsf{P}_{21}) = (\mathsf{C}_2 + \mathsf{P}_2 \mathsf{C}_1, \mathsf{P}_2 \mathsf{P}_1)$, and that the inverse is given by $(\mathsf{C},\mathsf{P})^{-1} = (-\mathsf{P}^{-1} \mathsf{C}, \mathsf{P}^{-1})$. 
In particular, translations of the origin are effected by elements $\mathsf{C} \in \mathbb{R}^{n\times 1}$ (understood as an abelian group under addition) acting by vector addition on $\mathbb{R}^{n\times 1}$ (understood as a set representing the points of $\mathbb{A}$), while basis changes of $\mathrm{V}$ are effected by the matrices $\mathsf{P} \in \mathrm{GL}(n)$.
For computational convenience, the action of $(\mathsf{C},\mathsf{P}) \in \mathrm{Aff}(n)$ acting on $n$-columns representing points of $\mathbb{A}$ can be represented by the linear transformations
\[
\begin{bmatrix} 1 \\ \mathsf{x} \end{bmatrix} = 
\begin{bmatrix} 1 & \mathsf{0} \\ \mathsf{C} & \mathsf{P} \end{bmatrix} 
\begin{bmatrix} 1 \\ \mathsf{x}' \end{bmatrix}
\]
acting on $(n+1)$-columns.

Of course, unlike points of $\mathbb{A}$---whose $n$-column representations transform according to Eq.~(\ref{eq:AffineTransformation}) under $\mathrm{Aff}(n)$---the $n$-column representations of vectors belonging to $\mathrm{V}$ are transformed by the linear part only.
In particular, for $\bm{v} \in \mathrm{V}$ represented by the $n$-column $\mathsf{v}$ relative to basis $\bm{\mathsf{B}}$, or by $\mathsf{v}'$ relative to basis $\bm{\mathsf{B}}' = \bm{\mathsf{B}}\, \mathsf{P}$,
\[
\bm{v} = \bm{\mathsf{B}}\, \mathsf{v} = \bm{\mathsf{B}}'\, \mathsf{v}' = \bm{\mathsf{B}}\, \mathsf{P} \,\mathsf{v}',
\]
where $\mathsf{P} \in \mathrm{GL}(n)$.
This implies
\begin{equation}
\mathsf{v} = \mathsf{P}\, \mathsf{v}',
\label{eq:VectorColumns}
\end{equation}
in contrast to Eq.~(\ref{eq:AffineTransformation}).

\subsection{Linear tensors}
\label{sec:LinearTensors}

Consider also the dual space $\mathrm{V}_*$ of $\mathrm{V}$.
This is the vector space of covectors, or linear forms on $\mathrm{V}$, such that $\bm{\omega}(\bm{v})$ is a real number for any $\bm{\omega} \in \mathrm{V}_*$ and $\bm{v} \in \mathrm{V}$.
Let $( \bm{e}_1, \dots, \bm{e}_n )$ be a basis of $\mathrm{V}$ and $( \bm{e}_*^1, \dots, \bm{e}_*^n )$ the basis of $\mathrm{V}_*$ dual to $( \bm{e}_1, \dots, \bm{e}_n )$ in the sense that 
\[
\bm{e}_*^i ( \bm{e}_j ) = \delta^i_j,
\] 
where $\delta^i_j$ is the Kronecker delta.
With the expansions $\bm{v} = \bm{e}_b \, v^b$ and $\bm{\omega} = \omega_a \, \bm{e}_*^a$, where $( \omega_1, \dots, \omega_n ) \in \mathbb{R}^n$ collects the components of $\bm{\omega}$,
\[
\bm{\omega}(\bm{v}) 
	= \omega_a \,  \bm{e}_*^a ( \bm{e}_b ) \, v^b
	= \omega_a \, v^a
\]
expresses the value of $\bm{\omega}$ on $\bm{v}$---also known as the canonical pairing of $\bm{\omega}$ and $\bm{v}$---in terms of their components.

Consider simultaneous changes of bases of $\mathrm{V}$ and $\mathrm{V}_*$ such that the dual basis relationship is preserved.
That is, let $( \bm{e}'_1, \dots, \bm{e}'_n ) = \left( \bm{P}( \bm{e}_1 ), \dots, \bm{P}( \bm{e}_n ) \right)$ for some $\bm{P} \in \mathrm{GL}(\mathrm{V})$, and $( \bm{e}_*'^1, \dots, \bm{e}_*'^n ) = \left( \bm{Q}_*( \bm{e}_*^1 ), \dots, \bm{Q}_*( \bm{e}_*^n ) \right)$ for the appropriate $\bm{Q}_* \in \mathrm{GL}(\mathrm{V}_*)$ such that $\bm{e}_*'^i ( \bm{e}'_j ) = \delta^i_j$.
Here the linear transformations $\bm{Q}_* \in \mathrm{GL}(\mathrm{V}_*)$ are the algebraic adjoints of the transformations $\bm{Q} \in \mathrm{GL}(\mathrm{V})$, defined by 
\begin{equation}
(\bm{Q}_*(\bm{\omega}))(\bm{v}) = \bm{\omega}(\bm{Q}(\bm{v}))
\label{eq:AlgebraicAdjoint}
\end{equation}
for any $\bm{\omega} \in \mathrm{V}_*$ and $\bm{v} \in \mathrm{V}$.
Then
\[
\bm{e}_*'^i ( \bm{e}'_j ) = (\bm{Q}_*(\bm{e}_*^i))(\bm{P}(\bm{e}_j)) 
						= \bm{e}_*^i((\bm{Q}\bm{P})(\bm{e}_j)).
\]
Thus imposing $\bm{e}_*'^i ( \bm{e}'_j ) = \bm{e}_*^i(\bm{e}_j) = \delta^i_j$ requires $\bm{Q} = \bm{P}^{-1}$, that is, that the dual covector basis transforms inversely (in this adjoint sense) to the vector basis.
Then, with expansions $\bm{v} = \bm{e}'_b \, v'^b$ and $\bm{\omega} = \omega'_a \, \bm{e}_*'^a$ in terms of the new bases,
\[
\bm{\omega}(\bm{v}) 
	= \omega_a \, v^a
	= \omega'_a \, v'^a,
\]
that is, not only the value but also the algebraic form of the canonical pairing in terms of components is preserved.

This introduction of the dual space is expressed in a manner suggestive of an extension of the matrix notation associated with $\mathrm{V}$ to embrace the dual space $\mathrm{V}_*$ as well.
Apply the algebraic adjoint relation of Eq.~(\ref{eq:AlgebraicAdjoint}) to elements of a basis of $\mathrm{V}$ and of its dual basis of $\mathrm{V}_*$:
\[
(\bm{Q}_*(\bm{e}_*^i))(\bm{e}_j) 
	= \bm{e}_*^i(\bm{Q}(\bm{e}_j)).
\]
Expanding the transformed basis elements in terms of the original bases as $\bm{Q}(\bm{e}_j) =  \bm{e}_b \, {Q^b}_j$ and $\bm{Q}_*(\bm{e}_*^i) =  {(Q_*)^i}_a \, \bm{e}_*^a$, this equation yields $ {(Q_*)^i}_j =  {Q^i}_j$. 
That is, the same coefficients appear in the transformations of a vector basis and a covector basis, but with the summations performed over opposite indices: in comparison and contrast with Eq.~(\ref{eq:TransformedBasisExpansion}) is 
\[
\bm{e}_*'^i = \bm{Q}_*(\bm{e}_*^i) =  {Q^i}_a \, \bm{e}_*^a.
\]
Thus, in a manner converse or dual to matrix representations associated with $\mathrm{V}$, a natural extension of the matrix notation to $\mathrm{V}_*$ begins with writing a basis of $\mathrm{V}_*$ as an $n$-column
\[
\bm{\mathsf{B}}_* = \begin{bmatrix} \bm{e}_*^1 \\ \vdots \\ \bm{e}_*^n \end{bmatrix},
\] 
and taking $\sfomega \in \mathbb{R}^{1 \times n}$ to be the $n$-row
\[
\sfomega = \begin{bmatrix} \omega_1 & \dots & \omega_n \end{bmatrix}
\]
collecting the components of a covector $\bm{\omega}$, so that $\bm{\omega} = \omega_a \, \bm{e}_*^a = \sfomega \, \bm{\mathsf{B}}_*$ is given by matrix multiplication.
The covector basis transforms according to $\bm{\mathsf{B}}'_* = \mathsf{Q} \, \bm{\mathsf{B}}_*$, where the matrix $\left[ {Q^i}_j \right] = \mathsf{Q} \in \mathrm{GL}(n)$ collects the transformation coefficients.
Then the dual basis relation $\bm{e}_*^i ( \bm{e}_j ) = \delta^i_j$ reads 
\[
\bm{\mathsf{B}}_* \, \bm{\mathsf{B}} = \mymathsf{1},
\] 
that is, the outer product of the $n$-column $\bm{\mathsf{B}}_*$ and the $n$-row $\bm{\mathsf{B}}$ (with element-by-element evaluation of basis covectors on basis vectors) yields the $n \times n$ identity matrix.
The demand that the dual basis relationship be preserved under simultaneous transformations $\bm{P} \in \mathrm{GL}(\mathrm{V})$ and $\bm{Q} \in \mathrm{GL}(\mathrm{V}_*)$ reads
\[
\mymathsf{1} = \bm{\mathsf{B}}'_* \, \bm{\mathsf{B}}' 
	= \mathsf{Q} \, \bm{\mathsf{B}}_* \, \bm{\mathsf{B}} \, \mathsf{P}  \\
	= \mathsf{Q} \, \mathsf{P},
\]
yielding the matrix relation $\mathsf{Q} = \mathsf{P}^{-1}$.
The simultaneous transformations 
\begin{equation}
\begin{aligned}
\bm{\mathsf{B}}' &= \bm{\mathsf{B}} \, \mathsf{P},  \\
\bm{\mathsf{B}}'_* &= \mathsf{P}^{-1} \, \bm{\mathsf{B}}_*
\end{aligned}
\label{eq:BasisChanges}
\end{equation}
of a vector basis and its dual covector basis will be adopted henceforth. 
For the covector $\bm{\omega} \in \mathrm{V}_*$ represented by the $n$-row $\sfomega$ relative to basis $\bm{\mathsf{B}}_*$, or by $\sfomega'$ relative to basis $\bm{\mathsf{B}}'_* = \mathsf{P}^{-1} \, \bm{\mathsf{B}}_*$,
\[
\bm{\omega} = \sfomega \, \bm{\mathsf{B}}_* = \sfomega' \, \bm{\mathsf{B}}'_* = \sfomega' \, \mathsf{P}^{-1} \, \bm{\mathsf{B}}_*,
\]
where $\mathsf{P} \in \mathrm{GL}(n)$ and therefore $\mathsf{P}^{-1} \in \mathrm{GL}(n)$ as well.
This implies
\begin{equation}
\sfomega = \sfomega' \, \mathsf{P}^{-1},
\label{eq:CovectorRows}
\end{equation}
to be compared and contrasted with Eq.~(\ref{eq:VectorColumns}) for the transformation of $n$-columns representing vectors.
The canonical pairing may be written alternatively as $\bm{\omega}(\bm{v}) = \sfomega \, \bm{\mathsf{B}}_* \, \bm{\mathsf{B}} \, \mathsf{v} = \sfomega \, \mathsf{v}$ or $\bm{\omega}(\bm{v}) = \sfomega' \, \bm{\mathsf{B}}'_* \, \bm{\mathsf{B}}' \, \mathsf{v}' = \sfomega' \, \mathsf{v}'$.
This is consistent with using Eqs.~(\ref{eq:VectorColumns}) and (\ref{eq:CovectorRows}) to write
\[
\bm{\omega}(\bm{v}) = \sfomega \, \mathsf{v} = \sfomega' \, \mathsf{P}^{-1} \, \mathsf{P} \, \mathsf{v}' = \sfomega' \, \mathsf{v}', 
\] 
which expresses the invariance of the canonical pairing under basis changes directly in terms of matrix products of $n$-rows and $n$-columns, without explicit reference to bases.

A natural isomorphism between $\mathrm{V}$ and the space $\mathrm{V}_{**}$ dual to $\mathrm{V}_*$ opens the way to more general tensors on $\mathrm{V}$.
The idea is to reverse the canonical pairing and regard vectors $\bm{v} \in \mathrm{V}$ as linear forms on covectors $\bm{\omega} \in \mathrm{V}_*$:
\[
\bm{v}(\bm{\omega}) 
	= v^a \,  \bm{e}_a ( \bm{e}_*^b ) \, \omega_b
	= v^a \, \omega_a
	= \omega_a \, v^a
	= \bm{\omega}(\bm{v}),
\]
with $\bm{e}_i ( \bm{e}_*^j ) = \delta_i^j$ expressing the dual basis relationship in reverse.
In matrix notation,
\[
\bm{v}(\bm{\omega}) 
	= ( \bm{\mathsf{B}} \, \mathsf{v} )^\mathrm{T} \, ( \sfomega \, \bm{\mathsf{B}}_* )^\mathrm{T}
	= \mathsf{v}^\mathrm{T} (\bm{\mathsf{B}}_* \, \bm{\mathsf{B}} )^\mathrm{T} \sfomega^\mathrm{T} 
	= \mathsf{v}^\mathrm{T} \sfomega^\mathrm{T}.
\]
Furthermore,
\[
\bm{v}(\bm{\omega}) 
	= \mathsf{v}^\mathrm{T} \sfomega^\mathrm{T} 
	= ( \sfomega \, \mathsf{v})^\mathrm{T}
	= \sfomega \, \mathsf{v}
	= \bm{\omega}(\bm{v}).
\]
A tensor on $\mathrm{V}$ of type $(r,s)$ is a multilinear function that accepts $r$ covectors and $s$ vectors as arguments and returns a real number; it is expressed in terms of basis elements that are tensor products of $r$ factors of $\bm{e}_i$ and $s$ factors of $\bm{e}_*^j$.
In particular, a vector $\bm{v} \in \mathrm{V}$ is a $(1,0)$ tensor, and a covector $\bm{\omega} \in \mathrm{V}_*$ is a $(0,1)$ tensor.

Matrix representations readily accommodate the various possible linear tensors of degree two.
For instance, the tensors $\bm{F} = F_{a b} \, \bm{e}_*^a \otimes \bm{e}_*^b$, $\bm{L} = {L^a}_b \, \bm{e}_a \otimes \bm{e}_*^b$, and $\bm{T} = T^{a b} \, \bm{e}_a \otimes \bm{e}_b$ are of respective types $(0,2)$, $(1,1)$, and $(2,0)$, with components $F_{i j} = \bm{F}( \bm{e}_i, \bm{e}_j)$, ${L^i}_j = \bm{L}( \bm{e}_*^i, \bm{e}_j)$, and $T^{i j} = \bm{T}( \bm{e}_*^i, \bm{e}_*^j)$.
They can be expressed in terms of the vector basis $n$-row $\bm{\mathsf{B}}$, the dual covector basis $n$-column $\bm{\mathsf{B}}_*$, and their matrix transposes as
\[
\begin{array}{rclcl}
\bm{F} &=& \left[ \bm{e}_*^a \right]^\mathrm{T}  \left[ F_{a b} \right]  \left[ \bm{e}_*^b \right]
	&=&  \bm{\mathsf{B}}_*^\mathrm{T} \, \mathsf{F} \, \bm{\mathsf{B}}_*, \\[5pt]
\bm{L} &=&  \left[ \bm{e}_a \right] \left[ {L^a}_ b \right] \left[ \bm{e}_*^b \right]
	&=&  \bm{\mathsf{B}} \, \mathsf{L} \, \bm{\mathsf{B}}_*,  \\[5pt]
\bm{T} &=& \left[ \bm{e}_a \right] \left[ T^{a b} \right] \left[ \bm{e}_b \right]^\mathrm{T}
	&=&  \bm{\mathsf{B}} \, \mathsf{T} \, \bm{\mathsf{B}}^\mathrm{T} 
\end{array}
\]
where element-by-element tensor products of basis elements are understood, and the $n \times n$ matrices $\mathsf{F} = \left[ F_{i j} \right]$, $\mathsf{L} = \left[ {L^i}_ j \right]$, and $\mathsf{T} = \left[ T^{i j} \right]$, all elements of $\mathbb{R}^{n \times n}$, collect the respective tensor components.
One sees immediately that these matrices transform as
\begin{align*}
\mathsf{F} &= \mathsf{P}^{-\mathrm{T}} \, \mathsf{F}' \, \mathsf{P}^{-1}, \\
\mathsf{L} &= \mathsf{P} \, \mathsf{L}' \, \mathsf{P}^{-1}, \\
\mathsf{T} &= \mathsf{P} \, \mathsf{T}' \, \mathsf{P}^\mathrm{T}
\end{align*} 
under the basis changes of Eq.~(\ref{eq:BasisChanges}), where $\mathsf{P}^{-\mathrm{T}} = \left(\mathsf{P}^{-1} \right)^\mathrm{T} = \left(\mathsf{P}^\mathrm{T} \right)^{-1}$. 
As seen above with the evaluation of covectors on vectors and vice-versa, the evaluation of tensors of degree two on their vector and covector arguments becomes a matter of matrix multiplication.
Let $\bm{u}, \bm{v} \in \mathrm{V}$, with $\mathsf{u}, \mathsf{v} \in \mathbb{R}^{n \times 1}$ the $n$-columns collecting their components with respect to basis $\bm{\mathsf{B}}$.
Let $\bm{\psi}, \bm{\omega} \in \mathrm{V}_*$, with $\sfpsi, \sfomega \in \mathbb{R}^{1 \times n}$ the $n$-rows collecting their components with respect to dual basis $\bm{\mathsf{B}}_*$.
Then for example
\begin{align*}
\bm{F}( \bm{u}, \bm{v} ) 
	&= \left( \bm{\mathsf{B}} \, \mathsf{u} \right)^\mathrm{T} 
		\left( \bm{\mathsf{B}}_*^\mathrm{T} \, \mathsf{F} \, \bm{\mathsf{B}}_* \right)
		\left( \bm{\mathsf{B}} \, \mathsf{v} \right) \\
	&= \mathsf{u}^\mathrm{T} 
		\left( \bm{\mathsf{B}}_* \, \bm{\mathsf{B}} \right)^\mathrm{T}
		\mathsf{F}
		\left( \bm{\mathsf{B}}_* \, \bm{\mathsf{B}} \right)
		\mathsf{v} \\
	&=	\mathsf{u}^\mathrm{T} \, \mathsf{F} \, \mathsf{v},
\end{align*} 
and similarly
\begin{align*}
\bm{L}( \bm{\omega}, \bm{v} )
	&= \sfomega \, \mathsf{L} \, \mathsf{v}, \\ 
\bm{T}( \bm{\omega}, \bm{\psi} )
	&= \sfomega \, \mathsf{T} \, \sfpsi^\mathsf{T}, 
\end{align*}
these last expressions bypassing explicit reference to bases.

When the `tensor slots' in question are unambiguous, an infix dot operator ($\cdot$) between two tensors will denote tensor evaluation, or contraction, via a natural `pairing between lower and upper indices'.
Examples with tensors from the previous paragraph include $\bm{\omega}( \bm{v} ) = \bm{\omega} \cdot \bm{v}$ and $\bm{F}( \bm{u}, \bm{v} ) = \bm{u} \cdot \bm{F} \cdot \bm{v}$.
This provides an additional notational link between coordinate-free and matrix notations.
In this work the dot operator will never denote a scalar product, if such exists, between two vectors; such will always be expressed explicitly in terms of the metric tensor defining the scalar product. 

\subsection{An affine space as a differentiable manifold}
\label{sec:AffineSpaceManifold}

Finally, note that an affine space $\mathbb{A}$ can be regarded as a differentiable manifold, with the one-to-one correspondence $\mathbf{x} \leftrightarrow ( x^1, \dots, x^n)$ between $\mathbb{A}$ and $\mathbb{R}^n$ established by Eq.~(\ref{eq:PointExpansion}) constituting an atlas with a single global chart (global coordinate system).
While innumerable other (collections of) coordinate systems are available via the maximal atlas, the ones of the type given by Eq.~(\ref{eq:PointExpansion}) in terms of the affine structure of $\mathbb{A}$ are special. 
This is because the coordinate curves of coordinate $x^i$, parametrized by $x^i$ itself and characterized by constant $x^j$ for $j \ne i$, are straight lines in $\mathbb{A}$:
because of Eq.~(\ref{eq:PointExpansion}), at every point $\mathbf{x} \in \mathbb{A}$ the coordinate basis $( \partial / \partial x^1, \dots, \partial / \partial x^n )$ of the tangent space $T_\mathbf{x}(\mathbb{A})$ is equal to the single basis 
$( \bm{e}_1, \dots, \bm{e}_n )$ of the vector space $\mathrm{V}$ underlying $\mathbb{A}$ itself.
The coordinate basis vectors not varying between the tangent spaces at neighboring points, the connection $\bm{\nabla}$ defining a covariant derivative is trivial, with vanishing connection coefficients.\footnote{It might be said that, rather than independent isomorphisms between $T_\mathbf{x}(\mathbb{A})$ and $\mathbb{R}^n$ for each point $\mathbf{x} \in \mathbb{A}$ as would be the case for arbitrary charts on arbitrary differentiable manifolds, for \textit{these} global charts on an \textit{affine space} regarded as a differentiable manifold, the tangent spaces $T_\mathbf{x}(\mathbb{A})$ and $\mathrm{V}$ are all related to $\mathbb{R}^n$ by the same isomorphism, such that in effect they can be simultaneously identified with $\mathbb{R}^n$ and therefore with each other.
In mathematical terms, the tangent bundle $T(\mathbb{A})$ is `parallelizable': it is a `trivial bundle', expressed not just locally but globally as the Cartesian product $\mathbb{A} \times \mathrm{V}$.}
Thus, even without a (pseudo-)metric and therefore not a (pseudo-)Riemann manifold, already by virtue of its affine structure $\mathbb{A}$ naturally possesses a connection $\bm{\nabla}$---normally called an `affine connection' for presumably understandable historical reasons that apparently here come full circle---and may be said to have vanishing curvature, and to be flat. 

A tensor field of type $(r,s)$ on $\mathbb{A}$ is a differentiable mapping that assigns to each point $\mathbf{x} \in \mathbb{A}$ a tensor on $\mathrm{V}$ of type $(r,s)$.
On occasion there may be a slippage of precision in which the distinction between a tensor on $\mathrm{V}$ and a corresponding tensor field on $\mathbb{A}$ is not carefully maintained, but in general this causes no mischief. 

When additional structure on $\mathbb{A}$ defines a preferred class of bases of $\mathrm{V}$, any such basis providing the same normalization, one can define a volume form $\bm{\varepsilon}$, an alternating form of top degree (an $n$-form).
Let the basis $\bm{\mathsf{B}} =  \begin{bmatrix} \bm{e}_1 & \dots & \bm{e}_n \end{bmatrix}$ be an instance of the preferred class.
For a tensor of degree $n$ on the $n$-dimensional vector space $\mathrm{V}$, the specification
\begin{equation}
\bm{\varepsilon}( \bm{e}_1, \dots, \bm{e}_n ) = 1
\label{eq:VolumeForm}
\end{equation}
completely defines the tensor because of the total antisymmetry.
Expanded in terms of the dual of the preferred basis, 
\[
\bm{\varepsilon} = \varepsilon_{a_1 \dots a_n} \, \bm{e}_*^{a_1} \otimes \dots \otimes \bm{e}_*^{a_n},
\]
or, in terms of the wedge product (antisymmetrized tensor product),
\begin{align*}
\bm{\varepsilon} &= \varepsilon_{0 \dots n} \, \bm{e}_*^{0} \wedge \dots \wedge \bm{e}_*^{n} \\[5pt]
	&= \frac{1}{n!} \, \varepsilon_{a_1 \dots a_n} \, \bm{e}_*^{a_1} \wedge \dots \wedge \bm{e}_*^{a_n}.
\end{align*}
With respect to this preferred dual basis, the components are 
\[
\varepsilon_{a_1 \dots a_n} = \left[ a_1 \dots a_n \right],
\] 
with the expression in brackets being the alternating (permutation) symbol of degree $n$.
Given a more general basis $\bm{\mathsf{B}}' = \bm{\mathsf{B}} \, \mathsf{P}$ of the same orientation ($\det \mathsf{P} > 0$), 
\begin{equation}
\bm{\varepsilon}( \bm{e}'_1, \dots, \bm{e}'_n ) = \det \mathsf{P},
\label{eq:VolumeFormTransformed}
\end{equation}
and the components become 
\[
\varepsilon'_{a_1 \dots a_n} = \det\mathsf{P} \, \left[ a_1 \dots a_n \right].
\]
A volume form $\bm{\varepsilon}$ is normally introduced in connection with a metric and called the `Levi-Civita tensor'; in this case the preferred class of bases is typically taken to be orthonormal with respect to the metric.
Note however that a suitable preferred class of bases, however it arises, is sufficient to define $\bm{\varepsilon}$, even in the absence of a metric.

While an affine space possesses a connection, in this paper exterior differentiation of alternating form fields will mostly suffice.
An alternating form field of degree $p$, or $p$-form, is a completely antisymmetric tensor field of type $(0,p)$.
Let $( \bm{e}_*^1, \dots, \bm{e}_*^n ) = ( \mathbf{d}x^1, \dots, \mathbf{d}x^n )$ be the 1-form basis dual to the coordinate basis $( \bm{e}_1, \dots, \bm{e}_n ) = ( \partial / \partial x^1, \dots, \partial / \partial x^n )$.
A $p$-form $\bm{F}$ can be expanded as
\begin{align*}
\bm{F} &= \sum_{a_1 < \dots < a_p} F_{a_1 \dots a_p} \, \bm{e}_*^{a_1} \wedge \dots \wedge \bm{e}_*^{a_p} \\[5pt] 
	&= F_{\underaccent{\rightharpoondown}{\mathscr{A}}} \, \bm{e}_*^\mathscr{A},
\end{align*}
where the second equality defines a multi-index notation.
The exterior derivative of $\bm{F}$ is a $(p+1)$-form whose components are ordinary partial derivatives of the components of $\bm{F}$:
\begin{align*}
\mathbf{d}\bm{F} &= \sum_{b = 1}^n \sum_{a_1 < \dots < a_p} \frac{\partial}{\partial x^b} F_{a_1 \dots a_p} \, \bm{e}_*^b \wedge \bm{e}_*^{a_1} \wedge \dots \wedge \bm{e}_*^{a_p} \\[5pt]
	&= \frac{\partial }{\partial x^b} F_{\underaccent{\rightharpoondown}{\mathscr{A}}} \, \bm{e}_*^b \wedge \bm{e}_*^\mathscr{A}.
\end{align*}
The exterior derivative satisfies $\mathbf{d} \, \mathbf{d} \bm{F} = 0$ or $\mathbf{d}^2 = 0$ thanks to the symmetry of mixed partial derivatives.
The exterior derivative operator can be written symbolically as
\begin{equation}
\mathbf{d} = \bm{e}_*^a \wedge \frac{\partial}{\partial x^a},
\label{eq:ExteriorDerivativeOperator}
\end{equation}
where the partial derivative acts only on the components and the wedge product with $\bm{e}_*^a$ passes through.
Given a volume form $\bm{\varepsilon}$ and a vector field $\bm{U}$, the expression
\begin{equation}
\mathbf{d}(\bm{U} \cdot \bm{\varepsilon}) = ( \bm{\nabla} \cdot \bm{U} ) \, \bm{\varepsilon}
\label{eq:DivergenceDefinition}
\end{equation}
defines $\bm{\nabla} \cdot \bm{U}$, the divergence of $\bm{U}$.

\begin{adjustwidth}{-\extralength}{0cm}

\reftitle{References}


\def\prd{Phys. Rev. D}

\PublishersNote{}
\end{adjustwidth}

\begin{thebibliography}{999}

\bibitem[{Penrose}(2004)]{Penrose2004The-Road-to-Rea}
{Penrose}, R.
\newblock {\em {The Road to Reality: A Complete Guide to the Laws of the
  Universe}}; Jonathan Cape: London,  2004.

\bibitem[{Gourgoulhon}(2013)]{Gourgoulhon2013Special-Relativ}
{Gourgoulhon}, E.
\newblock {\em {Special Relativity in General Frames}}; Graduate Texts in
  Physics, Springer-Verlag: Berlin Heidelberg,  2013.

\bibitem[Cardall(2023)]{Cardall2023Towards-full-Ga}
Cardall, C.Y.
\newblock {Towards full `Galilei general relativity': Bargmann-Minkowski and
  Bargmann-Galilei spacetimes}.
\newblock In Proceedings of the Geometric Science of Information; Nielsen, F.;
  Barbaresco, F., Eds., Cham,  2023; Vol. 14072, {\em Lecture Notes in Computer
  Science}, pp. 69--78,  \href{http://arxiv.org/abs/2305.18428}{{\normalfont
  [arXiv:physics.gen-ph/2305.18428]}}.
\newblock {\url{https://doi.org/10.1007/978-3-031-38299-4{\_}8}}.

\bibitem[{Weyl}(1922)]{Weyl1922Space---Time---}
{Weyl}, H.
\newblock {\em Space---Time---Matter}, 4th ed.; Methuen \& Co.: London,  1922.

\bibitem[Cartan(1923)]{Cartan1923Sur-les-variete}
Cartan, {\'E}.
\newblock {Sur les vari{\'e}t{\'e}s {\`a} connexion affine et la th{\'e}orie la
  relativit{\'e} g{\'e}n{\'e}ralis{\'e}e. (premi{\`e}re partie)}.
\newblock {\em Annales Sci. Ecole Norm. Sup.} {\bf 1923}, {\em 40},~325--412.

\bibitem[Cartan(1924)]{Cartan1924Sur-les-variete}
Cartan, {\'E}.
\newblock {Sur les vari{\'e}t{\'e}s {\`a} connexion affine et la th{\'e}orie de
  la relativit{\'e} g{\'e}n{\'e}ralis{\'e}e. (suite)}.
\newblock {\em Annales Sci. Ecole Norm. Sup.} {\bf 1924}, {\em 41},~1--25.

\bibitem[Cartan(1986)]{Cartan1986On-manifolds-wi}
Cartan, {\'E}.
\newblock {\em On manifolds with an affine connection and the theory of general
  relativity}; Bibliopolis: Napoli,  1986.

\bibitem[{Toupin}(1957)]{Toupin1957World-invariant}
{Toupin}, R.A.
\newblock {World invariant kinematics}.
\newblock {\em Archive for Rational Mechanics and Analysis} {\bf 1957}, {\em
  1},~181--211.

\bibitem[{Truesdell} and {Toupin}(1960)]{Truesdell1960The-Classical-F}
{Truesdell}, C.; {Toupin}, R.
\newblock {The Classical Field Theories}. In {\em {Principles of Classical
  Mechanics and Field Theory}}; {Fl\"ugge}, S., Ed.; Springer-Verlag: Berlin
  G\"ottingen Heidelberg,  1960; Vol. III/1, {\em Encyclopedia of Physcs}, pp.
  226--793.

\bibitem[Havas(1964)]{Havas1964Four-Dimensiona}
Havas, P.
\newblock {Four-Dimensional Formulations of Newtonian Mechanics and Their
  Relation to the Special and the General Theory of Relativity}.
\newblock {\em Rev. Mod. Phys.} {\bf 1964}, {\em 36},~938--965.
\newblock {\url{https://doi.org/10.1103/RevModPhys.36.938}}.

\bibitem[{Trautman}(1965)]{Trautman1965Foundations-and}
{Trautman}, A.
\newblock {Foundations and Current Problems of General Relativity}. In {\em
  {Lectures on General Relativity}}; {Trautman}, A.; {Pirani}, F.A.E.; {Bondi},
  H., Eds.; Prentice-Hall: Englewood Cliffs,  1965; pp. 1--248.

\bibitem[{Trautman}(1966)]{Trautman1966Comparison-of-N}
{Trautman}, A.
\newblock {Comparison of Newtonian and Relativistic Theories of Space-Time}. In
  {\em Perspectives in Geometry and Relativity: Essays in Honor of V\'aclav
  Hlavat\'y}; Hoffmann, B., Ed.; Indiana University Press: Bloomington,  1966;
  chapter~42, pp. 413--425.

\bibitem[{K{\"u}nzle}(1972)]{Kunzle1972Galilei-and-Lor}
{K{\"u}nzle}, H.P.
\newblock {Galilei and Lorentz structures on space-time: comparison of the
  corresponding geometry and physics}.
\newblock {\em Annales de l'I. H. P., section A} {\bf 1972}, {\em
  17},~337--362.

\bibitem[Bargmann(1954)]{Bargmann1954On-Unitary-ray-}
Bargmann, V.
\newblock {On Unitary Ray Representations of Continuous Groups}.
\newblock {\em Annals Math.} {\bf 1954}, {\em 59},~1--46.

\bibitem[{L{\'e}vy-Leblond}(1971)]{Levy-Leblond1971Galilei-Group-a}
{L{\'e}vy-Leblond}, J.M.
\newblock {Galilei Group and Galilean Invariance}. In {\em Group Theory and Its
  Applications}; Loebl, E.M., Ed.; Academic Press: New York,  1971; Vol.~II.

\bibitem[{L{\'e}vy-Leblond}(1974)]{Levy-Leblond1974The-pedagogical}
{L{\'e}vy-Leblond}, J.M.
\newblock {The pedagogical role and epistemological significance of group
  theory in quantum mechanics}.
\newblock {\em {Nuovo Cimento Rivista Serie}} {\bf 1974}, {\em 4},~99--143.

\bibitem[{L{\'e}vy-Leblond}(1976)]{Levy-Leblond1976Quantum-fact-an}
{L{\'e}vy-Leblond}, J.M.
\newblock {Quantum fact and classical fiction: Clarifying Land{\'e}'s
  pseudo‐paradox}.
\newblock {\em Am. J. Phys.} {\bf 1976}, {\em 44},~1130--1132.

\bibitem[Omote et~al.(1989)Omote, Kamefuchi, Takahashi, and
  Ohnuki]{Omote1989Galilean-Covari}
Omote, M.; Kamefuchi, S.; Takahashi, Y.; Ohnuki, Y.
\newblock {Galilean Covariance and the Schr{\"o}dinger Equation}.
\newblock {\em Fortschritte der Physik/Progress of Physics} {\bf 1989}, {\em
  37},~933--950.
\newblock {\url{https://doi.org/https://doi.org/10.1002/prop.2190371203}}.

\bibitem[Souriau(1970)]{Souriau1970Structure-des-s}
Souriau, J.M.
\newblock {\em Structure des syst{\`e}mes dynamiques}; Dunod: Paris,  1970.

\bibitem[Souriau(1997)]{Souriau1997Structure-of-Dy}
Souriau, J.M.
\newblock {\em Structure of Dynamical Systems: A Symplectic View of Physics};
  Vol. 149, {\em 149}, Birkh{\"a}user: Boston,  1997.

\bibitem[{Duval} et~al.(1985){Duval}, {Burdet}, {K{\"u}nzle}, and
  {Perrin}]{Duval1985Bargmann-struct}
{Duval}, C.; {Burdet}, G.; {K{\"u}nzle}, H.P.; {Perrin}, M.
\newblock {Bargmann structures and Newton-Cartan theory}.
\newblock {\em \prd} {\bf 1985}, {\em 31},~1841--1853.

\bibitem[{K\"unzle} and {Duval}(1986)]{Kunzle1986Relativistic-an}
{K\"unzle}, H.P.; {Duval}, C.
\newblock {Relativistic and non-relativistic classical field theory on
  five-dimensional spacetime}.
\newblock {\em Class. Quant. Grav.} {\bf 1986}, {\em 3},~957.
\newblock {\url{https://doi.org/10.1088/0264-9381/3/5/024}}.

\bibitem[Duval et~al.(1991)Duval, Gibbons, and
  Horv\'athy]{Duval1991Celestial-mecha}
Duval, C.; Gibbons, G.; Horv\'athy, P.
\newblock Celestial mechanics, conformal structures, and gravitational waves.
\newblock {\em Phys. Rev. D} {\bf 1991}, {\em 43},~3907--3922.
\newblock {\url{https://doi.org/10.1103/PhysRevD.43.3907}}.

\bibitem[de~Montigny et~al.(2003{\natexlab{a}})de~Montigny, Khanna, and
  Santana]{Montigny2003Lorentz-like-co}
de~Montigny, M.; Khanna, F.C.; Santana, A.E.
\newblock {Lorentz-like covariant equations of non-relativistic fluids}.
\newblock {\em Journal of Physics A: Mathematical and General} {\bf 2003}, {\em
  36},~2009.
\newblock {\url{https://doi.org/10.1088/0305-4470/36/8/301}}.

\bibitem[de~Montigny et~al.(2003{\natexlab{b}})de~Montigny, Khanna, and
  Santana]{Montigny2003Nonrelativistic}
de~Montigny, M.; Khanna, F.C.; Santana, A.E.
\newblock Nonrelativistic Wave Equations with Gauge Fields.
\newblock {\em International Journal of Theoretical Physics} {\bf 2003}, {\em
  42},~649--671.
\newblock {\url{https://doi.org/10.1023/A:1024485810807}}.

\bibitem[{de Saxc\'e} and {Vall\'ee}(2012)]{de-Saxce2012Bargmann-group-}
{de Saxc\'e}, G.; {Vall\'ee}, C.
\newblock {Bargmann group, momentum tensor and Galilean invariance of
  Clausius-Duhem inequality}.
\newblock {\em International Journal of Engineering Science} {\bf 2012}, {\em
  50},~216--232.

\bibitem[{de Saxc\'e} and {Vall\'ee}(2016)]{de-Saxce2016Galilean-Mechan}
{de Saxc\'e}, G.; {Vall\'ee}, C.
\newblock {\em Galilean Mechanics and Thermodynamics of Continua}; John Wiley
  \& Sons, Inc.: Hoboken,  2016.

\bibitem[{de Saxc\'e}(2017)]{de-Saxce20175-Dimensional-T}
{de Saxc\'e}, G.
\newblock {5-Dimensional Thermodynamics of Dissipative Continua}. In {\em
  Models, Simulation, and Experimental Issues in Structural Mechanics};
  {Fr\'emond}, M.; {Maceri}, F.; {Vairo}, G., Eds.; Springer: Cham,  2017;
  Vol.~8, {\em Springer Series in Solid and Structural Mechanics}, pp. 1--40.

\bibitem[Pinski(1968)]{Pinski1968Galilean-Tensor}
Pinski, G.
\newblock {Galilean Tensor Calculus}.
\newblock {\em Journal of Mathematical Physics} {\bf 1968}, {\em
  9},~1927--1930.
\newblock {\url{https://doi.org/10.1063/1.1664527}}.

\bibitem[{Cardall}(2019)]{Cardall2019Minkowski-and-G}
{Cardall}, C.Y.
\newblock {Minkowski and Galilei/Newton Fluid Dynamics: A Geometric 3+1
  Spacetime Perspective}.
\newblock {\em Fluids} {\bf 2019}, {\em 4},~1.
\newblock {\url{https://doi.org/10.3390/fluids4010001}}.

\bibitem[{Cardall}(2020)]{Cardall2020Combining-3-Mom}
{Cardall}, C.Y.
\newblock {Combining 3-Momentum and Kinetic Energy on Galilei/Newton
  Spacetime}.
\newblock {\em Symmetry} {\bf 2020}, {\em 12},~1775.
\newblock {\url{https://doi.org/10.3390/sym12111775}}.

\bibitem[Le~Bellac and L{\'e}vy-Leblond(1973)]{Le-Bellac1973Galilean-electr}
Le~Bellac, M.; L{\'e}vy-Leblond, J.M.
\newblock Galilean electromagnetism.
\newblock {\em Il Nuovo Cimento B (1971-1996)} {\bf 1973}, {\em 14},~217--234.
\newblock {\url{https://doi.org/10.1007/BF02895715}}.

\bibitem[K{\"u}nzle(1976)]{Kunzle1976Covariant-Newto}
K{\"u}nzle, H.P.
\newblock {Covariant Newtonian limit of Lorentz space-times}.
\newblock {\em General Relativity and Gravitation} {\bf 1976}, {\em
  7},~445--457.
\newblock {\url{https://doi.org/10.1007/BF00766139}}.

\bibitem[{Dixon}(1975)]{Dixon1975On-the-uniquene}
{Dixon}, W.G.
\newblock {On the uniqueness of the Newtonian theory as a geometric theory of
  gravitation}.
\newblock {\em Communications in Mathematical Physics} {\bf 1975}, {\em
  45},~167--182.

\bibitem[Dautcourt(1997)]{Dautcourt1997Post-Newtonian-}
Dautcourt, G.
\newblock {Post-Newtonian extension of the Newton - Cartan theory}.
\newblock {\em Classical and Quantum Gravity} {\bf 1997}, {\em 14},~A109.
\newblock {\url{https://doi.org/10.1088/0264-9381/14/1A/009}}.

\bibitem[Andringa et~al.(2011)Andringa, Bergshoeff, Panda, and
  de~Roo]{Andringa2011Newtonian-gravi}
Andringa, R.; Bergshoeff, E.; Panda, S.; de~Roo, M.
\newblock {Newtonian gravity and the Bargmann algebra}.
\newblock {\em Classical and Quantum Gravity} {\bf 2011}, {\em 28},~105011.
\newblock {\url{https://doi.org/10.1088/0264-9381/28/10/105011}}.

\bibitem[{Gourgoulhon}(2012)]{Gourgoulhon201231-Formalism-in}
{Gourgoulhon}, E.
\newblock {\em {3+1 Formalism in General Relativity: Bases of Numerical
  Relativity}}; Vol. 846, {\em Lecture Notes in Physics}, Springer: Berlin
  Heidelberg,  2012.

\bibitem[Geracie et~al.(2015)Geracie, Prabhu, and
  Roberts]{Geracie2015Curved-non-rela}
Geracie, M.; Prabhu, K.; Roberts, M.M.
\newblock {Curved non-relativistic spacetimes, Newtonian gravitation and
  massive matter}.
\newblock {\em Journal of Mathematical Physics} {\bf 2015}, {\em 56},~103505.
\newblock {\url{https://doi.org/10.1063/1.4932967}}.

\bibitem[Van~den Bleeken(2017)]{Van-den-Bleeken2017Torsional-Newto}
Van~den Bleeken, D.
\newblock {Torsional Newton--Cartan gravity from the large c expansion of
  general relativity}.
\newblock {\em Classical and Quantum Gravity} {\bf 2017}, {\em 34},~185004.
\newblock {\url{https://doi.org/10.1088/1361-6382/aa83d4}}.

\bibitem[Hansen et~al.(2019)Hansen, Hartong, and
  Obers]{Hansen2019Action-Principl}
Hansen, D.; Hartong, J.; Obers, N.A.
\newblock {Action Principle for Newtonian Gravity}.
\newblock {\em Phys. Rev. Lett.} {\bf 2019}, {\em 122},~061106.
\newblock {\url{https://doi.org/10.1103/PhysRevLett.122.061106}}.

\bibitem[Cariglia(2018)]{Cariglia2018General-theory-}
Cariglia, M.
\newblock {General theory of Galilean gravity}.
\newblock {\em Phys. Rev. D} {\bf 2018}, {\em 98},~084057.
\newblock {\url{https://doi.org/10.1103/PhysRevD.98.084057}}.

\bibitem[Hansen et~al.(2020)Hansen, Hartong, and
  Obers]{Hansen2020Non-relativisti}
Hansen, D.; Hartong, J.; Obers, N.A.
\newblock {Non-relativistic gravity and its coupling to matter}.
\newblock {\em Journal of High Energy Physics} {\bf 2020}, {\em 2020},~145.
\newblock {\url{https://doi.org/10.1007/JHEP06(2020)145}}.

\bibitem[{{Marek}, A. and {Dimmelmeier}, H. and {Janka}, {H.-T.} and
  {M{\"u}ller}, E. and {Buras},
  R.}(2006)]{Marek-A.-and-Dimmelmeier-H.-and-Janka-H.-T.-and-Muller-E.-and-Buras-R.2006Exploring-the-r}
{{Marek}, A. and {Dimmelmeier}, H. and {Janka}, {H.-T.} and {M{\"u}ller}, E.
  and {Buras}, R.}.
\newblock Exploring the relativistic regime with Newtonian hydrodynamics: an
  improved effective gravitational potential for supernova simulations.
\newblock {\em Astronomy \& Astrophysics} {\bf 2006}, {\em 445},~273--289.

\bibitem[{Frankel}(2012)]{Frankel2012The-Geometry-of}
{Frankel}, T.
\newblock {\em The Geometry of Physics: An Introduction}; Cambridge University
  Press: Cambridge,  2012.

\end{thebibliography}
\end{document}